\pdfoutput=1

\documentclass[11pt, reqno]{article}
\usepackage{epsfig}
\usepackage{amssymb}
\usepackage{amsmath}
\usepackage{mathrsfs}
\usepackage{cite}
\usepackage{hyperref}
\usepackage{multirow}

\newcommand{\be}{\begin{eqnarray}}
\newcommand{\ee}{\end{eqnarray}}

\newcommand{\smallminus}{{\rm\rule[2.4pt]{6pt}{0.65pt}}}
\newcommand{\smallplus}{\hspace{0.5pt}\text{{\small+}}\hspace{-0.5pt}}
\renewcommand{\cap}{\mathrm{\raisebox{0.75pt}{{$\,\bigcap\,$}}}}

\newcommand{\ab}[1]{\langle #1\rangle}

\makeindex
\begin{document}

\begin{titlepage}

\begin{center}

{\Large \bf The All-Loop Integrand For Scattering \\[1pt] Amplitudes in Planar ${\cal N}=4$ SYM\\[5pt]}

\vspace{0.3cm}

{\bf N. Arkani-Hamed$^a$, J. Bourjaily$^{a,b}$, F. Cachazo$^{a,c}$, S.
Caron-Huot$^a$, J. Trnka$^{a,b}$}

\vspace{.1cm}

{\it $^{a}$ School of Natural Sciences, Institute for Advanced Study, Princeton, NJ 08540, USA}

{\it $^{b}$ Department of Physics, Princeton University, Princeton, NJ 08544, USA}

{\it $^{c}$ Perimeter Institute for Theoretical Physics, Waterloo, Ontario N2J W29, CA}
\vspace{0.5cm}
\end{center}

\begin{abstract}

We give an explicit recursive formula for the all $\ell$-loop integrand for
scattering amplitudes in ${\cal N}=4$ SYM in the planar limit, manifesting the full Yangian symmetry of the theory. This
generalizes the BCFW recursion relation for tree amplitudes to all
loop orders, and extends the Grassmannian duality for leading
singularities to the full amplitude. It also provides a new physical picture for the
meaning of loops, associated with canonical operations for removing particles in a
Yangian-invariant way. Loop amplitudes arise from the ``entangled"
removal of pairs of particles, and are naturally presented as an integral over
lines in momentum-twistor space. As expected from manifest Yangian
invariance, the integrand is given as a sum over non-local terms, rather than the
familiar decomposition in terms of
local scalar integrals with rational coefficients. Knowing the integrands explicitly, it is straightforward to express them in local forms if desired; this turns out to be done most naturally using a novel basis of chiral, tensor integrals written in momentum-twistor space, each of which has unit leading singularities. As simple illustrative examples, we present a number of
new multi-loop results written in local form, including the 6- and 7-point 2-loop NMHV amplitudes. Very concise expressions are presented for all 2-loop MHV amplitudes, as well as
the 5-point 3-loop MHV amplitude. The structure of the loop
integrand strongly suggests that the integrals yielding the physical amplitudes are ``simple",
and determined by IR-anomalies. We briefly comment on extending these ideas to more general planar theories.

\end{abstract}

\bigskip
\bigskip

\end{titlepage}

\tableofcontents
\newpage

\section{The Loop Integrand for ${\cal N} = 4$ SYM Amplitudes}

Scattering amplitudes in gauge theories have extraordinary properties that are completely invisible in the textbook formulation of local quantum field theory. The earliest hint of this hidden structure was the remarkable simplicity of the Parke-Taylor formula for tree-level MHV amplitudes \cite{Parke:1986gb,Berends:1987me}. Witten's 2003 proposal of twistor string theory \cite{Witten:2003nn} gave a strong impetus to rapid developments in the field, inspiring the development of powerful new tools for computing tree amplitudes, including CSW diagrams \cite{Cachazo:2004kj} and BCFW  recursion relations \cite{Britto:2004ap, Britto:2005fq, Brandhuber:2008pf, ArkaniHamed:2008gz}. At one-loop, very efficient on-shell methods now exist \cite{Bern:2007dw, Berger:2008sj} and at higher-loop level generalizations of the unitarity-based method \cite{Bern:1994zx,Bern:1994cg,Bern:1996je,Bern:2004cz} have made a five-loop computation possible \cite{Bern:2007ct}, which should soon determine the five-loop cusp anomalous dimension \cite{fiveloopinProgress}. 

The BCFW recursion relations in particular presented extremely compact expressions for tree amplitudes using building blocks with both local and non-local poles. In a parallel development, an amazing hidden symmetry of planar ${\cal N}=4$ SYM---dual conformal invariance---was noticed first in multi-loop perturbative calculations \cite{Drummond:2006rz} and then at strong coupling \cite{Alday:2007hr}, leading to a remarkable connection between null-polygonal Wilson loops and scattering amplitudes \cite{Alday:2007hr,Alday:2007he,Brandhuber:2007yx,Drummond:2007aua,Drummond:2007cf,Drummond:2007bm,Drummond:2008aq,
Bern:2008ap,Drummond:2008vq,Alday:2009yn}. It was quickly realized that the BCFW form of the tree amplitudes manifested both full superconformal and {\it dual} superconformal invariance, which together close into an infinite-dimensional Yangian symmetry algebra \cite{Drummond:2009fd}.  Understanding the role of this remarkable integrable structure in the full quantum theory, however, was clouded by the IR-divergences that appear to almost completely destroy the symmetry at loop-level, leaving only the anomalous action of the (Bosonic) dual conformal invariance \cite{Drummond:2007au,Brandhuber:2009xz,Brandhuber:2009kh,Elvang:2009ya}.

\subsection{Grassmannian Duality for Leading Singularities}

In \cite{ArkaniHamed:2009dn}, a strategy for making progress on these questions was suggested. The idea was to find objects closely associated with scattering amplitudes which are completely free of IR-divergences; the action of the symmetries would be expected to be manifest on such objects, and they would provide ``data" that might be the output
of a putative dual theory of the S-Matrix.

The leading singularities of scattering amplitudes are precisely objects of this sort. Thinking of loop amplitudes as multi-dimensional complex integrals, leading singularities arise from performing the integration not on the usual non-compact `contours' over all real loop-momenta, but on compact contours `encircling' isolated (and generally complex) poles in momentum space.  As such, they are free of IR-divergences and well-defined at any loop order, yielding algebraic functions of the external momenta. Leading singularities were known to have strange inter-relationships and satisfy mysterious identities not evident in their field-theoretic definition. Morally they are also expected to be Yangian-invariant, although even this is not completely manifest\footnote{Indeed we will give a proof of this basic fact in the next section; a different argument for the same result is given in \cite{JJwip}.}.
Thus leading singularities seem to be prime candidates for objects to be understood and computed by a dual theory.
\newpage 
Such a duality was proposed in \cite{ArkaniHamed:2009dn}, connecting leading singularities of color-stripped,
$n$-particle N$^k$MHV scattering amplitudes in ${\cal N} = 4$ SYM to a simple
contour integral over the Grassmannian $G(k,n)$:\vspace{-0.4cm}
\begin{equation} \label{first} \mathcal{Y}_{n,k}({\cal Z}) = \frac{1}{{\rm vol} [{\rm GL}(k)]} \int
\frac{d^{\, k \times n} C_{\alpha a}}{(1 \cdots k) (2 \cdots
k\smallplus1) \cdots (n \cdots k\smallminus1)} \prod_{\alpha=1}^k
\delta^{4|4}(C_{\alpha a} {\cal Z}_a).\vspace{-0.1cm} \end{equation} Here $a=1,\cdots,n$ labels the external particles, and ${\cal Z}_a$ are variables in $\mathbb{CP}^{3|4}$. The original formulation of this object worked with twistor variables ${\cal W}_a = (W_a|
\widetilde \eta_a)$, and was given as ${\cal L}_{n,k+2}({\cal W}) = \mathcal{Y}_{n,k+2}({\cal W})$. This was quickly realized \cite{ArkaniHamed:2009vw} to be completely equivalent to a second form in \emph{momentum} twistor space
\cite{Mason:2009qx}, with ${\cal L}_{n,k+2}(\lambda,\widetilde \lambda, \widetilde \eta) = M^{\rm tree}_{{\rm MHV}} \times \mathcal{Y}_{n,k}({\cal Z})$.
Here the variables $\mathcal{Z}_a = (Z_a | \eta_a)$ are the ``momentum-twistors'' introduced by Hodges \cite{Hodges:2009hk},
which are the most natural variables with which to discuss {\it dual}
superconformal invariance.  Furthermore, these momentum twistors are simple algebraic functions of the external momenta, upon
which scattering amplitudes conventionally depend\footnote{
To quickly establish notation and conventions, the
momentum of particle $a$ is given by $p^\mu_a = x^\mu_{a+1} - x^\mu_a$, and the point $x^\mu_a$ in the dual co-ordinate space
is associated with the line $(Z_{a-1}\,Z_a)$ in the corresponding momentum-twistor space. This designation ensures that the lines $(Z_{a-1}\,Z_a)$ and
$(Z_a\,Z_{a+1})$ intersect, so that correspondingly, $x^\mu_{a+1} - x^\mu_a = p_a$ is null. (Bosonic) dual-conformal invariants are made with 4-brackets $
\ab{a\,\, b\,\, c\,\,d} = \epsilon_{IJKL} Z_a^I Z_b^J Z_c^K Z_d^L$. An important special case is $\ab{i\smallminus1\,\,i\,\,j\smallminus1\,\,j} = \ab{i\smallminus1\,\,i} \ab{j\smallminus1\,\,j} (x_j - x_i)^2$; 2-brackets $\ab{ij}$ are computed using the upper-two components of $Z_i,Z_j$ and cancel out in dual-conformal expressions.  For more detail see \cite{Hodges:2009hk,ArkaniHamed:2009vw,Mason:2009qx}.}.

Since the Grassmannian integral is invariant under both ordinary
and dual superconformal transformations, it enjoys the full Yangian symmetry of the theory, as has been
proven more directly in \cite{Drummond:2010qh}. In fact, it has been argued that
these contour integrals in $G(k,n)$ generates {\it all} Yangian invariants, \footnote{The residues of $G(k,n)$ are Yangian-invariant for generic momenta away from collinear limits. See \cite{Beisert:2010gn,Bargheer:2009qu} for important discussions of the fate of Yangian invariance in the presence of collinear singularities.}\cite{Drummond:2010uq, Korchemsky:2010ut}.

Leading singularities are associated with residues of the Grassmannian integral.
Residue theorems \cite{Griffiths:1978a} imply many non-trivial and otherwise mysterious
linear relations between leading singularities. These relations are associated with important physical properties such as locality and unitarity \cite{ArkaniHamed:2009dn}.

Further investigations \cite{ArkaniHamed:2009dg} identified a new principle, the Grassmannian ``particle interpretation", which determines the correct contour of integration yielding the BCFW form of tree amplitudes \cite{Drummond:2008cr}. Quite remarkably, a deformation of the integrand connects this formulation to twistor string theory \cite{NVW, ArkaniHamed:2009dg, Bourjaily:2010kw}. Furthermore, another contour deformation produces the CSW expansion of tree amplitudes \cite{ArkaniHamed:2009sx}, making the emergence of local space-time a derived consequence from the more primitive Grassmannian starting point.

The Grassmannian integral and Yangian-invariance go hand-in-hand and are essentially synonymous; indeed, the Grassmannian integral is the most concrete way of thinking about Yangian invariants, since not only the symmetries but also the non-trivial relationship between different invariants are made manifest; even connections to non-manifestly Yangian-invariant but important physical objects (such as CSW terms) are made transparent.

Given these developments, we are encouraged to ask again: is there an analogous structure underlying not just the leading singularities but the full loop amplitudes? Does Yangian-invariance play a role? And if so, how can we see this through the thicket of IR-divergences that appear to remove almost all traces of these remarkable symmetries in the final amplitudes?

\newpage
\subsection{The Planar Integrand}

Clearly, we need to focus again on finding well-defined objects associated with loop amplitudes. Fortunately, in {\it planar} theories, there is an extremely natural candidate: the loop {\it integrand} itself!

Now, in a general theory, the loop integrand is not obviously a well-defined object. Consider the case of 1-loop diagrams. Most trivially, in
summing over Feynman diagrams, there is no canonical way of
combining different 1-loop diagrams under the same integral sign, since
there is no natural origin for the loop-momentum space. The
situation is different in planar theories, however, and this ambiguity is
absent. This is easy to see in the dual $x$-space co-ordinates \cite{Drummond:2006rz}. The
ambiguity in shifting the origin of loop momenta is nothing other
than translations in $x$-space; but fixing the $x_1, \ldots, x_n$
of the external particles allows us to canonically combine all the
diagrams. Alternatively, in a planar theory it is possible to unambiguously define the loop momentum common to all diagrams to be the one which flows from particle ``1'' to particle ``2''.

At two-loops and above, we have a number of loop
integration variables in the dual space $x,y,\ldots,z$, and the well-defined loop integrand
is completely symmetrized in these variables.

So the loop integrand is well-defined in the planar limit, and any dual theory should be
able to compute it. All the symmetries of the theory should be
manifest at the level of the integrand, only broken by IR-divergences in actually carrying out the integration---the symmetries of the theory are broken only by the choice of integration contour.

\subsection{Recursion Relations for All Loop Amplitudes}

Given that the integrand is a well-defined, rational function of the loop variables and the external momenta, we should be able to determine it using BCFW recursion relations in the familiar way \footnote{We note that \cite{Sever:2009aa} have conjectured that the loop amplitudes can be determined by CSW rules, manifesting the superconformal invariance of the theory.}. At loop-level the poles have residues with different physical meaning. The first kind is the analog of tree-level poles and correspond to factorization channels. The second kind has no tree-level analog; these are single cuts whose residues are forward limits of lower-loop amplitudes. Forward limits are na\"{i}vely ill-defined operations but quite nicely they exist in any supersymmetric gauge theory, as was shown to one-loop level in \cite{CaronHuot:2010zt}.  There it was also argued that forward limits are well-defined to higher orders
in perturbation theory in ${\cal N}=4$ SYM. In principle, this is all we need for computing the loop integrand in ${\cal N}=4$ SYM to all orders in perturbation theory. However, our goal requires more than that. We would like to show that the integrand of the theory can be written in a form which makes all symmetries---the full Yangian---manifest. The Yangian-invariance of BCFW terms at tree-level becomes obvious once they are identified with residues of the Grassmannian integral, we would like to achieve the same at loop-level.

This is exactly what we will do in this paper. We will give an explicit recursive
construction of the all-loop integrand, in exact analogy to the BCFW
recursion relations for tree amplitudes, making the full
Yangian symmetry of the theory manifest.

The formulation also provides a new physical understanding of the
meaning of loops, associated with simple operations for
``removing" particles in a Yangian-invariant way. Loop amplitudes
are associated with removing pairs of particles in an ``entangled" way. We describe all these operations in momentum-twistor space, since this directly corresponds to familiar momentum-space loop integrals; presumably an ordinary twistor space description should also be possible.

As is familiar from the BCFW recursion relations at tree-level,
the integrand is expressed as a sum over non-local terms, in a form very different than the familiar ``rational function $\times$ scalar integral" presentation that is common in the literature. Nonetheless, the Yangian-invariance guarantees that every term in the loop amplitude has Grassmannian residues as its leading singularities.

The integrands can of course be expressed in a manifestly-local form if desired, and are most naturally written in momentum-twistor space \cite{Hodges:2010kq,Mason:2010pg}.
As we will see, the most natural basis of local integrands in which to express the answer is not composed of the familiar scalar loop-integrals, but is instead made up of chiral tensor integrals with unit leading-singularities, which makes the physics and underlying structure much more transparent.

Of course the integrand is a well-defined rational function which is computed in four-dimensions without any regulators. The regularization needed to carry out the integrations is a very physical one, given by moving out on the Coulomb branch \cite{Alday:2009zm} of the theory. This can be beautifully implemented, both conceptually and in practice, with the momentum-twistor space representation of the integrand \cite{Hodges:2010kq,Mason:2010pg}.

Quite apart from the conceptual advantages of this way of thinking about loops, our new formulation is also
completely systematic and practical, taking the ``art" out of the computation of multi-loop amplitudes in ${\cal N}=4$ SYM. As simple applications
of the general recursive formula, we present a number of new
multi-loop results, including the two-loop NMHV 6- and 7-particle integrands. We also include very concise, local expressions for all 2-loop
MHV integrands and for the 5-particle MHV integrand at 3-loops. All multiplicity results for the so-called ``parity even" part of two-loop amplitudes in the MHV sector were obtained by Vergu in \cite{Vergu:2009tu}, extending previous work done for 5-particles \cite{Bern:2007ct} and 6-particles \cite{Bern:2008ap,Cachazo:2008hp} in dimensional regularization. The ``parity even" part of the 6-particle amplitude in dimensional regularization has been computed in work in progress by Kosower, Roiban, and Vergu \cite{KRV}. Complete integrands have been computed at two-loop order for 5-particles in \cite{Bern:2007ct} using $D$-dimensional unitarity and for 5- and 6-particles in \cite{Cachazo:2008vp,Cachazo:2008hp} using the leading singularity technique developed in \cite{Buchbinder:2005wp,Cachazo:2008vp}. Also using the leading singularity technique, the 5-point 3-loop integrand was presented in \cite{Spradlin:2008uu}. Combining $D$-dimensional unitarity with a generalization of quadruple cuts to higher loop order \cite{Buchbinder:2005wp}, a method called maximal cuts was introduced in \cite{Bern:2007ct} and used for the computation of the 4-point 5-loop integrand. The 4-point amplitude integrand at $l=2,3,4$ loop-level were computed in \cite{Anastasiou:2003kj}, \cite{Bern:2005iz}, and \cite{Bern:2006ew}, respectively.  The method to be used in this paper is, however, very different both in philosophy and in practice from the leading singularity or generalized unitarity approaches.

In this paper, we give a brief and quite telegraphic outline of our arguments and results; we will present a much more detailed account of our methods and further elaborate on many of the themes presented here in upcoming work \cite{InPrep}. In \mbox{section \ref{canonical_operations_on_Yangian_invariants}}, we describe a number of canonical operations on Yangian invariants---adding and removing particles, fusing invariants---that generate a variety of important physical objects in our story. In \mbox{section \ref{loops_as_hidden_entanglement}} we describe the origin of Yangian-invariant loop integrals as arising from the ``hidden entanglement" of pairs of removed particles. In \mbox{section \ref{recursion_relations_for_loops}} we describe the main result of the paper: a generalization of the BCFW recursion relation to all loop amplitudes in the theory, and discuss some of its salient features through simple 1-loop examples. In \mbox{section \ref{local_loop_integrands}} we set the stage for presenting loop amplitudes in a manifestly local form by describing the most natural way of doing this in momentum-twistor space. In \mbox{section \ref{multiloop_examples}} we present a number of new multi-loop integrands computed using the recursion relation and translated into local form for the convenience of comparing with known results where they are available. We conclude in section \ref{outlook} with a discussion of a number of directions for future work. We discuss indications that not only the integrands but also the loop integrals should be ``simple". The idea of determining the loop integrand for planar amplitudes is a general one that can generalize well beyond maximally supersymmetric theories with Yangian symmetry, and we also very briefly discuss these prospects.

\section{Canonical Operations on Yangian Invariants}\label{canonical_operations_on_Yangian_invariants}

As a first step towards the construction of the all-loop integrand for ${\cal N}=4$ SYM in manifestly Yangian form, we study simple operations that can map Yangian invariants $Y_{n,k}({\cal Z}_1, \cdots, {\cal Z}_n)$ to other Yangian invariants. In this discussion it will not matter whether the ${\cal Z}$'s represent variables in twistor-space or momentum-twistor space; we will simply be describing mathematical operations that mapping between invariants. Combining these operations in various ways yields many objects of physical significance \cite{InPrep}. The same physical object will arise from different combinations of these operations in twistor-space vs. momentum-twistor space; we will content ourselves here by presenting mostly the momentum-twistor space representations.

As mentioned in the introduction, understanding these operations is not strictly necessary if we simply aim to find {\it a} formula for the integrand. The reason is that the BCFW recursion relations we introduce in section \ref{recursion_relations_for_loops} can be developed independently for theories with less supersymmetry, which do not enjoy a Yangian symmetry. Our insistence in keeping the Yangian manifest however will pay off in two ways. The first is conceptual: the Yangian-invariant formulation will introduce a new physical picture for meaning of loops. The second is  computational: the Yangian-invariant formulation gives a powerful way to compute the novel ``forward-limit" terms in the BCFW recursions in momentum-twistor space, using the Grassmannian language.

We will begin by discussing how to add and remove particles in a Yangian-invariant way. One motivation is an unusual feature of the Grassmannian integral--the space of
integration depends on the number $n$ of particles. It is natural to try and connect different $n$'s by choosing a contour of integration that allows a ``particle interpretation", by which we mean simply that the variety defining the contour for the scattering amplitudes of $(n+1)$ particles differs from the one for $n$ particles only by specifying the extra constraints associated with the new particle \cite{ArkaniHamed:2009dg}. Following this ``add one particle at a time"-guideline completely specifies the contour for all tree amplitudes \cite{ArkaniHamed:2009dg,Bourjaily:2010kw}, along the way exposing a remarkable connection with twistor string theory \cite{Witten:2003nn,Roiban:2004yf,NVW,SV,DG}. As we will see in this paper, loops are associated with interesting ``entangled" ways of {\it removing} particles from higher-point amplitudes. We will then move on to discuss how to ``fuse" two invariants together. Using these operations we demonstrate the Yangian invariance of all leading singularities, and discuss the important special case of the ``BCFW bridge" in some detail.

\subsection{Adding Particles}

Let us start with a general Yangian-invariant object \be Y_{n,k}({\cal Z}_1,
\ldots, {\cal Z}_n). \ee
We will first describe operations that will add a particle to lower-point invariants to get
higher-point invariants known as applying ``inverse soft factors" \cite{ABCCKT:2010}, which are so named because taking the usual soft limit of the resulting object returns the original object. This can be done preserving $k$ or increasing $k \mapsto k +1$. We can discuss these in both twistor- and momentum-twistor space; for the purposes of this paper we will describe these inverse-soft factor operations in momentum-twistor space.

The idea is that there are
residues in \mbox{$G(k,n)$} which are trivially related to residues in
\mbox{$G(k,n-1)$} or \mbox{$G(k-1,n-1)$}. The $k$-preserving operation \mbox{$Y_{n-1,k}\mapsto Y_{n,k}$} is particularly simple, being simply the identification \be
Y^\prime_{n,k}({\cal Z}_1, \ldots, {\cal Z}_{n-1},{\cal Z}_n)=Y_{n-1,k}({\cal Z}_1, \ldots, {\cal Z}_{n-1}); \ee that is, where we have
simply added particle $n$ as a label (but have not altered the functional form of $Y$ in any way); thanks to the momentum-twistor variables,
momentum conservation is automatically preserved.  The $k$-increasing inverse
soft factor is slightly more interesting. There is always a residue
of \mbox{$G(k,n)$} which has a $C$-matrix of the form
\be\left(\begin{array}{cccccccc}*&*&0&\cdots&0&*&*&1\\ * &\cdots&\cdots&\cdots&\cdots&\cdots&*&0\\\vdots&\ddots&\ddots&\ddots&\ddots&\ddots&*&\vdots\end{array}\right).\ee
Here, the non-zero elements in the top
row, $* \,*\,*\,*  \, 1$ correspond to particles
\mbox{$1,2,(n-2),(n-1),\,n$}, and we have generic non-zero entries in the
lower $(k-1) \times (n-1)$ matrix. The corresponding residue is easily seen to be associated with
\be Y^\prime_{n,k}(\ldots, {\cal Z}_{n-1}, {\cal Z}_n, {\cal Z}_1, \ldots)=[n\smallminus2\,\,\, n\smallminus1\,\, n \, 1 \, 2] \times Y_{n-1,k-1}(\ldots, \widehat{{\cal
Z}}_{n-1}, \widehat{{\cal Z}}_1, \ldots)\ee where \be\label{Rinv} [a \, b \,
c \, d \, e] = \frac{\delta^{0|4}(\eta_{a} \langle b \, c \,
d \, e \rangle + {\rm cyclic})}{\langle a \, b \, c \, d
\rangle \langle b \, c \, d \, e \rangle \ab{c\,d\,e\,a}\ab{d\,e\,a\,b} \langle e
\, a \, b \, c \rangle} \ee is the basic `NMHV'-like $R$-invariant\footnote{When two sets of the twistors are consecutive, these ``$R$-invariants'' are sometimes written \mbox{$R_{r;s,t}\equiv[r\,\,s\smallminus1\,\,s\,\,t\smallminus1\,\,t].$} These invariants were first introduced in \cite{Drummond:2008vq} in dual super-coordinate space.}
and the $\widehat{{\cal Z}}_{n-1,1}$ are deformed momentum twistor
variables. The Bosonic components of the deformed twistors have a very nice interpretation: $\widehat{Z}_1$ is simply the intersection of the line $(1\,2)$ with the plane $(n\smallminus2 \,\, n\smallminus1\,\, n)$, which we indicate by writing \mbox{$\widehat{Z_1}\equiv(n\smallminus2\,\,n\smallminus1\,\,n)\cap(1\,2)$}; and $\widehat{Z}_{n-1}$ is the intersection of the line $(n\smallminus2\,\,n\smallminus1)$ with the plane $(1 \, 2\,n)$, written \mbox{$\widehat{Z}_{n-1}\equiv(n\smallminus2\,\,n\smallminus1)\cap(n \, 1 \, 2)$}. Fully supersymmetrically, we have
\begin{equation}
\begin{split}\widehat{{\cal Z}}_1 &\equiv (n\smallminus2 \,\, n\smallminus1 \,\,n)\cap(1\,\,2) = \mathcal{Z}_1\ab{2\,\,n\smallminus2\,\,n\smallminus1\,\,n}+\mathcal{Z}_2\ab{n\smallminus2\,\,n\smallminus1\,\,n\,\,1}; \\
\widehat{{\cal Z}}_{n-1} &\equiv (n\smallminus2 \,\, n\smallminus1)\cap (n\,\,1\,\,2) = \mathcal{Z}_{n-2}\ab{n\smallminus1\,\,n\,\,1\,\,2}+\mathcal{Z}_{n-1}\ab{n\,\,1\,\,2\,\,n\smallminus2}.\end{split} \end{equation}

\subsection{Removing Particles}

We can also remove particles to get lower-point Yangian invariants from higher-point ones. This turns out to be more interesting than the inverse-soft factor operation, though physically one might think it is even more straightforward. After all, we can remove a particle simply by taking its soft limit. However, while this is a well-defined operation on {\it e.g.}~the full tree amplitude, it is {\it not} a well-defined operation on the individual residues (BCFW terms) in the tree amplitude. The reason is the presence of spurious poles: each term does not individually have the correct behavior in the soft limit.

Nonetheless, there {\it are} completely canonical and simple operations for removing particles in a Yangian-invariant way. One reduces
$k \mapsto k-1$, the other preserves $k$. The $k$-reducing
operation removes particle $n$ by integrating over its
twistor co-ordinate \be Y^\prime_{n-1,k-1}({\cal Z}_1, \ldots {\cal
Z}_{n-1})=\int\!\!d^{3|4} {\cal Z}_{n}\;Y_{n,k}({\cal Z}_1, \ldots, {\cal
Z}_{n-1}, {\cal Z}_{n}).  \ee This gives a Yangian-invariant for any
closed contour of integration---meaning that under the Yangian
generators for particles $1,\ldots,n-1$, this object transforms into
a total derivative with respect to ${\cal Z}_{n}$. This statement can be trivially verified by directly examining the action of the level-zero and level-one Yangian generators on the integral. It is also very easy to verify directly from the Grassmannian integral. Note that depending on the contour that is chosen, a given higher-point invariant can in general map to several lower-point invariants.

The $k$-preserving operation ``merges" particle $n$ with one of its
neighbors, $n-1$ or $1$. For example,
\be Y^\prime_{n-1,k}({\cal Z}_1, \ldots {\cal
Z}_{n-1})=Y_{n,k}({\cal Z}_1, \ldots, {\cal
Z}_{n-1}, {\cal Z}_{n}\mapsto{\cal Z}_{n-1}).  \ee
The Yangian-invariance of this operation is slightly less obvious to see by simply manipulating Yangian generators, but it can be verified easily from the Grassmannian formula.

We stress again that these operations are perfectly well-defined on {\it any} Yangian-invariant object, regardless of whether the standard soft-limits are well defined. Of course, they coincide with the soft limit when acting on {\it e.g.}~the tree amplitude.

\subsection{Fusing Invariants}

Finally, we mention a completely trivial way of combining two Yangian invariants to produce a new invariant. Start with two invariants
which are functions of a disjoint set of particles, which we can
label $Y_1({\cal Z}_1,\ldots,{\cal Z}_m)$ and $Y_2({\cal
Z}_{m+1},\ldots,{\cal Z}_n)$. Then, it is easy to see that the
simple product \be Y^\prime ({\cal Z}_1, \ldots, {\cal Z}_n) =
Y_1({\cal Z}_1,\ldots,{\cal Z}_m) \times Y_2({\cal Z}_{m+1}, \ldots,
{\cal Z}_n) \ee is also Yangian-invariant. Only the vanishing under the level-one generators requires a small comment. Note that the cross terms vanish because the corresponding level-zero generators commute and therefore the level-one generators cleanly splits into the smaller level-one generators.

\newpage
\subsection{Leading Singularities are Yangian Invariant}

Combining these operations builds new Yangian invariants from old
ones; all of which have nice physical interpretations. An immediate
consequence is a simple proof that all leading singularities are
Yangian invariant. For this subsection only, we work in ordinary twistor space.
Then we take any four Yangian invariants for disjoint sets of particles and we make a new
invariant by taking the product of all of them, $Y_1({\cal W}_1,
\ldots, {\cal W}_m) Y_2({\cal W}_{m+1}, \ldots, {\cal W}_l)
Y_3({\cal W}_{l+1}, \ldots, {\cal W}_p) Y_4({\cal
W}_{p+1},\ldots,{\cal W}_q)$. We then ``merge" $m$ and $m+1$, $l$
and $l+1$, $p$ and $p+1$, and $q$ with $1$. We then integrate over
$m,l,p,q$. This precisely yields the twistor-space expression for a
``1-loop" leading singularity topology \cite{Bullimore:2009cb,Kaplan:2009mh}.

\be
\includegraphics[scale=.45]{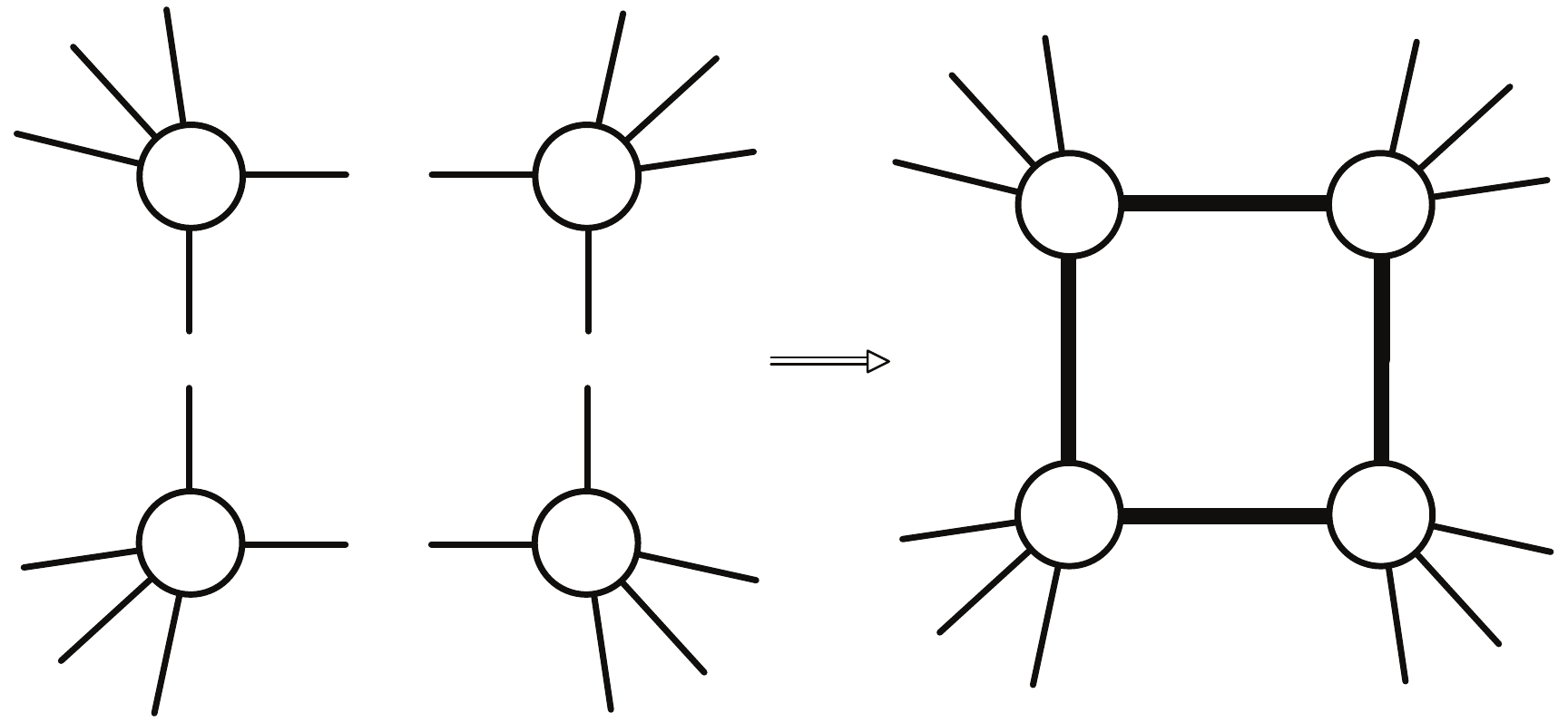} \nonumber
\ee
In the figure, a thick black line denotes the merging of the two particles at the ends of the line, and integrating over the remaining variable.
The generalization to all
leading singularities is obvious; for instance, starting with the ``1-loop" leading singularity we have already built, we can use the same merge and integrate operations to build ``2-loop" leading singularity topologies such as that shown below.
\be
\includegraphics[scale=.4]{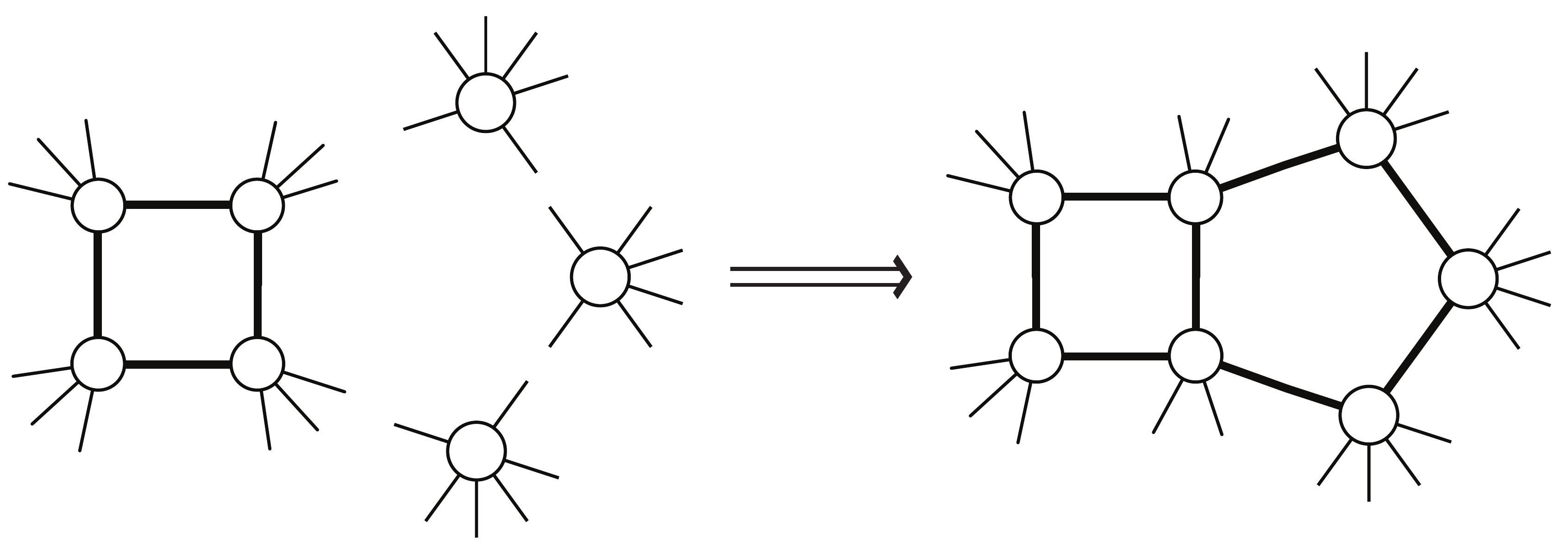} \nonumber
\ee
We conclude that all leading
singularities are Yangian invariant. Given that all Yangian
invariants are Grassmannian residues, this proves (in passing)
the original conjecture in \cite{ArkaniHamed:2009dn} that all leading singularities
can be identified as residues of the Grassmannian integral.

\newpage
\subsection{The BCFW Bridge}

A particularly important way of putting together two Yangian invariants to make a third is the ``BCFW bridge" \cite{ArkaniHamed:2009si,ArkaniHamed:2008yf,Brandhuber:2008pf}, associated with the
familiar ``two-mass hard" leading singularities drawn below in twistor space \cite{ArkaniHamed:2009si,ArkaniHamed:2008yf,Mason:2009sa,Skinner:2010cz}:
\be
\includegraphics[scale=.3]{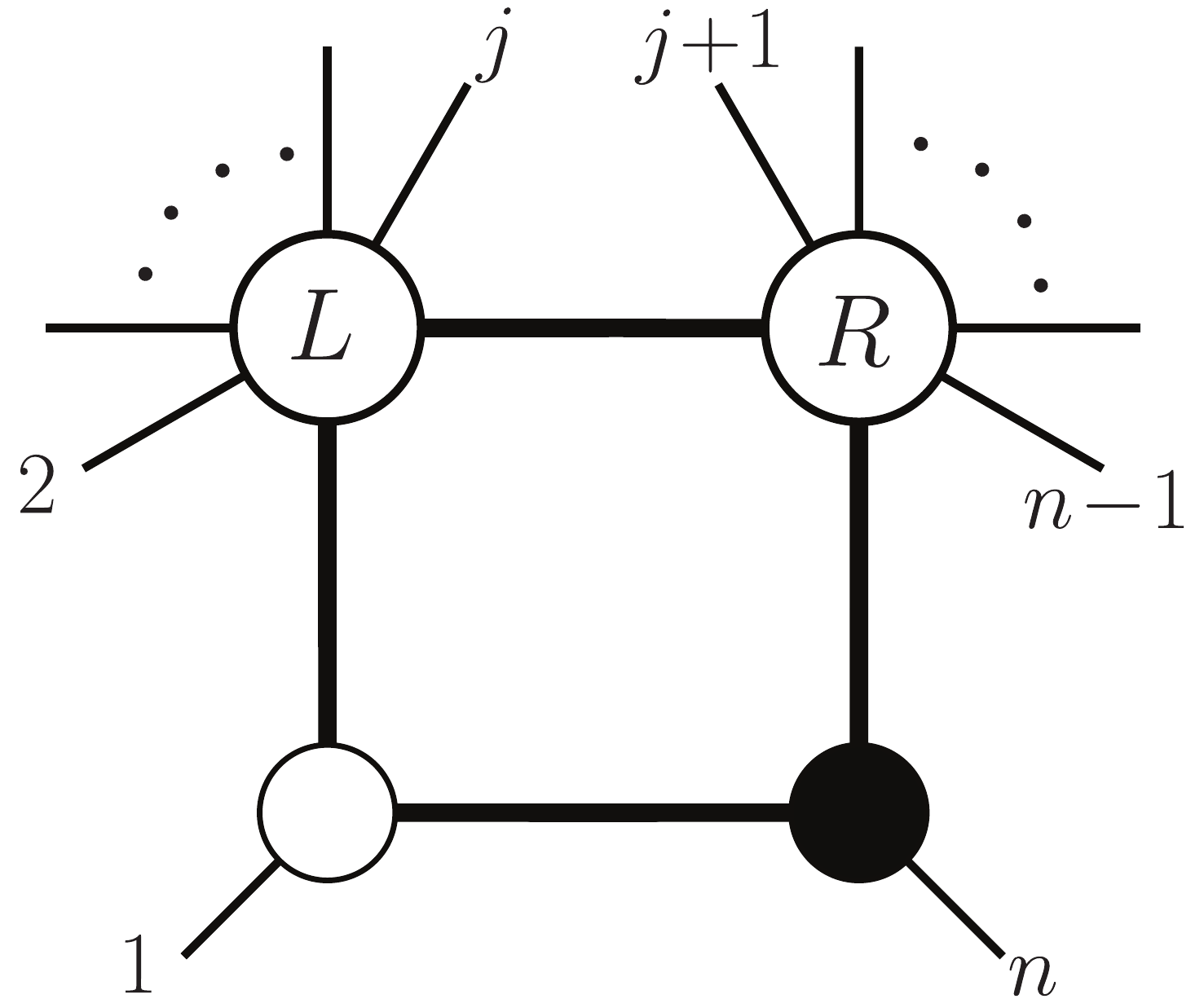} \nonumber
\ee
Here, the open and dark circles respectively denote MHV and $\overline{{\rm MHV}}$ three-particle amplitudes, respectively. We remark in passing that the inverse-soft factor operations mentioned above are special cases of the BCFW bridge where a given Yangian invariant is bridged with an $\overline{{\rm MHV}}$ three-point vertex \mbox{(for the $k$-preserving case)} or an MHV three-point vertex (for the $k$-increasing case).

We will find it useful to also see the bridge expressed as a composition of our basic operations in momentum-twistor space, as

\be
\includegraphics[scale=.4]{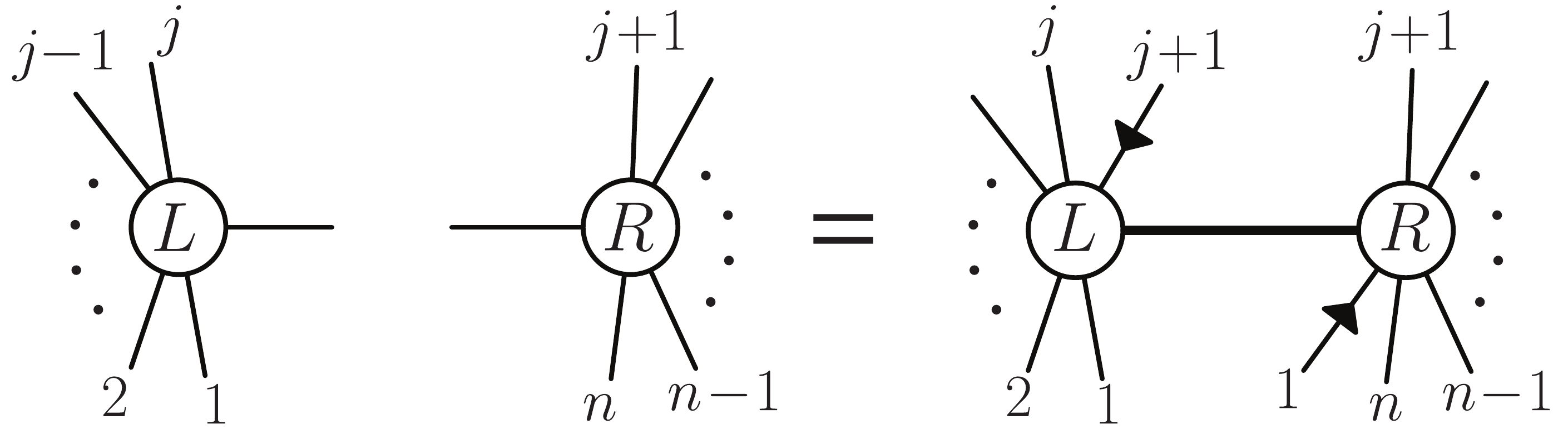} \nonumber
\ee
This is a pretty object since it uses all of our basic operations in an interesting way. In the figure, the solid arrows pointing inward indicate that particle-``$1$'' is added as an $k$-increasing inverse soft factor on $Y_L$, and $j\smallplus1$ is added as a $k$-increasing inverse soft factor on $Y_R$. We are also using the merge operation to identify the repeated ``1'' and ``$j\smallplus1$'' labels across the bridge. The internal line, which we label as ${\cal Z}_I$,  is integrated over. The contour of integration is chosen to encircle the $\ab{n\smallminus1 \,\, n \,\, 1\,\, I}$-pole from the $[n\smallminus1\,\,n\,\,1\,\,I\,\,j\smallplus1]$-piece of the inverse-soft factor on $Y_L$, and the $\ab{1\,\,I\,\,j\smallplus1\,\,j}$- and $\ab{I\,\,j\smallplus1\,\,j\,\,j\smallminus1}$-poles from the $[1\,\,I\,\,j\smallplus1\,\,j\,\,j\smallminus1]$-piece of the inverse soft factor on $Y_R$. The deformation on ${\cal Z}_n$ induced by the inverse-soft factor adding particle-$1$ on $Y_L$ is of the form
\be
{\cal Z}_n \mapsto \widehat{{\cal Z}}_n = {\cal Z}_n + z {\cal Z}_{n-1}, \quad {\rm where} \quad \ab{\widehat{Z}_n Z_1 Z_j Z_{j+1}} = 0.
\ee
This is the momentum-twistor space version of the BCFW deformation, which corresponds to deforming $\lambda_n, \widetilde \lambda_1$ in momentum-space. We remind ourselves of this deformation by placing the little arrow pointing from $n \mapsto n-1$ in the figure for the bridge. The momentum-twistor space geometry associated with this object is
\begin{equation}\vspace{-0.2cm}
\includegraphics[scale=.7]{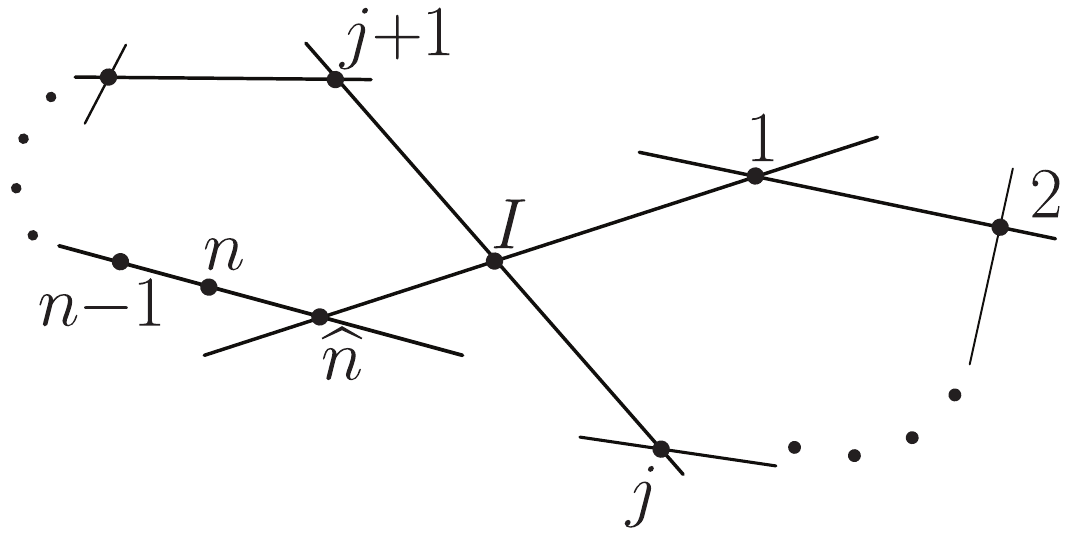} \nonumber
\end{equation}
which precisely corresponds to the expected BCFW deformation and the corresponding factorization channel.

We leave a detailed derivation of this picture to \cite{InPrep}, but in fact the momentum-twistor structure of the BCFW bridge can be easily understood. Note that $Y_L,Y_R$ have $k$-charge $k_L,k_R$, while $Y_L \otimes Y_R$ has $k$-charge $k_L + k_R + 1$; given that the ${\cal Z}_I$ decreases the $k$-charge by 1,  we must start with $Y_L$ and $Y_R$ and get objects with $k$-charge $(k_L + 1)$ and $(k_R+1)$ on the left and right. This can be canonically done by acting with $k$-increasing inverse soft factors; the added particle on $Y_L$ must be adjacent to $n$ in order to affect a deformation on ${\cal Z}_n$. Finally, the data associated with the ``extra" particles introduced by the inverse soft factor must be removed in the only way possible, by using the merge operation. Explicitly, the final result for $Y_L \!\underset{\mathrm{BCFW}}{\otimes}\! Y_R$ is
\begin{equation}
 \Big(Y_L\!\underset{\mathrm{BCFW}}{\otimes}\!
Y_R\Big)(1,\ldots,n) = [n\smallminus1\,\,n\,\, \, 1 \, j \,\, j\smallplus1]\times Y_R\big(1,\ldots, j,I\big) \times Y_L\big(I,j+1,\ldots, n\smallminus1,\widehat n\big)
\end{equation}
with\vspace{-0.3cm}
\begin{equation}
 \widehat n = (n\smallminus1\,\,n) \cap (j\,\,j\smallplus1\,\,1), \quad\mathrm{and}\quad I = (j\,\,j\smallplus1) \cap (n\smallminus1\,\,n\,\,1).
\end{equation}

Starting with the tree amplitude $M_{n,k,{\rm tree}}$ \footnote{We remind the reader that we are working in momentum-twistor space, so that what
we are calling $M_{{\rm tree}}$ here is obtained after stripping off the MHV tree-amplitude factor from the full amplitude in momentum space.}, the BCFW deformation ${\cal Z}_n \mapsto {\cal Z}_n + z {\cal Z}_{n-1}$ can
be used to recursively construct tree amplitudes in the familiar
way: by writing, \be M_{n,k,{\rm tree}} = \oint\!\!\frac{dz}{z}\;
\widehat{M}_{n,k,{\rm tree}}(z), \ee it is clear that the desired
amplitude $ \widehat{M}_{n,k,{\rm tree}}(z)$ is obtained by summing-over all the  residues of the RHS {\it except} the pole at origin
$z=0$. Notice that there is a non-zero pole at infinity in this
deformation: as $z \to \infty$, ${\cal Z}_n \to {\cal Z}_{n-1}$
projectively, and so the tree amplitude gets a contribution from
\mbox{$M_n({\cal Z}_1, \ldots,{\cal Z}_{n-1},{\cal Z}_n) \to
M_{n-1}({\cal Z}_1, \ldots,{\cal Z}_{n-1})$}
\footnote{Note that $z \to \infty$ here does {\it not} correspond to going to infinity in the familiar momentum-space
version of BCFW. The  pole at infinity in ordinary momentum space here corresponds to a pole involving the infinity twistor $\langle
Z_n(z)\,I\,Z_1\rangle=0$. Of course we do not expect such a pole to arise in a dual-conformal invariant theory, not only at tree-level, but at all-loop order, as will be relevant to
our subsequent discussion. A direct proof of this fact, not assuming dual conformal invariance, should follow from the ``enhanced spin-lorentz symmetry" arguments of \cite{ArkaniHamed:2008yf}.}. The pole at $z \to \infty$ corresponds to
the term in the usual momentum-space BCFW formula using an $\overline{\rm MHV}$
three-point vertex bridged with $M_{n-1}$, which simply acts as a
$k$-preserving inverse-soft factor The remaining physical poles are of the form $\ab{i \,\, i\smallplus1 \,\,
j \,\, j\smallplus1}$. Under ${\cal Z}_n \mapsto {\cal Z}_n + z
{\cal Z}_{n-1}$, we only access the poles where $\ab{Z_n(z)Z_1 Z_j
Z_{j\smallplus1}} \to 0$, and the corresponding residues are
computed by the BCFW bridge indicated above, with $Y_L,Y_R$ being
the lower-point tree amplitudes.

\section{Loops From Hidden Entanglement}\label{loops_as_hidden_entanglement}

Let's imagine starting with some scattering amplitude or Grassmannian residue, and begin removing particles. The operation that decreases $k$ in particular demands a choice for the contour of integration. If we remove particle ${\cal Z}_A$ by integrating over it as $\int\!d^{3|4}{\cal Z}_A$, it is natural to choose a $T^3$-contour of integration for the Bosonic $d^3 Z_A$ integral and compute a simple residue\footnote{Residues of rational functions in $m$ complex variables are computed by choosing $m$ polynomial factors $f_i$'s from the denominator and integrating along a particular $T^m$-contour, {\it i.e.}~the product of $m$ circles given as the solutions of $|f_i|=\epsilon$ with $\epsilon \ll 1$ and near a common zero of the $f_i$'s. See \cite{Griffiths:1978a} for more details.}.

We can then proceed to remove a subsequent particle either by merging, or performing further integrals $\int\!d^{3|4}{\cal Z}_B$ and so on. In this way we will simply proceed from higher-point Grassmannian residues to lower-point ones. In particular, if these operations are performed on a higher-point tree amplitude, we arrive at lower-point tree amplitudes, and don't encounter any new objects.

But we can imagine a more interesting way of removing not just one but a pair of particles.
Consider removing particle $A$ and subsequently removing the adjacent particle $B$. Instead of first integrating-out $A$ and then $B$ on separate $T^3$'s,
let's consider an ``entangled" contour of integration, which we will discover to yield, instead of lower-point Grassmannian residue, a loop integral.

Consider as a simple example removing two particles from the 6-particle
N$^2$MHV = $\overline{{\rm MHV}}$ tree amplitude, $M_{6,4,\ell=0}(1234AB)$. Performing the $d^{0|4}\eta_A,d^{0|4}\eta_B$ integrals is
trivial, and this gives \be \int\!\!d^3 z_A d^3 z_B\;\frac{\langle 1234
\rangle^3}{\langle 2 3 4 z_A \rangle \langle 34 z_A z_B \rangle \langle 4 z_A z_B 1
\rangle \langle z_A z_B 12 \rangle \langle z_B123 \rangle} \ee where we have chosen to label the Bosonic momentum twistors with lower-case $z$'s
for later convenience.
As we have
claimed, on any closed contour, these integrals should give a
Yangian-invariant answer. Indeed, computing the $z_B$ integral by residue on any contour
 leaves us with \be \int\!\! d^3 z_A \;\frac{\langle 1 2 3 4
\rangle^3}{\langle z_A 123 \rangle \langle z_A 234 \rangle \langle z_A 341
\rangle \langle z_A 412 \rangle} \ee and computing any of the simple residues of this
remaining $z_A$ integral gives $1$, which is of course the only Yangian invariant for MHV
amplitudes.

We will now see that starting with exactly the same integrand but
choosing a different contour of integration yields, instead of ``1",
the 4-particle 1-loop amplitude. Geometrically, the points $z_A,z_B$
determine a line in momentum-twistor space, which is interpreted as a point in the dual $x$-space,
or equivalently, a loop-integral's four-momentum. We will first integrate
over the positions of $z_A,z_B$ on the line $(AB)$, and then integrate
over all lines $(AB)$.

This contour can be described explicitly by parametrizing $z_{A,B}$ as
\be z_A = \left( \begin{array}{c} \lambda_A \\ x  \lambda_A
\end{array} \right), \, \, z_B = \left( \begin{array}{c} \lambda_B \\ x
\lambda_B
\end{array} \right) \ee
where $x$ will be the loop momentum.  The measure is \be d^3 z_A d^3 z_B = \langle \lambda_A
d \lambda_A \rangle \langle \lambda_B d \lambda_B \rangle \langle
\lambda_A \lambda_B \rangle^2 d^4 x.
\ee
The $\lambda_A,\lambda_B$ integrals will be treated as contour integrals on $\mathbb{CP}^1 \times \mathbb{CP}^1$, while the $x$-integral will be over real points
in the (dual) Minkowksi space.

Using that $\langle z_A z_B\,\,j\smallminus1\,\,j\rangle = \langle \lambda_A \lambda_B\rangle \langle j\smallminus1 \,\, j
\rangle (x {-} x_j)^2$ our integral becomes \be \int
d^4 x \frac{x_{13}^2 x_{24}^2}{(x{-}x_1)^2(x{-}x_2)^2(x{-}x_4)^2}
\int \frac{\langle 1234\rangle \langle 23\rangle \langle \lambda_A d \lambda_A \rangle \langle \lambda_B d \lambda_B \rangle}
  {\langle z_A 123 \rangle \langle 234 z_B \rangle \langle \lambda_A \lambda_B \rangle}.
\ee
The factor $\langle z_A 234\rangle$ is linear in the projective variable $\lambda_A$ while the factor $\langle 123z_B\rangle$ is linear in $\lambda_B$.
This implies that there is a unique way to perform the $\lambda_A$ and $\lambda_B$ integrals by contour integration, which gives us
\be \int  d^4 x \frac{x_{13}^2 x_{24}^2}{(x-x_1)^2(x-x_2)^2(x-x_3)^2(x-x_4)^2}.
\ee
This is precisely the 1-loop MHV amplitude!

We have thus seen that, removing a pair of particles with this ``entangled" contour of integration, where we first integrate over the position of two points along the line joining them and then integrate over all lines, naturally produces objects that look like loop integrals.

There is a nicer way of characterizing this ``entangled" contour that is also more convenient for doing calculations, let us describe it in detail.
Given $z_A,z_B$, a general GL(2)-transformation on the 2-vector $(z_A,z_B)$ moves $A,B$ along the line $(AB)$. Thus, in integrating over $d^3z_A d^3z_B$, we'd like to ``do the GL(2)-part of the integral first" to leave us with an integral that only depends on the line $(AB)$:\vspace{-0.5cm}

\be \includegraphics[scale=.5]{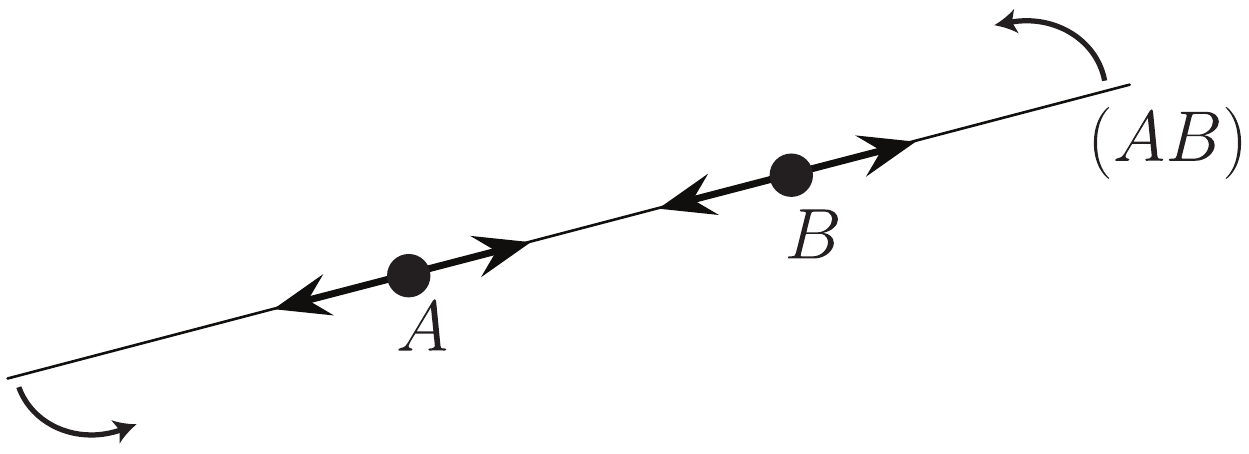} \nonumber \ee We
can do this explicitly by writing \be \left(\begin{array}{c} z_A \\
z_B \end{array} \right) = \left(\begin{array}{cc} c_A^{(A)} & c_A^{(B)} \\
c_B^{(A)} & c^{(B)}_B \end{array} \right) \left(\begin{array}{c} Z_A \\
Z_B
\end{array} \right); \ee then \be d^3 z_A d^3 z_B =  \ab{c_A d c_A}
\ab{c_B d c_B} \ab{c_A c_B}^2 \left[\frac{d^4 Z_A d^4 Z_B}{{\rm
vol[GL(2)]}}\right], \ee and our integral becomes---this time writing
it out fully: \be \int \left[\frac{d^4 Z_A d^4 Z_B}{{\rm
vol [GL(2)]}}\right] \frac{\langle 1 2 3 4 \rangle^3}{\langle AB 12
\rangle \langle AB 34 \rangle \langle AB 41 \rangle} \int
\frac{\langle c_A d c_A \rangle \langle c_B d c_B \rangle}{\langle
c_A c_B \rangle \langle c_A \psi_A \rangle \langle c_B \psi_B
\rangle}, \ee where \be \psi_A = \left(\begin{array}{c} \ab{A234} \\
\ab{B234} \end{array} \right), \, \psi_B = \left(\begin{array}{c}
\ab{A123} \\ \ab{B123} \end{array} \right). \ee The $c_A,c_B$
integral is naturally performed on a contour `encircling' $c_A =
\psi_A, c_B = \psi_B$, yielding $\frac{1}{\langle \psi_A \psi_B
\rangle} = \frac{1}{\langle AB 23 \rangle \langle 1234 \rangle}$.
More generally, if ``234" and ``123" in the definitions of
$\psi_A,\psi_B$ were to be replaced by arbitrary ``$abc$" and
``$xyz$", $\ab{\psi_A \psi_B} = \ab{A xyz} \ab{Babc} - \ab{Aabc}
\ab{Bxyz} \equiv \ab{AB|(abc)\cap(xyz)}$ where $(abc) \cap (xyz)$ is
the line corresponding to the intersection of the planes $(abc)$ and
$(xyz)$. We are then left with \be \int \left[\frac{d^4Z_A
d^4Z_B}{{\rm vol[GL(2)]}}\right] \frac{\langle 1234
\rangle^2}{\langle AB 12 \rangle \langle AB 23 \rangle \langle AB 34
\rangle \langle AB 41 \rangle}, \ee where the integration region is
such that the line $(AB)$ corresponds to a real point in the (dual)
Minkowski space-time. We recognize this object as the 1-loop MHV
amplitude, exactly as above.

We can clearly perform this operation starting with any Yangian invariant object $Y[{\cal Z}_A, {\cal Z}_B, {\cal Z}_1,\ldots]$, which we will graphically denote as:\vspace{-0.5cm}
\begin{equation}\includegraphics[scale=.45]{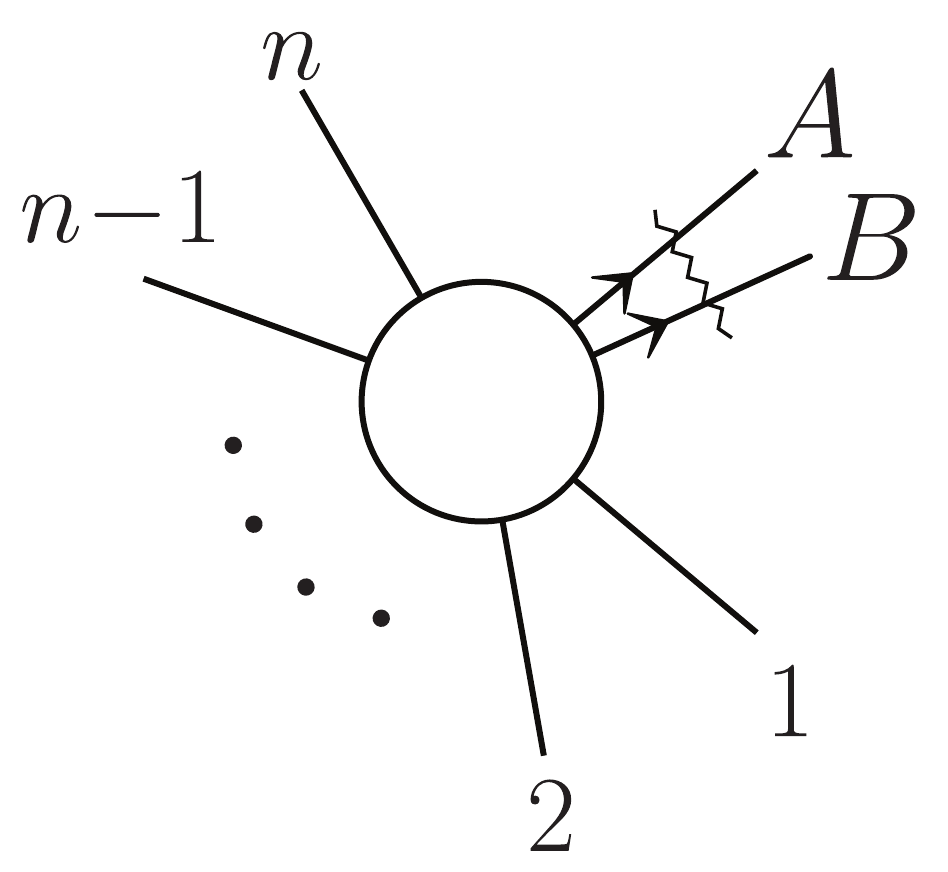} \nonumber\vspace{-0.5cm}
\end{equation}
and write as
\be
\int\limits_{{\rm GL(2)}} Y[\ldots,{\cal Z}_n,{\cal Z}_A, {\cal Z}_B, {\cal Z}_1,\ldots]
\ee
This object is formally Yangian-invariant, in the precise sense that the integrand will transform into a total derivative under the action of the Yangian generators for the external particles. Of course, such integrals may have IR-divergences along some contours of integration, which is how Yangian-invariance is broken in practice.

The usual way of writing the loop amplitudes as \mbox{``leading singularity $\times$ scalar integral"} ensures that the leading singularities of the individual terms are Yangian-invariant, but the factorized form seems very un-natural, and there is no obvious action of the symmetry generators on the integrand. By contrast, the loop integrals we have defined, as we will see, will not take the artificial \mbox{``residue $\times$ integral"} form, but of course their leading singularities are automatically Grassmannian residues. The reason is that a leading singularity of the $(AB)$-integral can be computed as a simple residue of the underlying $d^{3|4}z_A d^{3|4}z_B$ integral, which is free of IR divergences and guaranteed to be Yangian-invariant.

\section{Recursion Relations For Arbitrary Loop Amplitudes}\label{recursion_relations_for_loops}

Having familiarized ourselves with the basic operations on Yangian invariants, we are ready to discuss the recursion relations for loops in the most transparent way. The loop integrand is a rational function of both the loop
integration variables and the external kinematical variables. Just as
the BCFW recursion relations allow us to compute a rational
function from its poles under a simple deformation, the loop
integrand can be determined in the same way. Consider the $l$-loop integrand
$M_{n,k,\ell}$, and consider again the
(supersymmetric) momentum-twistor deformation
\be {\cal
Z}_n \mapsto {\cal Z}_n + z {\cal Z}_{n-1}.
\ee
Then
\be
M_{n,k,\ell} = \oint\!\! \frac{dz}{z} \;\widehat{M}_{n,k,\ell}(z)
\ee
and we sum over all the residues of the RHS away from the origin, all of which can be determined from lower-point/lower-loop amplitudes. This recursion relation can be derived in a large class of theories and is not directly tied to  ${\cal N}=4$ SYM or Yangian-invariance.
However our experience with building Yangian-invariant objects will help us to understand (and compute) the terms in the recursion relations in a transparent way, and also easily recognize them as manifestly Yangian-invariant objects.

As in our discussion of the BCFW bridge at tree-level, the pole at
infinity is simply the lower-point integrand with particle $n$
removed. All the rest of the poles in $z$ also have a simple interpretation: in
general, all the poles arise either from $\ab{Z_n(z)\,Z_1\,
Z_j\,Z_{j\smallplus1}} \to 0$ or $\ab{(AB)_q \,\,Z_n(z)\,Z_1} \to 0$,
where $(AB)_q$ denotes the line in momentum twistor space associated
with the $q^{\mathrm{th}}$ loop-variable. The first type of pole
simply corresponds to factorization channels, and the corresponding
residue is computed by the BCFW bridges between
lower-loop/lower-point amplitudes:
\be
\includegraphics[scale=0.6]{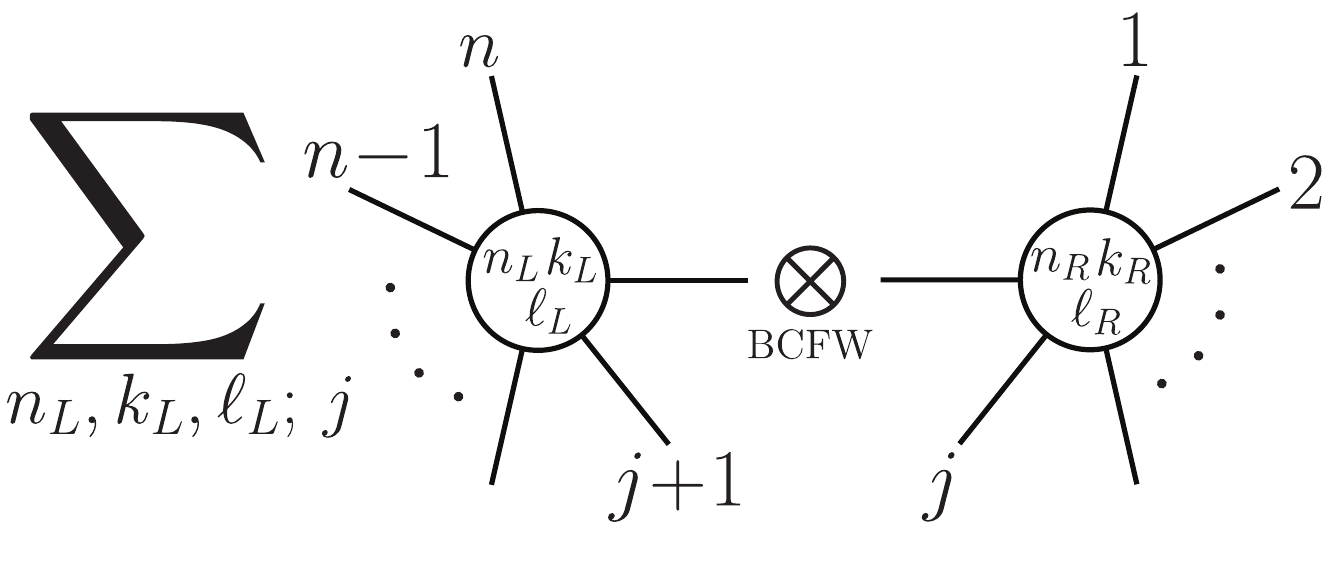} \nonumber
\ee
where $n_L+ n_R=n+2$, $k_L+k_R=k-1$, $\ell_L+\ell_R=\ell$. Note that we treat all the poles \mbox{(including the pole at infinity)} on an equal footing by declaring the term with $j=2$ to be given by
the $k$-preserving inverse soft-factor acting on lower-point amplitude.

This is the most obvious generalization of the BCFW recursion relation from trees to loops, but it clearly can't be the whole story, since it would allow us to recursively reduce loop amplitudes to the 3-particle loop amplitude, which vanishes! Obviously, at loop-level, a ``source" term is needed for the recursive formula.

\newpage\subsection{Single-Cuts and the Forward-Limit}

This source term is clearly provided by the second set of poles,
arising from $\ab{(AB)_q\,\,Z_n(z)\,\,Z_1} \to 0$. For simplicity of
discussion let's first consider the 1-loop amplitude. This pole
corresponds to cutting the loop momentum running between $n$ and
$1$, and is therefore given by a tree-amplitude with two additional
particles sandwiched-between $n,1$, with momenta $q,-q$, summing-over
the multiplet of states running around the loop. These single-cuts
associated with ``forward-limits" of lower-loop integrands are
precisely the objects that make an appearance in the context of the
Feynman tree theorem \cite{CaronHuot:2010zt}. The geometry of the
forward limit is shown below for both in the dual $x$-space and momentum-twistor space: \begin{equation}
\includegraphics[scale=0.4]{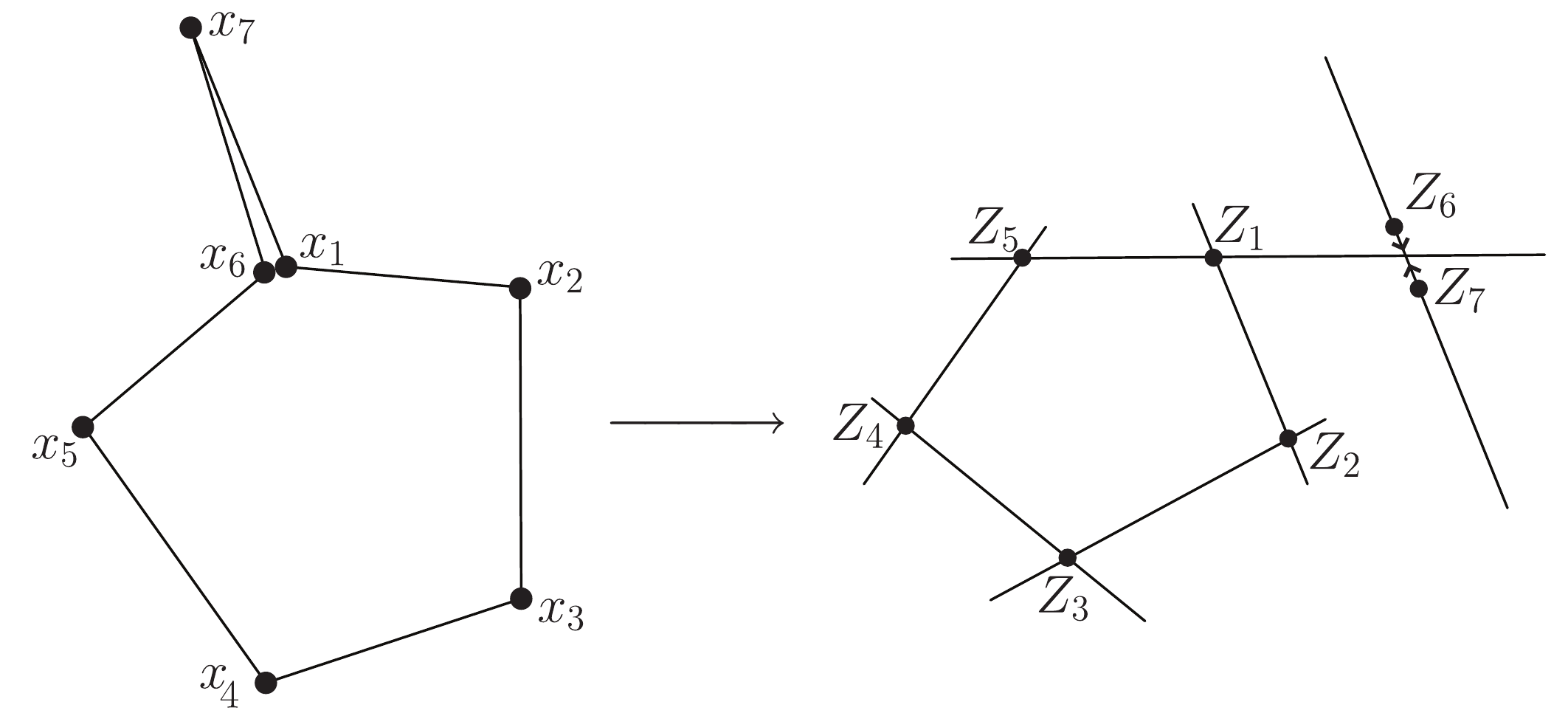} \nonumber
\end{equation}
Here, between particles $5$ and $1$, we have particles $6,7$ with momenta $q^\mu,-q^\mu$, where $q^\mu = x^\mu_1 - x^\mu_7$ is a null vector.
In momentum-twistor space, the null condition means that
the line $(76)$ intersects $(15)$, while in the forward limit both $Z_6$ and $Z_7$ approach the intersection point $(76) \cap (15)$.

In a generic gauge theory, the forward limits of tree amplitudes suffer from collinear divergences and are not obviously well-defined.
However remarkably, as pointed out in \cite{CaronHuot:2010zt}, in supersymmetric theories
the sum over the full multiplet makes these objects completely well-defined and equal to single-cuts!

Indeed, we can go further and express this single-cut ``forward limit" term
in a manifestly Yangian-invariant way. It turns out to to be a beautiful object, combining the entangled removal of two particles with the ``merge" operation:
\begin{equation}
\includegraphics[scale=0.4]{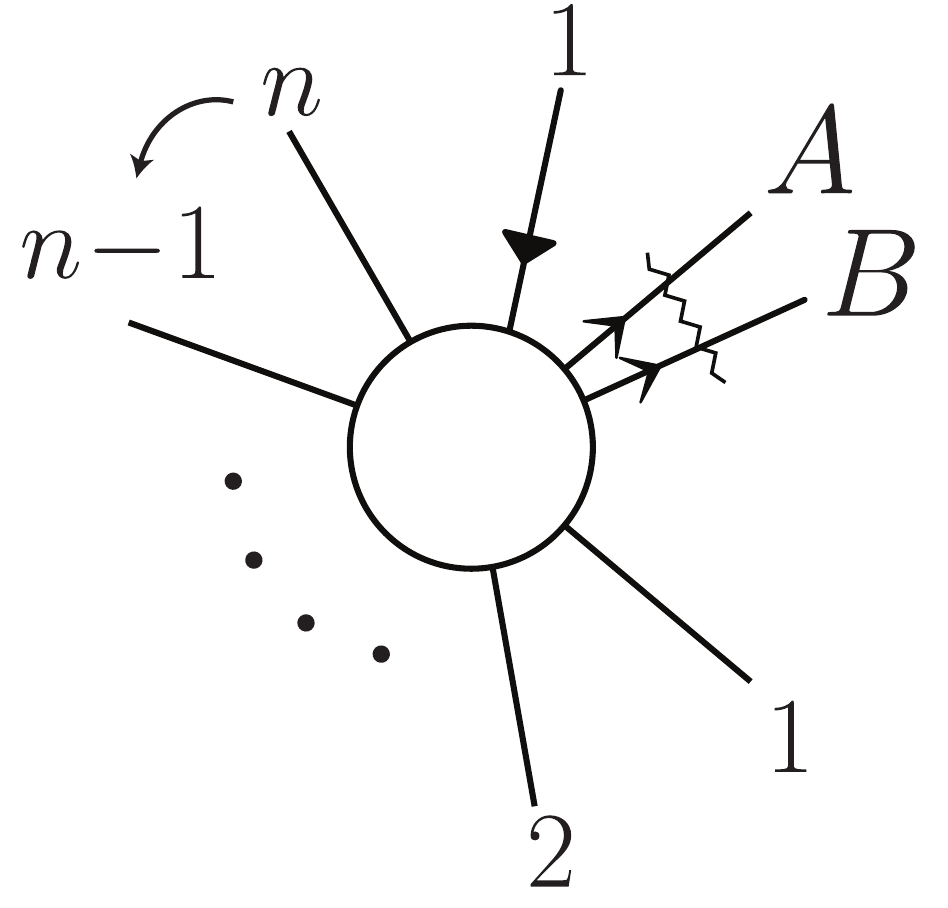} \nonumber
\end{equation}
Here a particle $(n+1)$ is added adjacent to $A,B$ as a $k$-increasing inverse soft factor, then $A,B$ are removed by entangled integration. The GL(2)-contour is chosen to encircle points where both points $A,B$ on the line $(AB)$ are located at the intersection of the line $(AB)$ with the plane $(n\smallminus1\,\,n\,\,1)$. Note that there is no actual integral to be done here; the GL(2)
integral is done on residues and is computed purely algebraically. Finally, the added particle $(n+1)$ is merged with $1$.

As in our discussion of the BCFW bridge, this form can be easily understood by looking at the deformations induced by the $``1"$ inverse soft factors; the associated momentum-twistor geometry turns out to be
\begin{equation}\vspace{-0.3cm}
\includegraphics[scale=0.6]{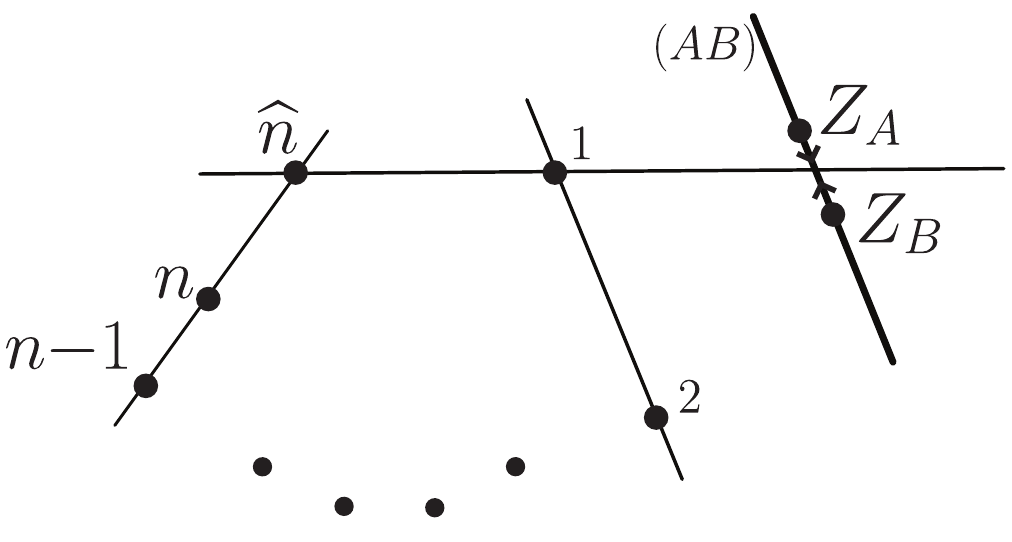} \nonumber
\end{equation}
exactly as needed. The picture is the same for taking the single cut of any Yangian-invariant object.

Note that we were able to identify the BCFW terms in a straightforward way since the residues of the poles {\it of the integrand} have obvious ``factorization" and ``cut" interpretations.
This is another significant advantage of working with the integrand, since as is well known, the full loop amplitudes (after integration) have more complicated factorization properties 
\cite{Bern:1995ix}. This is due to the IR divergences which occur when the loop momenta becomes collinear to external particles, when the integration is performed.


\subsection{BCFW For All Loop Amplitudes}

Putting the pieces together, we can give the recursive definition for all loop integrands in planar ${\cal N}=4$ SYM as
\begin{equation}
\hspace{-0.5cm}\includegraphics[scale=.6]{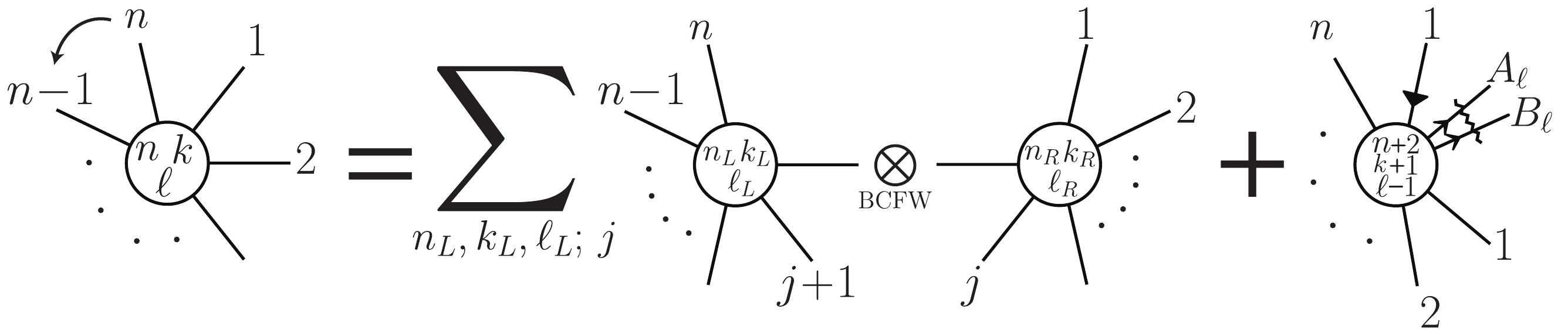} \nonumber
\end{equation}
To be fully explicit, the recursion relation is
\begin{equation}
\begin{split}
\hspace{-1.65cm}M_{n,k,\ell}(1,\ldots,n)=&\phantom{\,+\,}M_{n{-}1,k,\ell}(1,\ldots,n-1)\\&+\sum_{n_L,k_L,\ell_L; j}  [j\,\,j\smallplus1\,\,n{-}1\,\,n\,\,1] ~
   M^R_{n_R,k_R,\ell_R} (1,\ldots,j,I_j) \times
   M^L_{n_L,k_L,\ell_L} (I_j,j\smallplus 1,\ldots,\widehat n_{j})\\
&+\int\limits_{\textrm{GL(2)}} [AB\,\,n\smallminus1\,\,n\,\,1] \times M_{n{+}2,k{+}1,\ell{-}1} (1,\ldots,\widehat{n}_{AB}, \widehat{A}, B).   \label{loop_level_BCFW}
\end{split}
\end{equation}
where $n_L+ n_R=n+2$, $k_L+k_R=k-1$, $\ell_L+\ell_R=\ell$ and the shifted momentum (super-)twistors that enter are
\begin{equation}\begin{split}
 &\widehat{n}_{j} = (n\smallminus1\,\,n) \cap (j\,\,j\smallplus1\,\,1),\qquad\quad I_j = (j\,\,j\smallplus1) \cap (n\smallminus1\,\,n\,\,1); \\
 &\widehat{n}_{AB} = (n\smallminus1\,\,n) \cap (AB\,\,1), \quad\quad\,\;\;\; \widehat{A}= (AB)\cap (n\smallminus1\,\,n\,\,1).\end{split}
\end{equation}
Beyond 1-loop, it is understood that this expression is to be fully-symmetrized with equal weight in all the loop-integration variables $(AB)_{\ell}$; it is easy to see that this correctly captures the recursive combinatorics.
Recall again that GL(2)-integral is done on simple residues and is thus computed purely algebraically; the contour is chosen so that
the points $A,B$ are sent to $(AB) \cap (n\smallminus1\,\,n\,\,1)$. As we will show in \cite{InPrep}, recursively using the BCFW form for
the lower-loop amplitudes appearing in the forward limit allows us to carry
out the GL(2)-integral completely explicitly, but the form we have given here will suffice for this paper.


\subsection{Simple Examples}
In \cite{InPrep}, we will describe the loop-level BCFW computations in detail; here we will just highlight some of the results for some simple cases, to illustrate some of the important properties of the recursion and the amplitudes that result. We start by giving the BCFW formula for all one-loop MHV amplitudes.

In this case the second line in the above formula vanishes, and the recursion relation trivially reduces to a single sum.
To compute the NMHV tree amplitudes which enters through the third line, it is convenient to use the
tree BCFW deformation $\widetilde {\cal Z}_B = {\cal Z}_B+z{\cal \widehat Z}_A$ which leads to \be
M^{{\rm 1-loop}}_{{\rm MHV}} = \int\left[\frac{d^{4|4}{\cal Z}_A d^{4|4}{\cal Z}_B}{{\rm vol[GL(2)]}}\right] \int\limits_{GL(2)}
 \sum_j [AB\,\,j\,\,j\smallplus1\,\,1] \times \left(\sum_{i<j} [\widehat{A}B\,\, 1\,\, i \,\,i\smallplus1] + \ldots\right),
\ee
where the omitted terms are independent of ${\cal Z}_B$ and vanish upon Fermionic-integration.
The GL(2) and Fermion integrals are readily evaluated, as explained above, reducing this to
\be
M^{{\rm 1-loop}}_{{\rm MHV}} = \int \left[\frac{d^4Z_A d^4Z_B}{{\rm vol[GL(2)]}}\right] \sum_{i<j} \frac{\ab{AB| \,(1 \, \,i \,\, i\smallplus1) \cap (1 \,\, j \,\, j\smallplus1)}^2}{\ab{AB \, 1 \, i}\ab{AB \,\, i \,\, i\smallplus1} \ab{AB \,\, i\smallplus1 \,\, 1}  \ab{AB \,\, 1 \,\, j}\ab{AB \,\, j \,\, j\smallplus1} \ab{AB \,\, j\smallplus1 \,\, 1}}.  \label{MHV_1loop_amplitude}
\ee
This is the full one-loop integrand for MHV amplitudes.

As expected on general grounds from Yangian-invariance, and also as familiar from BCFW recursion at tree-level, the individual terms in this formula contain both local and non-local poles. We will graphically denote a factor $\ab{AB xy}$ in the denominator by drawing a line $(xy)$; the numerators of tensor integrals (required by dual conformal invariance) will be denoted by wavy- and dashed-lines---the precise meaning of which will be explained shortly. In this notation, this result is
\be
\includegraphics[scale=.65]{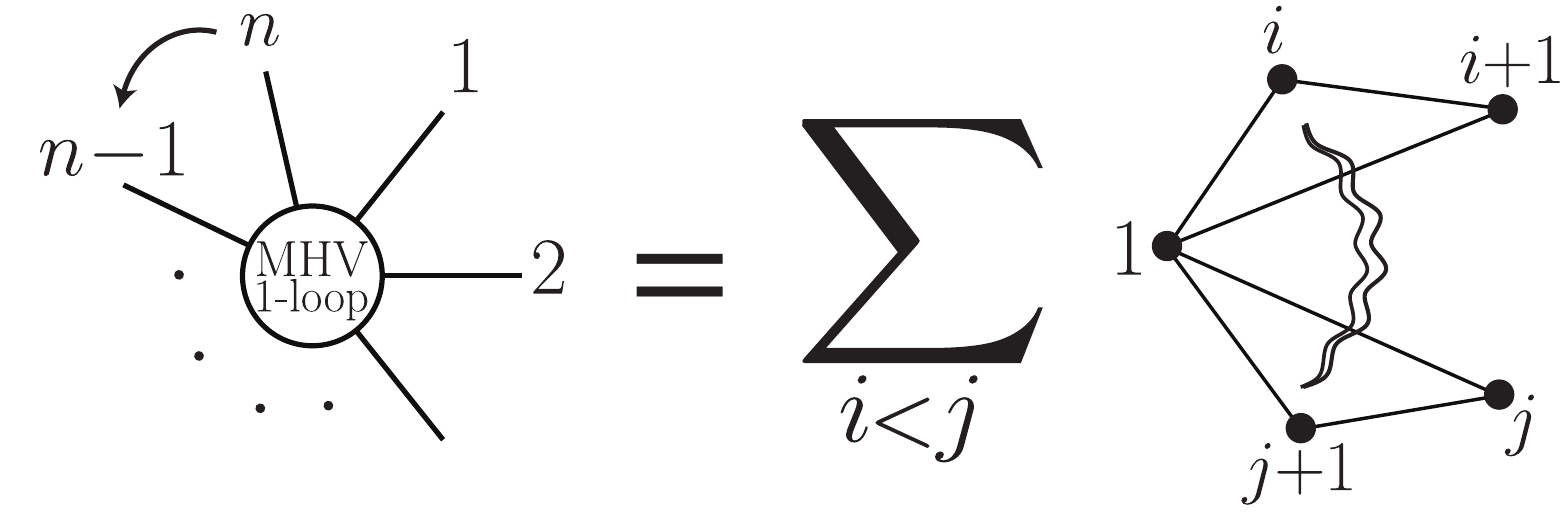} \nonumber
\ee
Notice that all the terms have 6 factors in the denominator, and hence by dual conformal invariance we must have two factors containing $AB$ in the numerators. These are denoted by the wavy lines: the numerator is $\ab{AB| (1\,\,i\,\,i\smallplus1) \cap (1\,\,j\,\,j\smallplus1)}^2\equiv(\ab{A \,\, 1 \,\, i \,\, i\smallplus1} \ab{B\, \, 1 \,\, j \,\, j\smallplus1} - \ab{B \,\,1 \,\, i\, \, i\smallplus1} \ab{A\, \, 1\, \, j\, \, j\smallplus1})^2$, where the power of 2 has been indicated by the line's multiplicity.

Notice that when $i+1 = j$, the numerator cancels the two factors $\ab{AB\,\,1\,\,j}^2$ in the denominator: by a simple use of the Schouten identity it is easy to see that \be\left[\ab{A\,\,1\,\,j\smallminus1\,\,j}\ab{B\,\,1\,\,j\,\,j\smallplus1}-\ab{A\,\,1\,\,j\,\,j\smallplus1}\ab{B\,\,1\,\,j\smallminus1\,\,j}\right]^2=\left[\ab{AB\,\,1\,\,j}\ab{1\,\,j\smallminus1\,\,j\,\,j\smallplus1}\right]^2.\ee In general, all of these terms contain both physical as well as spurious poles. Physical poles are denominator factors of the form $\langle AB\,\,i\,\,i\smallplus1\rangle$ and $\langle i\,\,i\smallplus1\,\,j\,\,j\smallplus1\rangle$ while spurious poles are all other denominator factors. We often refer to physical poles as local poles and to spurious poles as non-local. A small explanation for the ``non-local" terminology is in order. Consider the 5-particle amplitude as an example, where there are three terms in the integrand.
These three terms are \be \begin{split}
  &\frac{\ab{1234}^2}{ \ab{ AB12}\ab{AB 23}\ab{AB34}\ab{AB14}} +
     \frac{\ab{AB| \,(12 3) \cap (1  4  5)}^2}{\ab{AB12}\ab{AB23} \ab{AB  31}  \ab{AB14}\ab{AB  4  5}\ab{AB  5  1} }
\\
 & \frac{\ab{3451}^2}{ \ab{ AB34}\ab{AB 45}\ab{AB51}\ab{AB31}}. \end{split} \ee
The spurious poles are $\ab{AB14}$ and $\ab{AB13}$. The line defined by $Z_1$ and $Z_3$ corresponds to a complex point, but what makes $\ab{AB13}$ non-local? The reason is that in field theory $1/\ab{AB13}$ could only come from a loop integration, {\it e.g.}~it is generated by a local one-loop integral of the form
\be
\int\left[\frac{d^4Z_C d^4Z_D}{{\rm vol[GL(2)]}}\right]
\frac{\langle CD|(512)\cap (234) \rangle}{\langle CD AB\rangle \langle CD 51
\rangle \langle CD 12 \rangle \langle CD 23 \rangle\langle CD 34 \rangle}.
\ee
(This is also the secret origin of the non-local poles in BCFW at tree-level.)

Back to the 5-particle example, $\ab{AB14}$ and $\ab{AB31}$ occur each in two of the three terms and they cancel in pairs. Indeed upon
collecting denominators we find, after repeated uses of the Schouten identity, the result for the sum
\be - \frac{\ab{AB12}\ab{2345}\ab{1345} + \ab{AB23}\ab{1345}\ab{1245} + \ab{AB13}\ab{1245}\ab{3245} + \ab{AB45}\ab{1234}\ab{1235}}
{\ab{ AB12}\ab{AB 23}\ab{AB34}\ab{AB45}\ab{AB51}}.
\ee
This is furthermore cyclically-invariant,
albeit in a nontrivial way involving Schouten identities.

Let us also briefly discuss the 6-particle NMHV amplitude at 1-loop. The full integrand has 16 terms which differs even more sharply from familiar local forms of writing the amplitude. As we will review in the next section, the usual box decomposition of 1-loop amplitudes does not match the full integrand (only the ``parity-even" part of the integrand); even so, there is a natural generalization of the basis of integrals that can be used to match the full integrand in a manifestly dual conformal invariant form. Any such representation, however, will have the familiar form ``leading singularity/Grassmannian residue $\times$ loop integral". However, this is {\it not} the form we encounter with loop-level BCFW. Instead, the supersymmetric $\eta$-variables are entangled with the loop integration variables in an interesting way. For instance, one of the terms from the forward limit contribution to the 6-particle NMHV amplitude integrand is the following,
\begin{align*}
&\hspace{-0.5cm}\frac{\delta^{0|4}\left(\begin{array}{llllll}& \eta_1\ab{AB1|(23)\!\cap\!(456)}&+& \eta_2\ab{4561}\ab{AB31}&+& \eta_3\ab{4561} \ab{AB12}\\+& \eta_4\ab{AB|(123)\!\cap\!(561)}&+& \eta_5\ab{AB1|(46)\!\cap\!(123)}&+& \eta_6\ab{AB1|(123)\!\cap\!(45)}\end{array}\right)}{\ab{4561} \ab{AB45} \ab{AB61}\ab{AB12}\ab{AB23} \ab{AB13} \ab{AB41}\ab{AB|(123)\!\cap\!(456)} \ab{AB|(123)\!\cap\!(561)}}
\end{align*}
The full expression is given in appendix \ref{bcfw_6pt}.
Note the presence of the explicit $AB$-dependence in the argument of the Fermionic $\delta$-function. Seemingly miraculously, when the residues of this integral are computed on its leading singularities, the $\eta$-dependence precisely reproduces the standard NMHV $R$-invariants. Of course this miracle is guaranteed by our general arguments about the Yangian-invariance of these objects.

\subsection{Unitarity as a Residue Theorem}

The BCFW construction of tree-level amplitudes make Yangian-invariance manifest, but are not manifestly cyclic-invariant. The statement of cyclic-invariance is then a remarkable identity between rational functions. Of course one could say that the field theory derivation of the recursion relation gives a proof of these identities, but this is quite a circuitous argument. One of the initial striking features of the Grassmannian picture for tree amplitudes was that these identities were instead a direct consequence of the global residue theorem applied to the Grassmannian integral. This observation ultimately led to the ``particle interpretation" picture for the tree contour, giving a completely autonomous and deeper understanding of tree amplitudes, removed from the crutch of their field theory origin.

In complete analogy with BCFW at tree-level, the BCFW construction of the loop integrand is not manifestly cyclically-invariant. Again cyclic-invariance is a remarkable identity between rational functions, and again this identity can be thought of as a consequence of the field theory derivation of the recursion relation. But of course we strongly suspect that there is an extension of the ``particle interpretation" picture that gives a completely autonomous and deeper understanding of loop amplitudes, independent of any field theoretic derivation.

Just as at tree-level, a first step in this direction is to find a new understanding of the cyclic-invariance identities. To whit, we have understood how the cyclic-identity for the 1-loop MHV amplitude can be understood as a residue theorem; we very briefly outline the argument here, deferring a detailed explanation to \cite{InPrep}. The idea is to identify the terms appearing in the MHV 1-loop formulas as the residues of a new Grassmannian integral. All the terms in the MHV 1-loop formula can actually be thought of as arising from $\int d^{3|4} {\cal Z}_A d^{3|4} {\cal Z}_B Y_{n+2,k=2}({\cal Z}_A,{\cal Z}_B, \ldots)$, where $Y_{n+2,k=2}$ is computed from the $G(2,n+2)$ Grassmannian integral. Note that ${\cal Z}_A,{\cal Z}_B$ appear in the delta functions of the integral in the combination $C_{\beta A} {\cal Z}_A + C_{\beta B} {\cal Z}_B$, so the GL(2)-action on $({\cal Z}_A,{\cal Z}_B)$ also acts on $(C_{\beta A},C_{\beta B})$. Performing the $\eta_{A,B}$ and GL(2)-integrals leaves us with a new Grassmannian integral:
\be\label{auxGras}
\int d^{2 \times (n+2)}C_{\beta a} \frac{\delta^4(C_{\beta i} Z_i + C_{\beta A} Z_A + C_{\beta B} Z_B) (AB)^2}{(12)(23)\cdots(n1)}.
\ee
By construction, this integral has a GL(2)-invariance acting on columns $(A,B)$ and $(Z_A,Z_B)$, and hence all of is residues are only a function of the line $(Z_A Z_B)$. In particular all terms appearing in the MHV 1-loop formula, after GL(2) integration, are particular residues of this Grassmannian integral.

As we will discuss at greater length in \cite{InPrep}, the equality of cyclically-related BCFW expressions of the 1-loop amplitude follows from a residue theorem applied to this integral. In fact, it can be shown that the {\it only} combination of these residues that is free of spurious poles is the physical 1-loop amplitude.

At tree level, the cyclic-identity applied to {\it e.g.}~NMHV amplitudes ensures the absence of spurious poles. The same is true at 1-loop level. Since the BCFW formula manifestly guarantees that one of the single cuts is correctly reproduced, cyclicity guarantees that {\it all} the single cuts are correct. Having all correct single cuts, automatically ensures that all higher cuts---and in particular unitarity cuts---are correctly reproduced. Unitarity then finds a deeper origin in this residue theorem.

\section{The Loop Integrand in Local Form}\label{local_loop_integrands}

We have seen that the loop integrand produced by BCFW consists of a sum over non-local terms. In order to present the results in a more familiar form, and also as a powerful check on the formalism, it is interesting to instead re-write the integrand in a manifestly local way (which will of course spoil the Yangian-invariance of each term). We will do this for a number of multi-loop examples in the next section, but first we must describe a new basis of local loop integrals which differs in significant ways from the standard scalar integrals, but which will greatly simplify the results and make the physics much more transparent.

Loop amplitudes are normally written as scalar integrals\footnote{Here we abuse terminology and use the term ``scalar", which is appropriate at one-loop, to refer to possibly tensor integrals at higher-loop order where the tensor structure is the product of ``local" factors, {\it i.e.}, of the form $\langle (AB)_{\ell}\,\,i\,\,i\smallplus1\rangle$ and $\langle (AB)_{\ell} (AB)_{k}\rangle$.} with rational coefficients. Obviously this form can not match the full loop integrand, since scalar integrals are even under parity but the amplitude is chiral. Let's consider one-loop integrals to begin the discussion. In the usual way of discussing the integral reduction procedure, manipulations at the level of the integrand reduces integrals down to pentagons \cite{vanNeerven:1983vr}. The final reduction to the familiar boxes uses the fact that the parity-odd parts of the integrand integrate to zero.

We are instead interested in the full integrand, however, and since the amplitudes aren't parity symmetric, there is no natural division between ``parity-odd" and ``parity-even". In fact,
for the purpose of writing recursion relations, it is crucial to know both. Furthermore, the BCFW recursion relation guarantees that the loop integrand is dual conformally invariant and thus most usefully discussed in momentum-twistor space. We are then led to construct a novel basis of naturally chiral integrals written directly in momentum-twistor space, as we now briefly describe. These issues will be discussed at much greater length in \cite{InPrep}.

Let's look at a few quick examples of local integrals written in momentum-twistor space. We have encountered the simplest example
already; the zero mass integral at 1-loop \be \int \left[\frac{d^4 Z_A d^4
Z_B}{{\rm vol[GL(2)]}}\right] \frac{\langle 1 2 3 4 \rangle^2}{\langle AB 12
\rangle \langle AB 23 \rangle \langle AB 34 \rangle \langle AB 41
\rangle}. \ee Henceforth, we will drop the integration measure and only write the integrand.
The most general $1$-loop integrand is of the form
\be
\frac{\ab{ABY_1} \ldots \ab{ABY_{n-4}}}{\ab{AB12} \ab{AB23} \cdots \ab{ABn1}},
\ee
where $Y^{IJ}_1, \ldots, Y^{IJ}_{n-4}$ are general $4\times4$ antisymmetric matrices or `bitwistors'; with 6 independent components.
Momentum-twistors make integral reduction trivial. Suppose there are 6 or more local propagator factors including $\ab{ABj_1 j_1 \smallplus 1} \cdots \ab{ABj_6 j_6\smallplus1}$ in the denominator. We can always expand all the $Y^{IJ}$'s in a basis of the 6 bitwistors $Z^{[I}_{j_1} Z^{J]}_{j_1\smallplus1}, \ldots, Z^{[I}_{j_6} Z^{J]}_{j_6\smallplus1}$. Inserting this expansion into the integrand, each term knocks-out a propagator from the denominator. Thus we can reduce any integral down to pentagons.

These will contain 5 $``AB"$ factors in the denominator and a single
$``AB"$ factor in the numerator. In the literature, $x$-space
loop integrals are written with numerator factors like $(x -
x_j)^2$, which in momentum-twistor space correspond to $\langle AB \,\,j \,\, j\smallplus1 \rangle$. However, we will find more general numerators to be more natural. For instance, a typical pentagon integrand we consider takes the
form \be \frac{\langle AB 14 \rangle \ab{5123}\ab{2345}}{\langle AB 12 \rangle \langle AB 23
\rangle \langle AB 34 \rangle \ab{AB 45} \langle AB 51
\rangle}.\label{pentagon_integral}\ee

We can trivially translate this integral into
$x$-space; the numerator is proportional to $(x -
x_{14})^2$, where $x_{14}$ is a complex point associated with the
line $(14)$ in momentum-twistor space; specifically, the pentagon-integral (\ref{pentagon_integral}) is given by
\begin{equation} \frac{\langle 14\rangle\langle 23\rangle}{\langle 12\rangle\langle 34\rangle}\int\!\!d^4x \frac{(x-x_{14})^2x_{13}^2x_{35}^2}{(x-x_1)^2(x-x_2)^2(x-x_3)^2(x-x_4)^2(x-x_5)^2},
\quad\mathrm{with}\quad
x_{14}\equiv\frac{|1\rangle x_4|4\rangle-|4\rangle x_{1}|1\rangle}{\ab{14}}.
\end{equation}
The complex point $x_{14}$ is null-separated from $x_1,x_2,x_4$ and $x_5$; the second point sharing this property
is its parity conjugate which will be described shortly.
These complex points play an important role in the story, and it is most convenient to discuss them
on an equal footing with the rest of the points by working directly
with momentum-twistor space integrands.

Notably, unlike standard scalar
integrals, this pentagon integral is {\it chiral}. Like any pentagon
integral, it has 5 quadruple cuts and twice as many leading singularities. But unlike a generic pentagon integral, with
this special numerator, half of the leading singularities vanish, and the
others are all equal up to sign---hence, we say that this integral has ``unit leading
singularities". All of the local integrals we consider have this quite remarkable feature.

Local momentum-twistor space integrals can be drawn in exactly the same way as familiar planar integrals in $x$-space; we introduce a new bit of notation to denote the numerator factors. The pentagon integral we just discussed is drawn as,
\be
\raisebox{-1.25cm}{\includegraphics[scale=0.375]{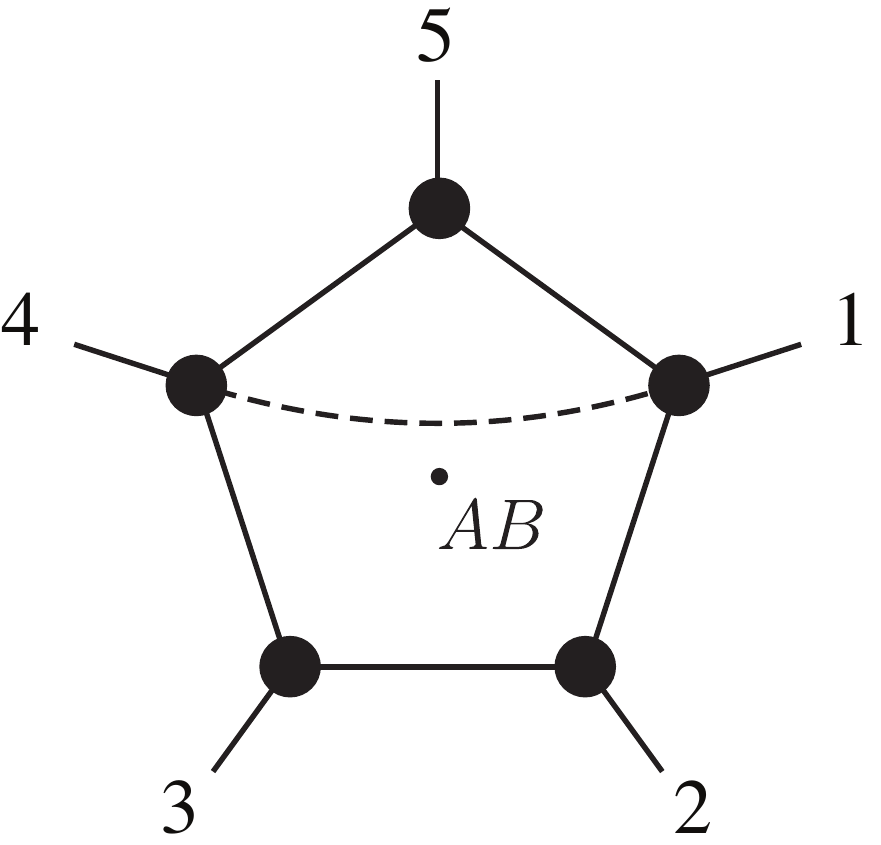}}\ee
where the dashed line connecting $(1,4)$ denotes the numerator factor $\ab{AB14}$. We will also have recourse to use the parity conjugates of these lines. The point $Z_i$ in momentum twistor space is naturally paired with its projective-dual plane $W_i = (i\smallminus1\,\, i \,\, i\smallplus1)$, and the parity conjugate of a line $(ij)$ is the line which is the intersection of the corresponding planes $(\overline{ij}) \equiv (i\smallminus1 \,\, i \,\, i\smallplus1) \cap (j\smallminus1 \,\, j \,\, j\smallplus1)$. The numerator factor \mbox{$\ab{AB \overline{ij}}\equiv \ab{A\,\,i\smallminus1 \,\, i \,\, i\smallplus1} \ab{B\,\,j\smallminus1\, \, j \,\, j\smallplus1} - \ab{B\,\,i\smallminus1\, \, i \,\, i\smallplus1}\ab{A\,\,j\smallminus1 \,\, j \,\, j\smallplus1} $} will be denoted by a wavy-line connecting $i,j$.

With this notation we can nicely write the integrand for $n$-particle 1-loop MHV amplitudes as
\be\frac{1}{n}\left(\!\!\!\!\!\begin{array}{c@{~}c@{~}c@{~}c@{~}c@{~}c@{~}c@{~}c@{}}\raisebox{-1.25cm}{\includegraphics[scale=0.65]{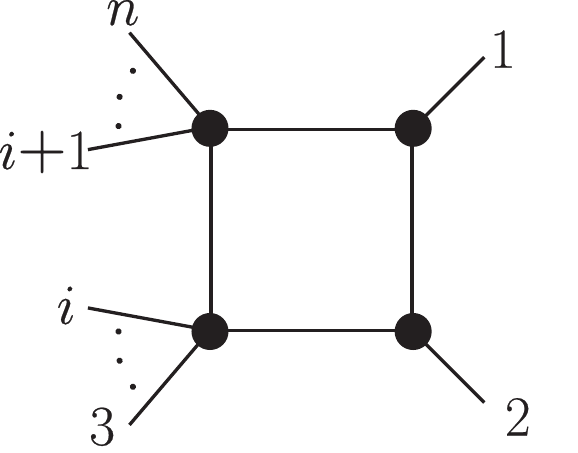}}&+\quad&\raisebox{-1.55cm}{\includegraphics[scale=0.75]{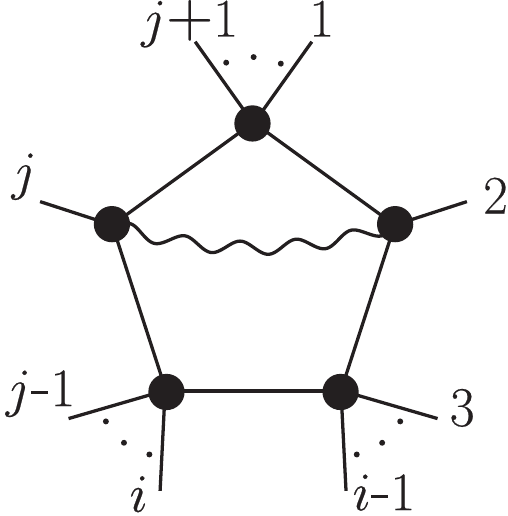}}&\hspace{-0.6cm}+\;\mathrm{cyclic}\\
\begin{array}{c}\ab{n\,1\,2\,3} \ab{1\,2\,\,i\,\,i\smallplus1}\\~\\2<i<n\end{array}&&\begin{array}{c}\ab{2\,\,j\,\,i\smallminus1\,\,i}\\\times\ab{AB|(123)\cap(j\smallminus1\,\,j\,\,j\smallplus1)}\phantom{\times}\\3<i<j\leq n\end{array}\end{array}\!\!\!\right).
\ee
In this expression we sum over all cyclic integrands, including duplicates, which is related to the presence of the $1/n$ pre-factor.

For definiteness, we have indicated the numerator factor beneath the corresponding picture. Recall the familiar form of the MHV amplitude as a sum over all 2-mass easy boxes; it is amusing that in our form the only boxes are 2-mass hard.  The algorithm by which this form was deduced will be explained shortly.

We pause to point out that the full integrand for some MHV amplitudes have been computed in the literature, in the context of using the leading singularity method to determine the integrand \cite{Cachazo:2008vp}. A peculiarity in these papers was that the set of integrals that were used to match all the leading singularities did not appear to be manifestly dual conformal invariant---which is particularly ironic, given that the leading singularities themselves are fully Yangian-invariant! This led some authors to the conclusion that the parity-odd parts of the amplitude are somehow irrelevant, since they not only integrate to zero on the real contour but are also not dual conformal invariant. Of course, nothing could be further from the truth: we have seen very clearly that the {\it full} integrand is determined recursively and exhibits the Yangian symmetry of the theory; the decomposition into parity even and odd parts is artificial. The problem is quite simple, the basis of scalar integrals has only parity even elements! Therefore, one is trying to model the full integrand with a very inappropriate basis.

From the momentum-twistor viewpoint, the source of the previous difficulties can be seen quite explicitly. We have seen that all 1-loop integrals can be reduced to pentagons, but these are {\it tensor} pentagons, {\it i.e.}~with factors of $AB$ in the numerator. Now, it {\it is} possible to further reduce a pentagon with numerator $\ab{AB Y}$, with $Y$ corresponding to a real line or not, to a scalar pentagon integral, by expanding $Y$ in a basis of the 5 bitwistors appearing in the denominators, together with the infinity twistor $I^{IJ}$. But this breaks manifest dual conformal invariance! Thus the integrands obtained in \cite{Cachazo:2008vp,Cachazo:2008hp,Spradlin:2008uu} are indeed dual conformal invariant, but the symmetry was obscured by insistence to use scalar integrals.

Let's give an example of an interesting two-loop integrand using our notation:
\be
\frac{\ab{1345} \ab{5613} \ab{AB46} \ab{CD|(234) \cap (612)}}{\ab{CD61} \ab{CD12} \ab{CD23} \ab{CD34} \ab{ABCD} \ab{AB34} \ab{AB45} \ab{AB56} \ab{AB61}}\ee
which we draw as
\begin{equation}\vspace{-0.2cm}
\includegraphics[scale=0.45]{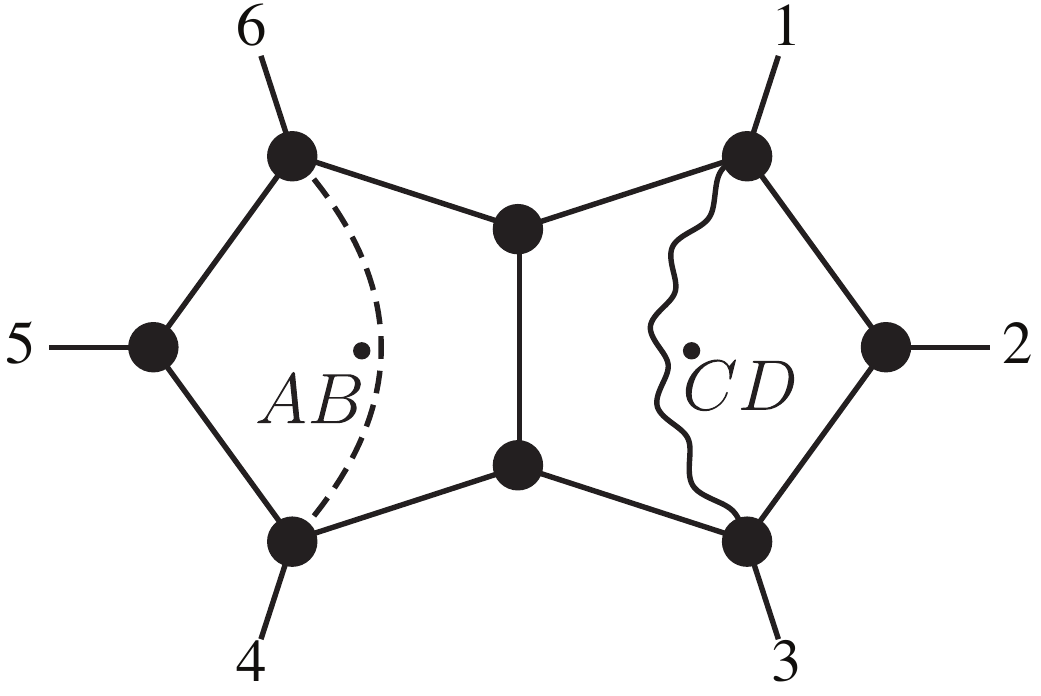} \nonumber
\end{equation}
At two-loops, there are generally 4 solutions to cutting any eight propagators, and so this integral has $9 \times 4 = 36$ different (non-composite) leading singularities. However, the integral is maximally chiral: putting any choice of eight propagators on shell will have only {\it one} solution with a non-vanishing residue. Moreover, the non-vanishing residues are equal up to a sign. This non-trivial fact can be understood as following from the global residue theorem applied to the integral. All the tensor integrals we write in this paper are {\it chiral} in this sense, and the overall normalization of each has been chosen so that all its non-vanishing leading singularities are equal to $\pm1$.

These chiral momentum-twistor integrals have another remarkable feature: they are less IR-divergent than generic loop integrals; indeed, many of them are completely IR-finite. Infrared divergences arise when the loop momenta become collinear with the external momenta $p_j$. In the dual co-ordinate space, this happens when a loop-integration variable $x$ lies on the line connecting $x_j$ and $x_{j+1}$. In momentum-twistor space, this corresponds to configurations where the associated line $(AB)$ passes through the point $Z_j$ while lying in the plane
$(j\smallminus1 \,\, j \,\, j\smallplus1)$. An integral is IR-finite if the numerator factors have a zero in the dangerous configurations. There are an infinite class of IR-finite integrals at any loop order; for instance, it is easy to see that the two-loop example above is IR-finite. Further discussion of these objects and their role in determining IR-finite parts of amplitudes like the remainder \cite{Alday:2007he} and ratio \cite{Drummond:2008bq} functions will be carried out in \cite{InPrep}. Of course we expect that IR finite quantities, such as the ratio function, are manifestly finite already at the level of the integrand.

It is interesting that the naively ``hardest" multi-loop integrands can be reduced to finite integrals plus simpler integrals. Consider for instance a general double pentagon integrand for six particles, of the form 
\be 
\frac{\ab{ABY_1} \ab{CDY_2}}{\ab{CD61} \ab{CD12} \ab{CD23} \ab{CD34} \ab{ABCD} \ab{AB34} \ab{AB45} \ab{AB56} \ab{AB61}}.
\ee
We can expand $Y_1$ in terms of the 6 bitwistors $(Z_3 Z_4),(Z_4 Z_5),(Z_5 Z_6),(Z_6 Z_1)$ as well as the bitwistors corresponding to $(46)$ and 
its parity conjugate $(\overline{46})$. Similarly we can expand $Y_2$ in terms of $(Z_1 Z_2),(Z_2 Z_3),(Z_3 Z_4), (Z_6 Z_1)$ as well as $(31)$ and $(\overline{31})$. Doing this reduces the integral to finite double-pentagon integrals, plus simpler pentagon-box and double-box integrals. 

Finally, let us describe the general algorithm which we used to find local forms of the loop integrands.
The first step is to construct an algebraic basis of dual conformal-invariant integrals, over which the integrand is to be expanded.
It turns out, quite remarkably, that for at least 1- and 2-loops an (over-complete) algebraic basis can be
constructed which contains exclusively integrals with unit leading singularities, in the sense just defined. We have explicitly constructed such
a bases at 1- and 2-loops and arbitrary $n$ \cite{InPrep}.
The second step is to match the integrand as generated by equation~(\ref{loop_level_BCFW}) with a linear combination of the basis integrals.
Since the loop integrand is a well-defined function of external momenta and loop momenta,
this can be done by simply evaluating it at sufficiently many random points.  Numerical evaluation of the integrand is itself quite fast.
Finally, this procedure is greatly facilitated by the fact that, when using our particular integral basis,
the coefficients are guaranteed to be pure numbers (or multiple of leading singularities, for arbitrary N$^k$MHV), as opposed to arbitrary rational functions of the external momenta.

\section{Multi-Loop Examples}\label{multiloop_examples}
The recursion relation for loops gives a completely systematic way of determining the integrand for amplitudes with any $(n,k,\ell)$. All the required operations are completely algebraic and can be easily automated. In this section we use the recursion relation to present a number of multi-loop results.

As we have stressed repeatedly, the individual terms in the BCFW expansion of the loop integrand have spurious poles and are also not manifestly cyclically-invariant; thus as a very strong consistency check on our results, necessary for a local form to exist, we verify that the integrand is free of all spurious poles: the only poles in the integrand should be of the form $\ab{i\smallminus1\,\,i\,\,j\smallminus1\,\,j}, \ab{(AB)_{\ell}\,\,j\smallminus1 \,\, j}, \ab{(AB)_{\ell_1}(AB)_{\ell_2}}$. We also explicitly check cyclic-invariance. Recall that the absence of spurious poles and cyclicity guarantees that all single-cuts of the amplitude are reproduced, and thus {\it all} cuts are automatically correctly matched. While preparing this paper we have explicitly checked that our recursive determination of the integrand passes these checks up to 14 pt N$^4$MHV amplitudes at 1-loop, 22-pt MHV amplitudes at 2-loops, 8 pt NMHV amplitudes at 2-loops and 5-pt MHV amplitude at 3 loops.

We can expand the integral in a local basis of chiral momentum-space integrals with unit leading singularities using the algorithm briefly described in the previous section. While the BCFW form of the integrand is almost always more concise than the local form, the local form is more familiar, so we will present the results in this way. Indeed, the (modestly) non-trivial work here is only in determining the natural basis for local integrands. While this is a straightforward exercise using momentum-twistor machinery, the result is non-trivial, yielding a canonical basis of multi-loop integrals, which we have constructed explicitly for all $n$ up to 2-loops. In order to present a tree-loop result, we also found the 5pt basis at three-loops, deferring a complete discussion to \cite{InPrep}. Given the basis of local integrals with unit leading singularities, generating the integrand and finding its expansion in the basis is not difficult. The natural basis is over-complete and so the results can be expressed in a number of equivalent forms. We will choose the forms that seem canonical and reveal patterns.  As we will see, somewhat surprisingly, the local forms are also often remarkably simple.

\subsection{All 2-loop MHV Amplitudes}

The two-loop amplitude for 4- and 5-particles is given by, respectively,
\be~\hspace{1.75cm}~\begin{array}{c@{~}c@{~}c@{~}c@{~}c@{~}c@{~}c@{~}c@{}}\raisebox{-1.25cm}{\includegraphics[scale=0.45]{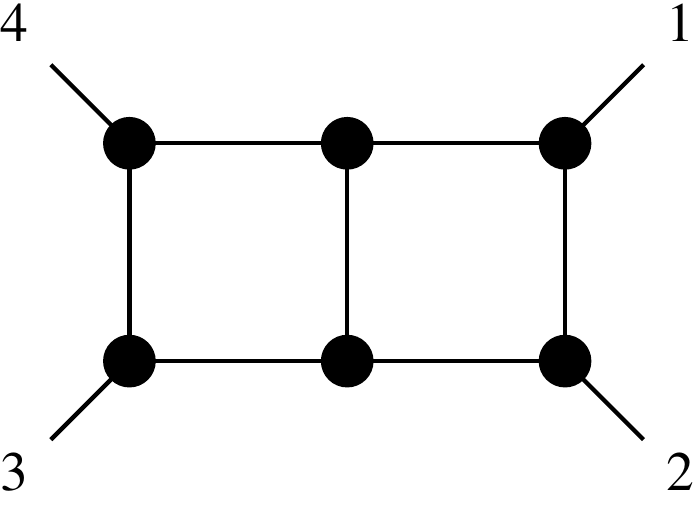}}&+\begin{array}{c}\\\mathrm{cyclic}\\\mathrm{(no~repeat)}\end{array}\\
\begin{array}{c}\ab{2341} \ab{3412}\ab{4123}\end{array}\end{array}
\ee
and\be~\hspace{0.75cm}~
\begin{array}{c@{~}c@{~}c@{~}c@{~}c@{~}c@{~}c@{~}c@{}}\raisebox{-1.25cm}{\includegraphics[scale=0.45]{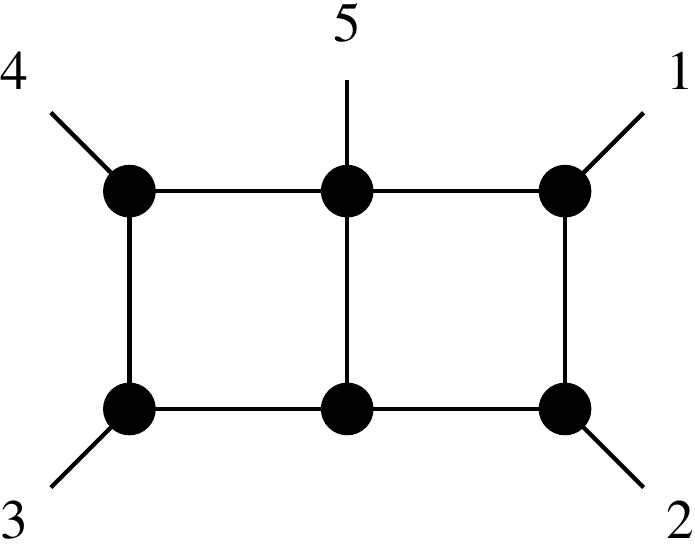}}&+&\raisebox{-1.25cm}{\includegraphics[scale=0.45]{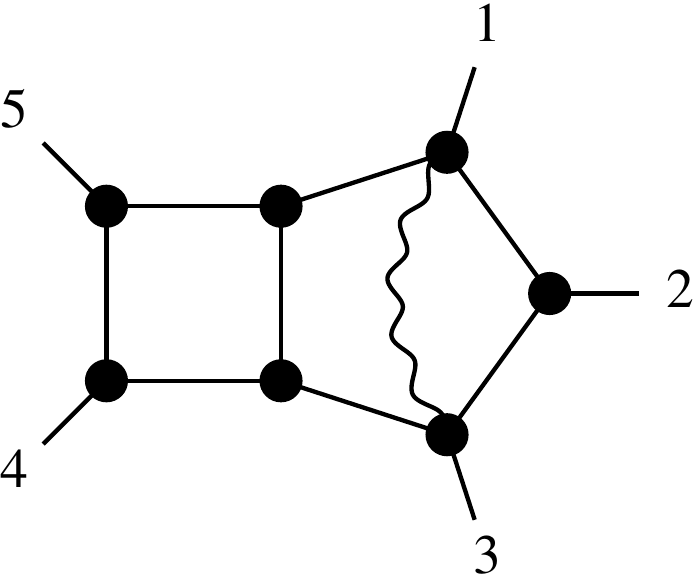}}&+\begin{array}{c}\\\mathrm{cyclic}\\\mathrm{(no~repeat)}\end{array}\\\begin{array}{c}\ab{2345} \ab{5123}\ab{3412}\\~\end{array}&&\hspace{-0.5cm}\begin{array}{c}\ab{3451} \ab{4513}\\\times \ab{AB|(512)\!\cap\!(234)}\phantom{\times}\end{array}\end{array}\label{five_pt_MHV_two_loops}
\ee
while the 6-particle amplitude is
\be\begin{split}
&\begin{array}{c@{~}c@{~}c@{~}c@{~}c@{~}c@{~}c@{~}c@{}}\raisebox{-1.05cm}{\includegraphics[scale=0.45]{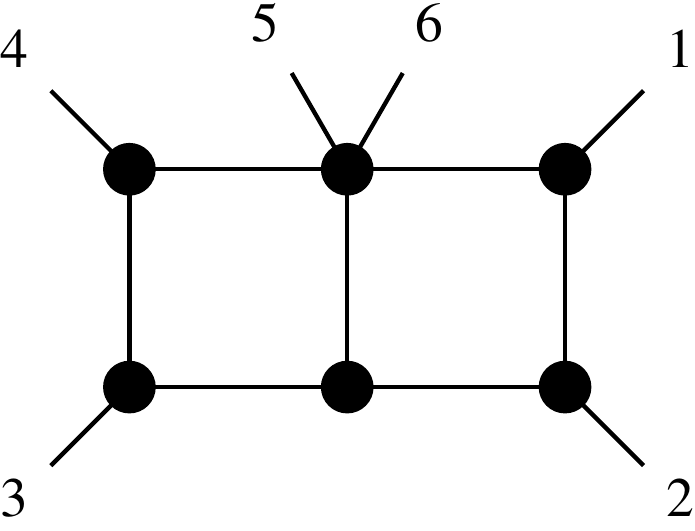}}&+&\raisebox{-1.25cm}{\includegraphics[scale=0.45]{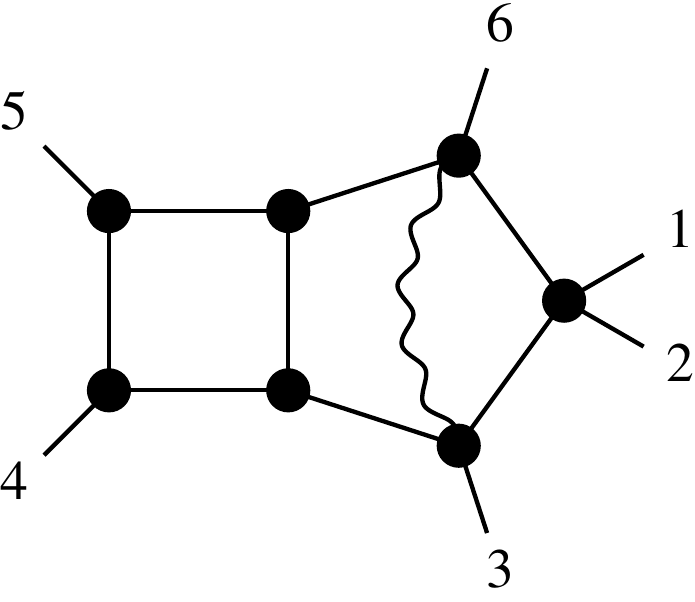}}&+&\raisebox{-1.25cm}{\includegraphics[scale=0.45]{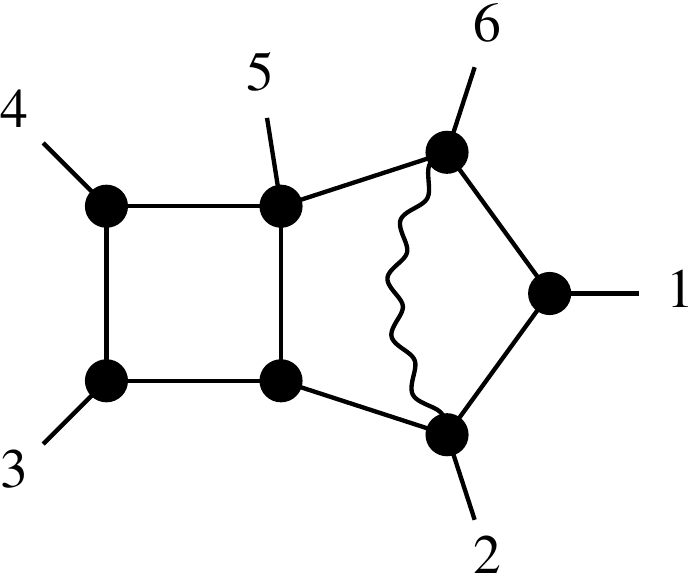}}&+&\raisebox{-1.25cm}{\includegraphics[scale=0.45]{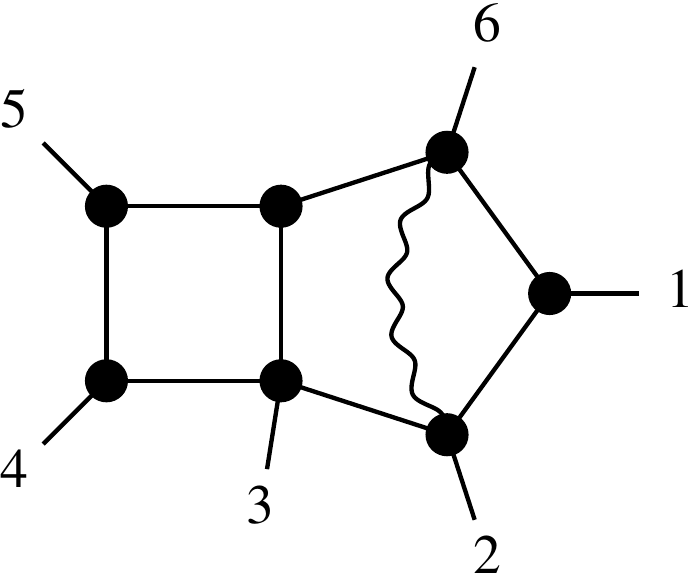}}\\\begin{array}{c}\ab{2345}\ab{6123}\ab{3412}\\~\end{array}&&\hspace{-0.5cm}
\begin{array}{c}\ab{3456} \ab{4563}\\\times\ab{AB|(561)\cap(234)} \phantom{\times}\end{array}&&\hspace{-0.5cm}\begin{array}{c}\ab{2345} \ab{3462}\\\times \ab{AB|(561)\cap(123)}\phantom{\times}\end{array}&&\hspace{-0.5cm}\begin{array}{c}\ab{3456} \ab{4562}\\\times \ab{AB|(561)\cap(123)}\phantom{\times}\end{array}\end{array}\\&\hspace{2.2cm}\begin{array}{c@{~}c@{~}c@{~}c@{~}c@{~}c@{~}c@{~}c@{}}+&\;\;\raisebox{-1.25cm}{\includegraphics[scale=0.45]{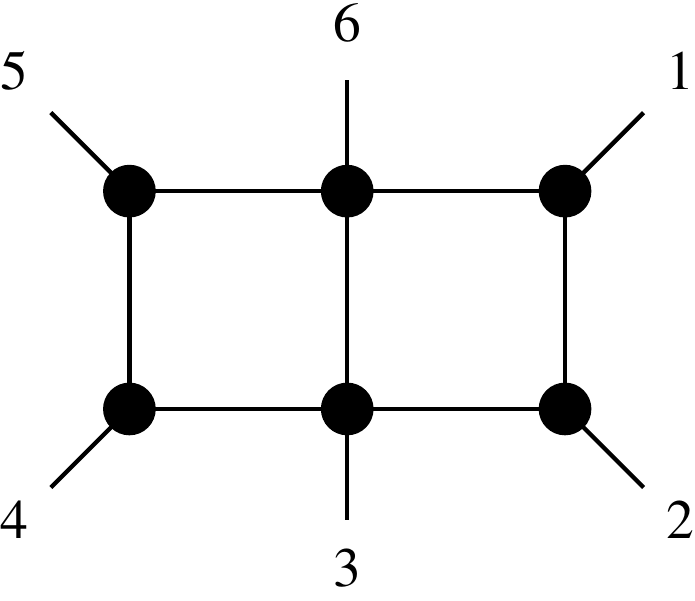}}&+&\raisebox{-1.205cm}{\includegraphics[scale=0.45]{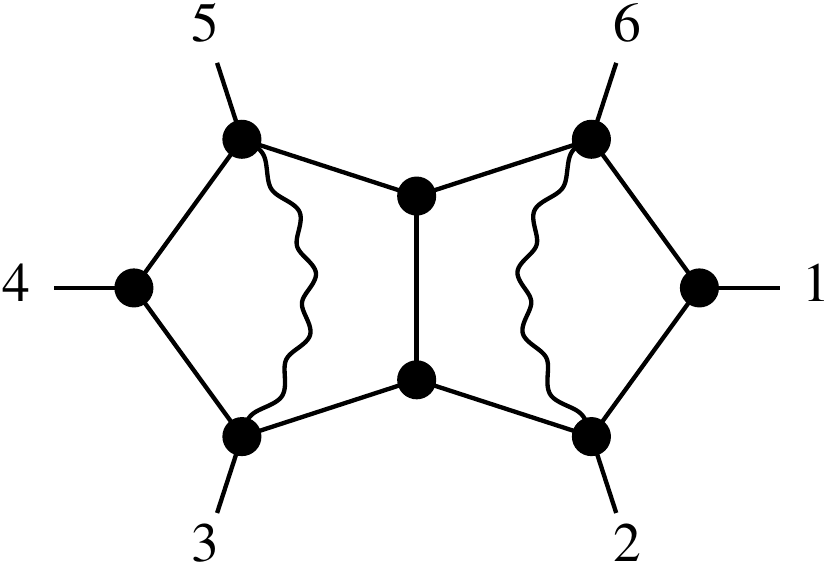}}&\qquad+\;\begin{array}{c}\\\mathrm{cyclic}\\\mathrm{(no~repeat)}\end{array}\\&\begin{array}{c}\ab{3456}\ab{6123}\ab{4512}\\~\\~\end{array}&&\begin{array}{c}\ab{6235}\\\times\ab{AB|(234)\cap(456)}\phantom{\times}\\\times\ab{CD|(561)\cap(123)}\phantom{\times}\end{array}&&\end{array}\end{split}
\ee
To be completely explicit, we have written the numerator factors accompanying each given term under its corresponding picture.

What about higher-points? The parity-even part of the integrand has been computed in \cite{Vergu:2009tu}, though the expressions are lengthy and do not expose a discernable pattern. However, looking at the \emph{full} (non-parity invariant) integrand for 4-, 5- and 6-particles in momentum-twistor space  reveals a clear pattern: the structure looks like the ``square" of the 1-loop objects, with double-box, pentagon-box and double-pentagon topologies. This motivates a simple conjecture for all 2-loop MHV amplitudes:
\be\begin{split}
&\begin{array}{c@{~}c@{~}c@{~}c@{~}c@{~}c@{~}c@{~}c@{}}\raisebox{-1.25cm}{\includegraphics[scale=0.5]{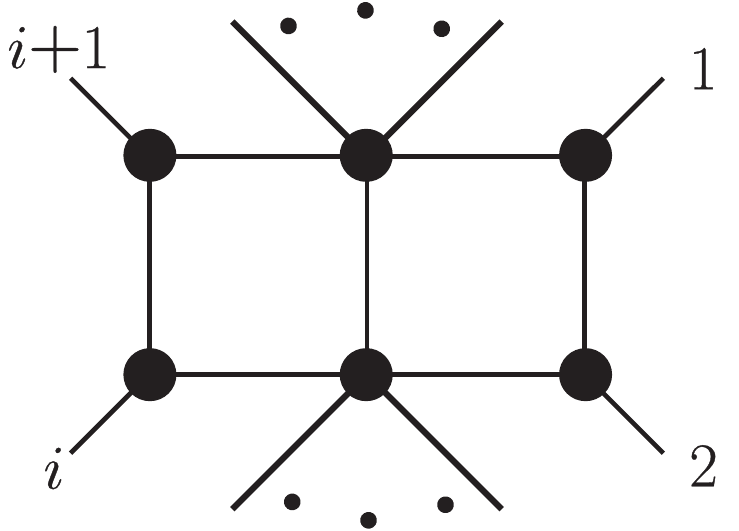}}&+&\hspace{-0.4cm}\raisebox{-1.5cm}{\includegraphics[scale=0.5]{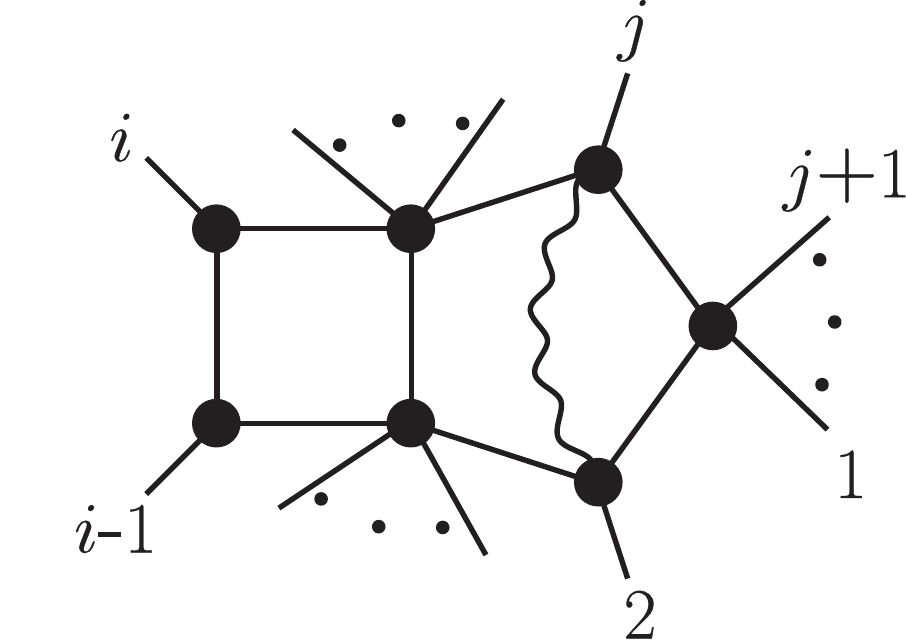}}&+&\raisebox{-1.5cm}{\includegraphics[scale=0.5]{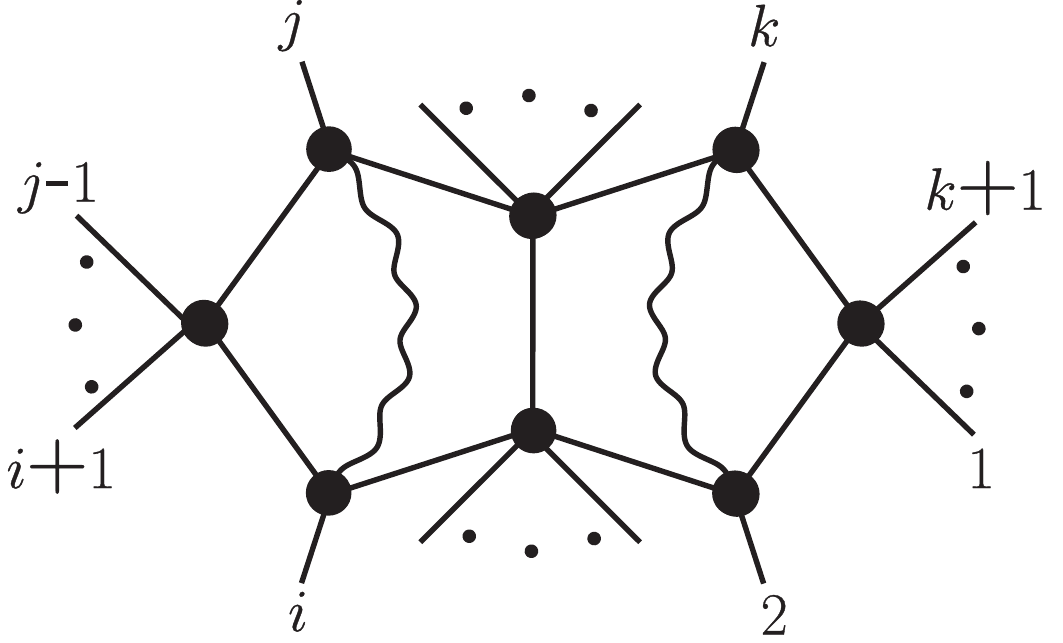}}\\\hspace{-0.5cm}
\begin{array}{c} \phantom{\times}\ab{n\,\,1\,\,2\,\,3} \times\\ \ab{1\,\,2\,\,i\,\,i\smallplus1} \ab{i\smallminus1\,\,i\,\,i\smallplus1\,\,i\smallplus2}\\~\\2<i<n\end{array}&&\hspace{-0.25cm}\begin{array}{c}\ab{2\,\,j\,\,i\smallminus1\,\,i} \ab{i\smallminus2\,\,i\smallminus1\,\,i\,\,i\smallplus1}\\\times \ab{AB|(123)\cap(j\smallminus1\,\,j\,\,j\smallplus1)}\phantom{\times}\\~\\3<i<j\leq n\end{array}&&\hspace{-0.0cm}\begin{array}{c}\ab{2\,\,i\,\,j\,\,k}\\\times \ab{AB|(123)\cap(k\smallminus1\,\,k\,\,k\smallplus1)}\phantom{\times}\\\times\ab{CD|(i\smallminus1\,\,i\,\,i\smallplus1)\cap(j\smallminus1\,\,j\,\,j\smallplus1)}\phantom{\times}\\2<i<j-1<k-1<n\end{array}\\\end{array}\end{split}
\ee
We checked numerically that this matches the 2-loop MHV integrand as calculated by BCFW directly.
Because the recursion relations are easily automated, this can be verified for any number of particles. 
We have checked this explicitly for up to 22 particles.  It is worth emphasizing that independent of verifying the local-ansatz, the cancellation of spurious poles (and propagators) is a particularly strong consistency check for the recursion relations. For instance, for the 22-point 2-loop MHV amplitude, there are exactly $49,590$ terms in the BCFW recursion, each riddled with spurious poles that cancel in the sum. Even a single sign-mistake would have spoiled this miracle. 

It is interesting to note that the na\"{i}vely ``hardest" integrals that appear here---the double pentagons---have a numerator which renders them completely finite. 

\subsection{2-loop NMHV Amplitudes}
Although structurally identical to the 2-loop 5-particle MHV amplitude, it is worth writing explicitly the 2-loop 5-particle NMHV amplitude; it is,\vspace{-0.25cm}
\begin{equation}\hspace{-1cm}[1\,2\,3\,4\,5]\left(\begin{array}{c@{~}c@{~}c@{~}c@{~}c@{~}c@{~}c@{~}c@{}}\raisebox{-1.25cm}{\includegraphics[scale=0.45]{522loopFig_1.pdf}}&+&\raisebox{-1.25cm}{\includegraphics[scale=0.45]{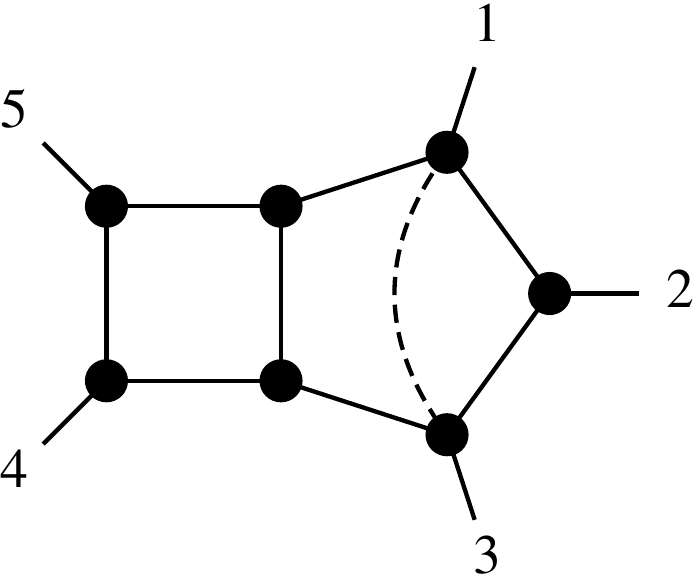}}&+\begin{array}{c}\\\mathrm{cyclic}\\\mathrm{(no~repeat)}\end{array}\\\begin{array}{c}\ab{2345} \ab{5123}\ab{3412}\\~\end{array}&&\hspace{-0.5cm}\begin{array}{c}\ab{2345}\ab{3451}\ab{4512}\\\times \ab{AB31}\phantom{\times}\end{array}\end{array}\right)\label{five_pt_NMHV_two_loops}\vspace{-0.35cm}
\end{equation}
Notice how this answer highlights the role played by parity: equations (\ref{five_pt_NMHV_two_loops}) and (\ref{five_pt_MHV_two_loops}) differ only by the parity of the numerator in the tensor-integral---and one can be obtained from the other simply by exchanging wavy- for dashed-lines.
\newpage 
Next we present the 6-particle 2-loop NMHV amplitude, written in the manifestly-cyclic form, \vspace{-0.2cm}
\begin{equation}
(1) I_1 + \, {\rm cyclic},\vspace{-0.2cm}\end{equation}
where $(1)$ is the Grassmannian residue given by the $R$-invariant $[2\,3\,4\,5\,6]$ written explicitly in equation (\ref{Rinv}). Below, we show the coefficient $I_1$ of residue $(1)$.
\begin{table}[t!]\centering\caption{Coefficients of residue $(1)=[2\,3\,4\,5\,6].$ Here, ``$g$'' rotates each figure by $g:i\mapsto i\smallplus1$, and $P$ exchanges wavy- and dashed-lines (together with each figure's corresponding normalization).\vspace{-0.00cm}}\mbox{\vspace{-7cm}\hspace{-0cm}\scriptsize\begin{tabular}{|@{~}l@{\hspace{-0.85cm}}c@{~}|@{~}l@{\hspace{-0.65cm}}c@{~}|@{~}l@{\hspace{-0.95cm}}c@{~}|}\hline
\multirow{2}{*}{\mbox{$  1$}}  & \raisebox{-1.5cm}{\includegraphics[scale=0.45]{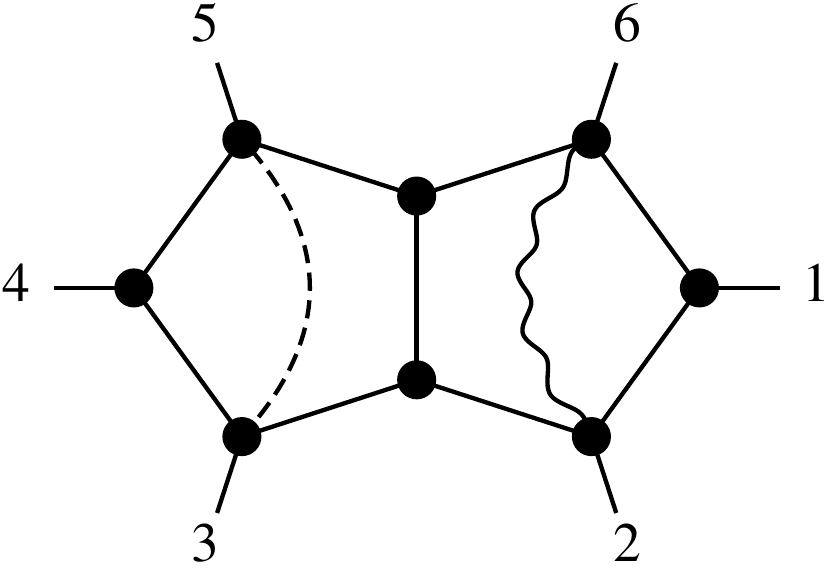}} & \multirow{2}{*}{\mbox{$\begin{array}{l}\phantom{\!-}1+g^3\\\!-g(1\!-\!g)(1\!-\!P)\end{array}$}\hspace{0.3cm}} &  \raisebox{-1.5cm}{\includegraphics[scale=0.405]{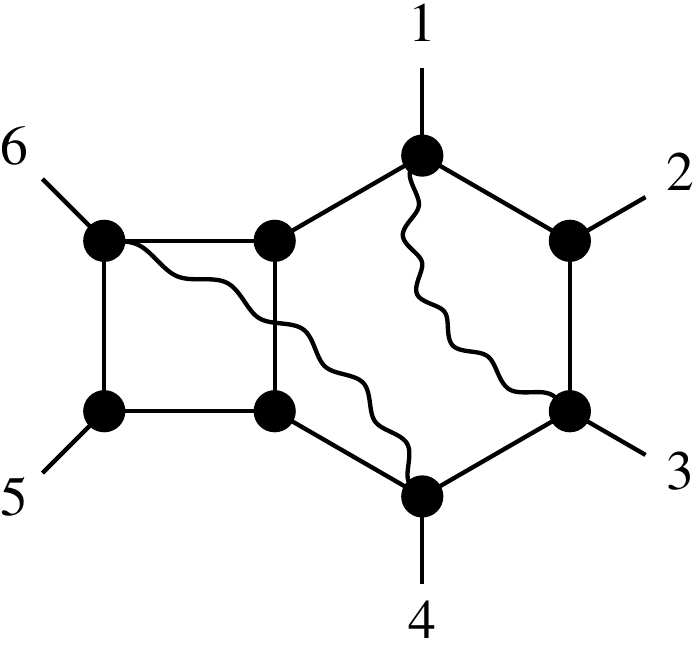}} &\multirow{2}{*}{\mbox{$(1+g^3P)$\hspace{1cm}}}  &  \raisebox{-1.5cm}{\includegraphics[scale=0.405]{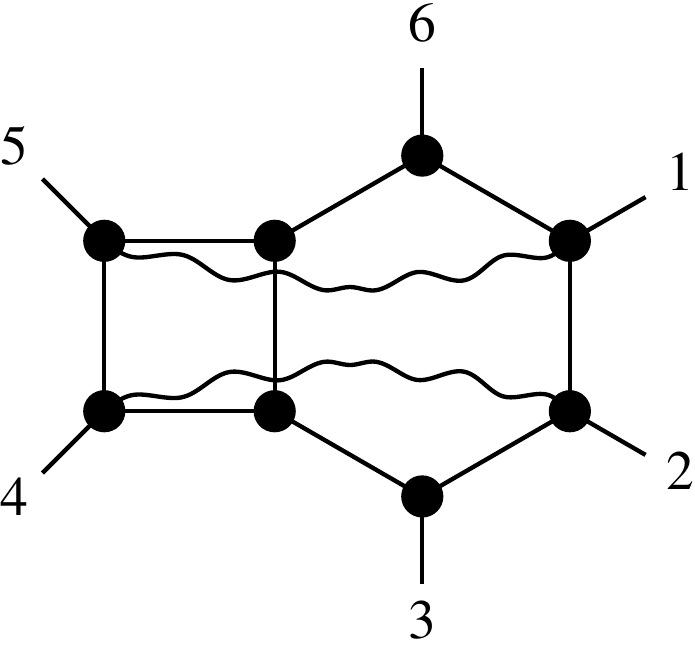}} \\[-5pt]&  \raisebox{-0.4cm}{\mbox{$\begin{array}{c}\ab{6234} \ab{6245}\\\times\ab{AB53} \ab{CD|(123)\!\cap\!(561)}\phantom{\times}\end{array}  $}} && \raisebox{-0.2cm}{\mbox{$\begin{array}{l}\phantom{\times}\ab{4561}\\\times\ab{AB|(345)\!\cap\!(561)}\phantom{\times}\\\times\ab{AB|(612)\!\cap\!(234)}\phantom{\times}\end{array}  $}}&&\raisebox{-0.2cm}{\mbox{$ \begin{array}{l}\phantom{\times}\ab{3456}\\\times\ab{AB|(123)\cap(345)}\phantom{\times}\\\times\ab{AB|(456)\cap(612)}\phantom{\times}\end{array}$}}  \\
\hline \multirow{2}{*}{\mbox{$(1+g^3P)$}} &  \raisebox{-1.5cm}{\includegraphics[scale=0.45]{632loopFig_4.pdf}}&\multirow{2}{*}{\mbox{$-(1+g^3P)$}}  &  \raisebox{-1.5cm}{\includegraphics[scale=0.45]{632loopFig_5.pdf}} &\multirow{2}{*}{\mbox{$\begin{array}{l}\phantom{\times\,}(1\!+\!g^3P)\\\times(1\!+\!g\!-\!g^3)\end{array}\hspace{1cm}$}} &  \raisebox{-1.5cm}{\includegraphics[scale=0.45]{632loopFig_6.pdf}} \\[-5pt]&  \hspace{-0.75cm}\mbox{$\begin{array}{c}\ab{3456} \ab{4563}\\\times\ab{AB|(561)\!\cap\!(234)} \phantom{\times}\end{array}$} &&  \hspace{-0.75cm}\mbox{$\begin{array}{c}\ab{2345} \ab{3462}\\\times \ab{AB|(561)\!\cap\!(123)}\phantom{\times}\end{array}  $}&&\hspace{-0.75cm} \mbox{$\begin{array}{c}\ab{3456} \ab{4562}\\\times \ab{AB|(561)\!\cap\!(123)}\phantom{\times}\end{array}  $} \\
\hline\multirow{2}{*}{\mbox{$(1\!-\!g\!+\!g^2)$}\hspace{0.4cm}}  &  \raisebox{-1.5cm}{\includegraphics[scale=0.375]{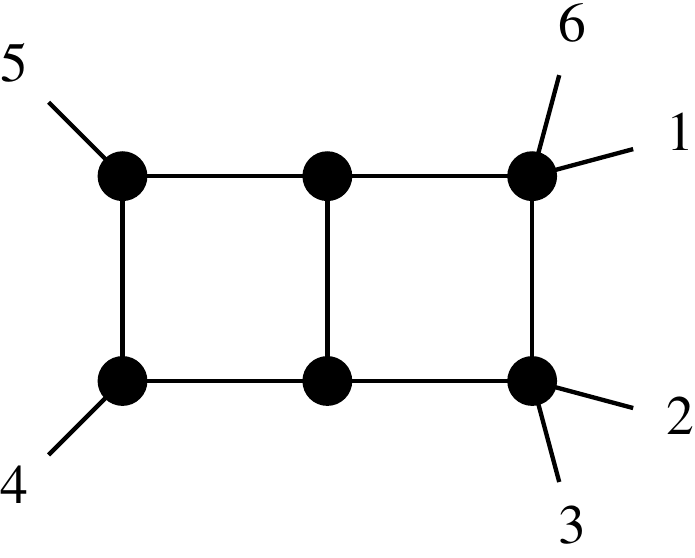}} &\multirow{2}{*}{\mbox{$(1\!+\!g^2\!+\!g^4)$}} &  \raisebox{-1.5cm}{\includegraphics[scale=0.375]{632loopFig_8.pdf}} &\multirow{2}{*}{\mbox{$  \frac{1}{2}\left(1\!+\!g^2\!+\!g^4\right)$}}  &  \raisebox{-1.5cm}{\includegraphics[scale=0.35]{632loopFig_9.pdf}} \\[-2.5pt]&  \mbox{$  \ab{3456}^{2} \ab{4512}  $}  && \mbox{$\ab{2345} \ab{3412} \ab{6123}  $} &&\mbox{$  \ab{3456} \ab{4512} \ab{6123}  $} \\\hline
\end{tabular}\vspace{1cm}}\end{table}

We next move to the 7-particle NMHV amplitude, which will be presented in the form,\begin{equation}
\left[(7)(1) I_{7,1} +{\rm cyclic} \right] + \left[(7)(2) I_{7,2} + {\rm cyclic} \right] + \left[(7)(3) I_{7,3} + {\rm cyclic}\right]
\end{equation}
where $(i)(j)$ is the Grassmannian residue given by the $R$-invariant defined by the complement of $\{i,j\}$ in $\{1,2,\ldots ,7\}$. The expressions for $I_{7,1},I_{7,2},I_{7,3}$ are given in appendix \ref{local_7pt}.

\newpage
\subsection{3-loop MHV Amplitudes}
The four-point three-loop amplitude is given by the cyclic-sum of the following two classes of integrands:\vspace{-0.2cm}
\begin{equation}
\hspace{1.75cm}\begin{array}{c@{~}c@{~}c@{~}c@{~}}\raisebox{-0.855cm}{\includegraphics[scale=0.425]{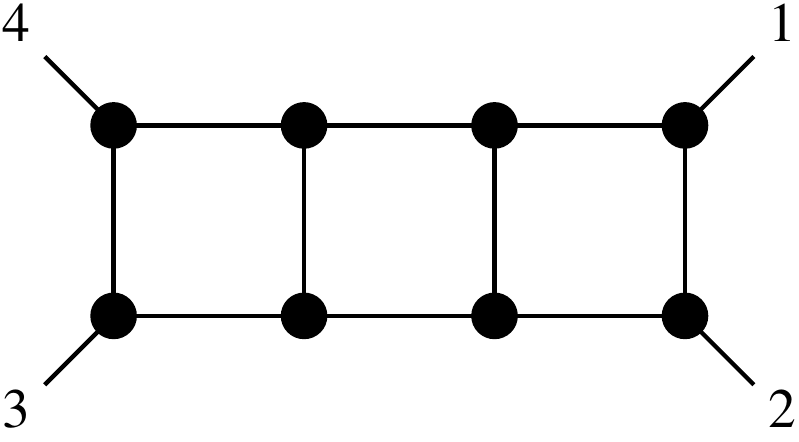}}&\quad+\;&\raisebox{-1.15cm}{\includegraphics[scale=0.4]{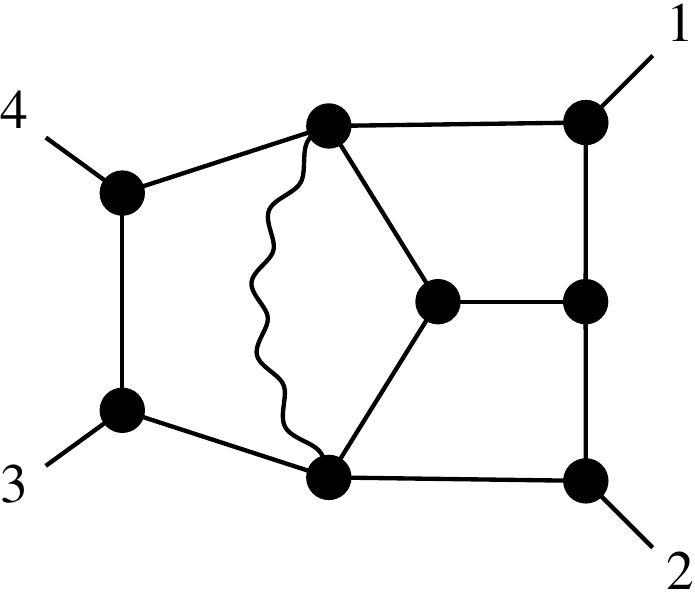}}&+\begin{array}{c}\\\mathrm{cyclic}\\\mathrm{(no~repeat)}\end{array}\\\ab{2341}^3\ab{3412}&&\begin{array}{c}\ab{2341}\ab{3412}\\\times\ab{AB|(412)\cap(123)}\phantom{\times}\end{array}\end{array}\vspace{-0.3cm}
\end{equation}
Although perhaps visually unfamiliar, the second integral above is commonly referred to as the ``tennis-court'' because of the way it is usually drawn. We have drawn it the way we have to highlight the presence of the pentagon sub-integral and the role played by the tensor-integral's numerator (which should be read as connecting to vertices ``1'' and ``2'').

Finally, we give the integrals contributing to the full 3-loop MHV amplitude for 5 particles. It is given by the following cyclic-sum of the integrands,\vspace{-0.3cm}
{\small\begin{equation}\begin{split}
&\hspace{-0.15cm}\begin{array}{c@{~}c@{~}c@{~}c@{~}c@{~}c@{~}c@{~}c@{}}\raisebox{-1.075cm}{\includegraphics[scale=0.35]{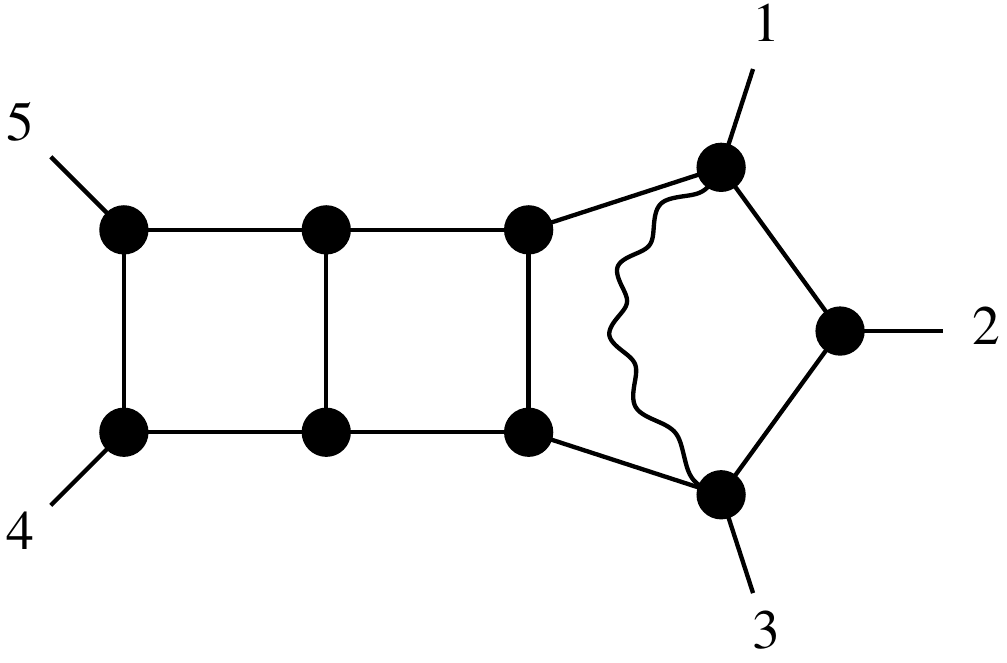}}&+&\raisebox{-1.075cm}{\includegraphics[scale=0.4]{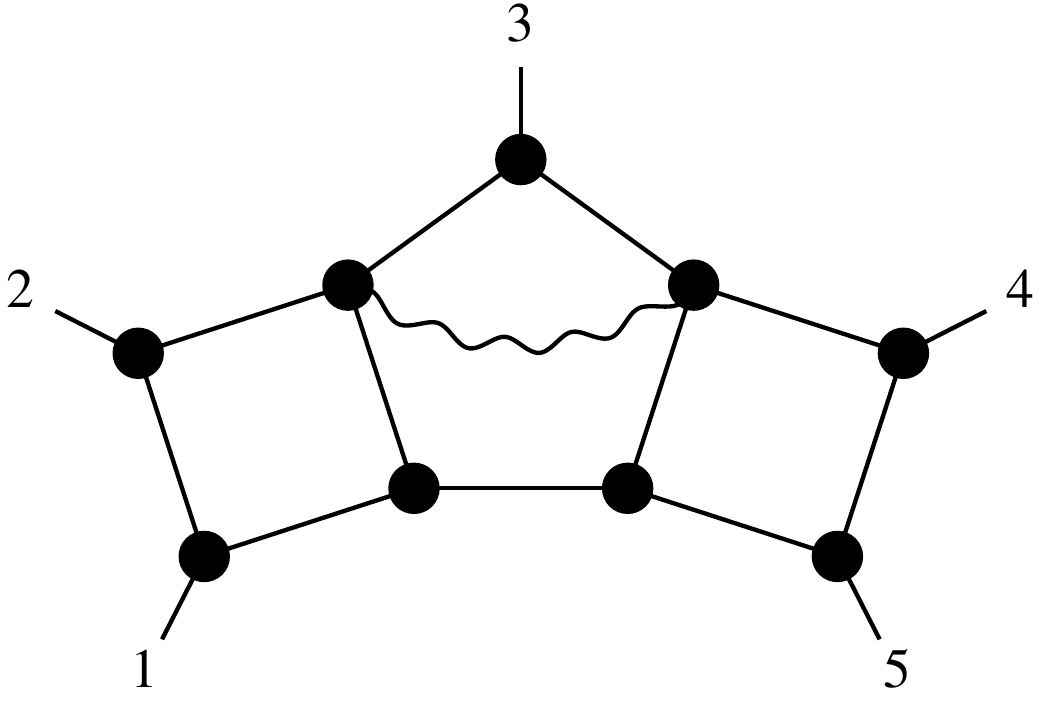}}&+&\raisebox{-1.325cm}{\includegraphics[scale=0.405]{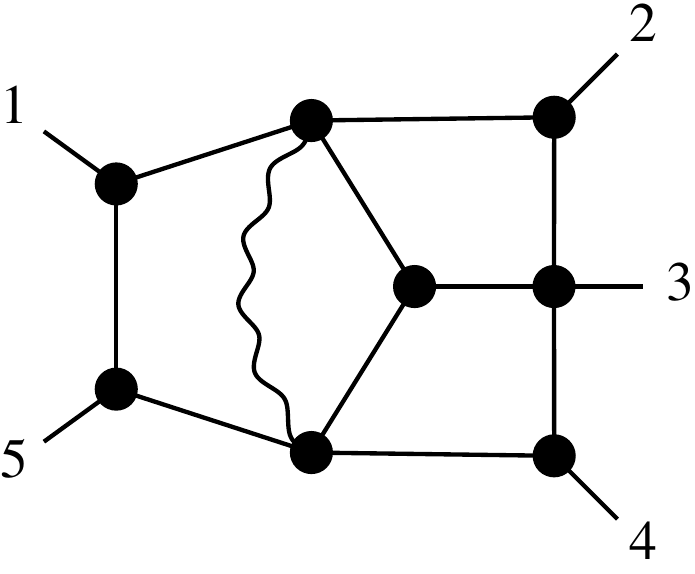}}&+&\raisebox{-1.105cm}{\includegraphics[scale=0.4]{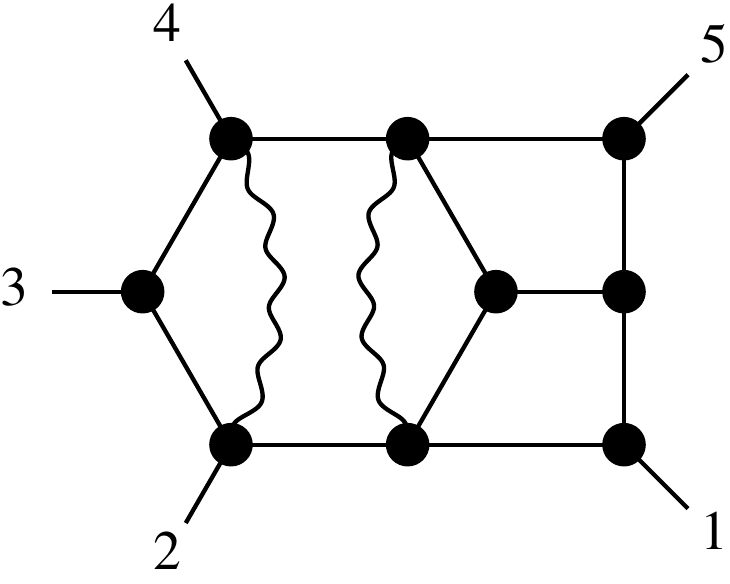}}\\\begin{array}{c}\ab{3451}^3\\\times\ab{AB|(234)\cap(512)}\phantom{\times}\\~\end{array}&&\begin{array}{c}\ab{5123}\ab{4512}\ab{3451}\\\times\ab{AB|(123)\cap(345)}\phantom{\times}\\~\end{array}&&\begin{array}{c}\ab{4512}^2\\\times\ab{AB|(345)\cap(123)}\phantom{\times}\\~\end{array}&&\begin{array}{c}\ab{4512}\\\times\ab{AB|(451)\cap(512)}\phantom{\times}\\\times\ab{AB|(345)\cap(123)}\phantom{\times}\end{array}\end{array}\\[-0.2cm]
&\hspace{-1.5cm}(1+r)\left(\!\!\!\!\!\begin{array}{c@{~}c@{~}c@{~}c@{~}c@{~}c@{~}c@{~}c}\raisebox{-0.875cm}{\includegraphics[scale=0.425]{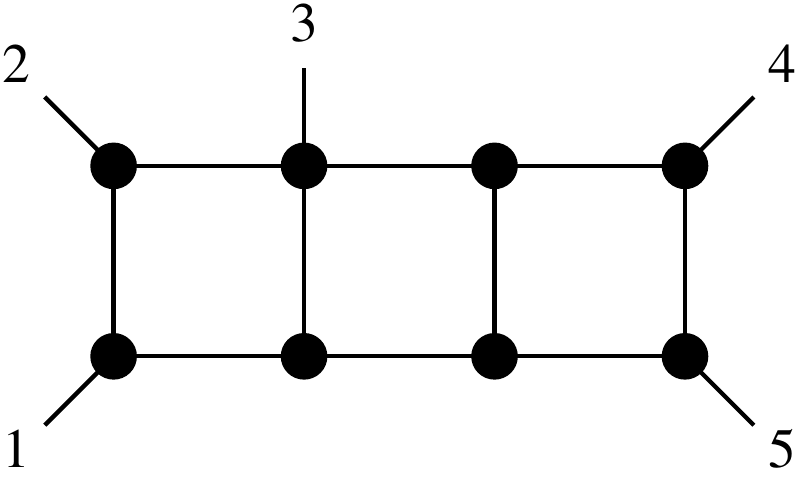}}&+&\raisebox{-1.075cm}{\includegraphics[scale=0.4]{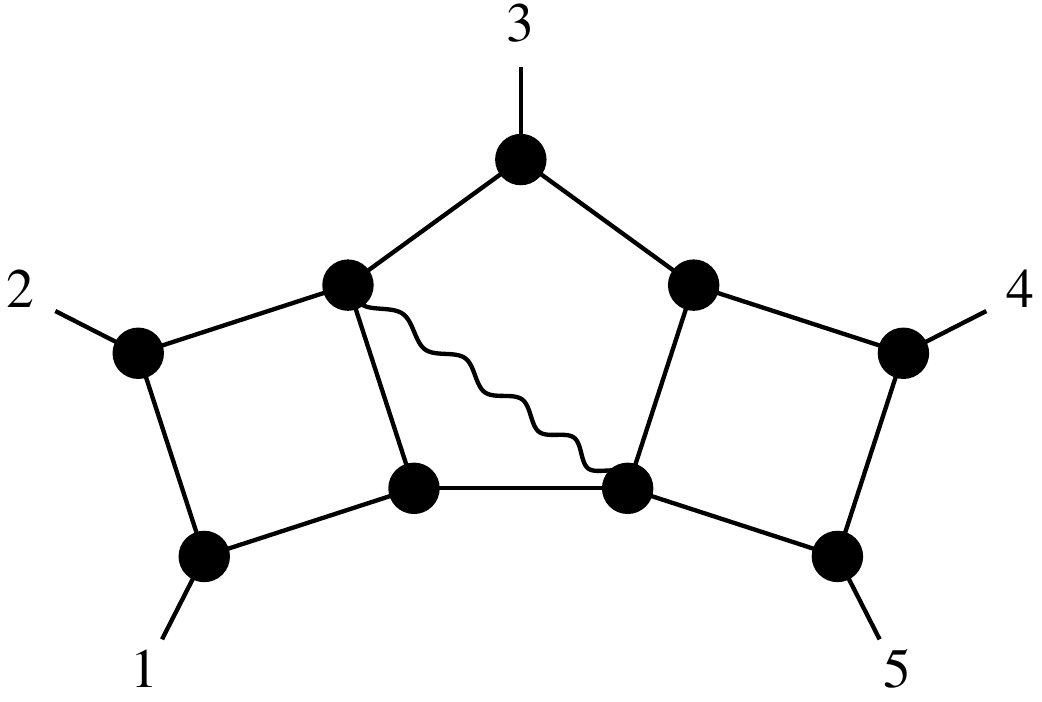}}&+&\raisebox{-1.075cm}{\includegraphics[scale=0.405]{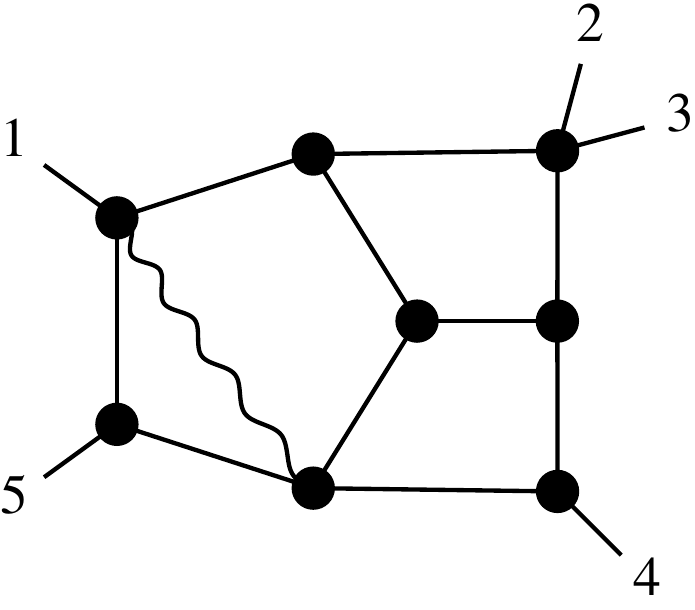}}&+&\raisebox{-1.05cm}{\includegraphics[scale=0.4]{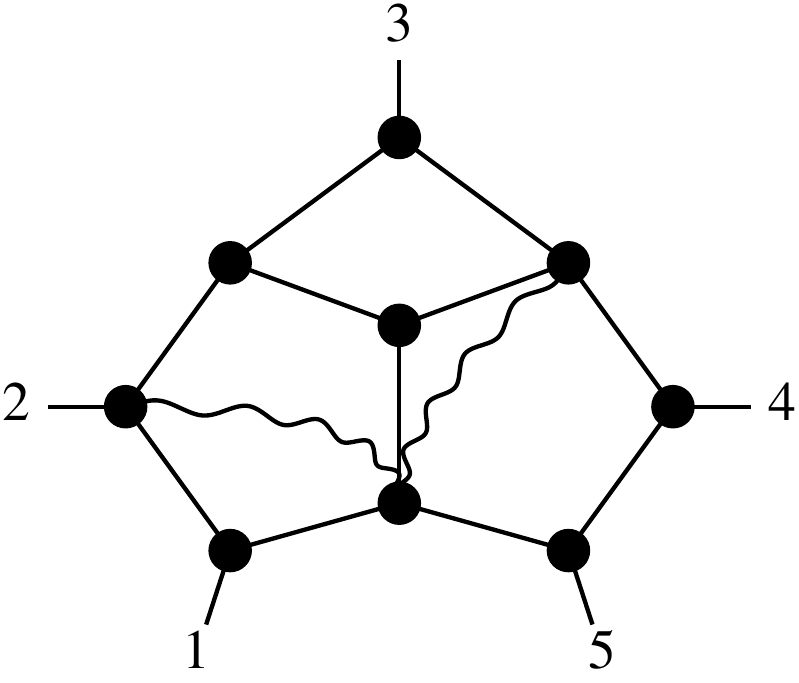}}\\\begin{array}{c}\ab{5123}\ab{4512}\ab{3451}^2\\~\\~\end{array}&&\begin{array}{c}\ab{5123}\ab{3451}\ab{2345}\\\times\ab{AB|(123)\cap(451)}\phantom{\times}\\~\end{array}&&\begin{array}{c}\ab{3451}\ab{4512}\ab{1234}/\ab{5123}\\\times\ab{AB|(345)\cap(512)}\phantom{\times}\\~\end{array}&&\begin{array}{c}\ab{2345}\ab{3451}/\ab{4512}\\\times\ab{AB|(123)\cap(451)}\phantom{\times}\\\times\ab{CD|(234)\cap(512)}\phantom{\times}\end{array}\end{array}\!\!\!\!\!\!\!\right)\nonumber;
\end{split}\end{equation}}here, $r$ is the reflection operation that maps $i \mapsto(6 - i)$.
Notice that deriving this three-loop amplitude using the loop-level recursion requires both the 1-loop 9-particle N$^2$MHV integrand, and the 2-loop 7-particle NMHV integrand; and so the success of getting a manifestly-cyclic and spurious-pole-free, local object is an indirect check of the validity of the whole structure at lower-loops and higher points.

We conclude this quick tour of some simple multi-loop integrands by stressing again a remarkable feature of all these results. The integrals that appear are special objects with unit leading singularities---they are thus the most natural basis of local integrals with which to match the singularities of the theory. As a consequence the coefficients are also simple objects: ``$\pm 1$" for MHV amplitudes, and Grassmannian residues with integer coefficients for more general amplitudes. These objects should be thought of as the correct building blocks for the local integrand, just as the BCFW terms provide the building blocks for the integrand in Yangian-invariant form. As we will discuss below, it is also likely that carrying out the integration will yield ``simple" results for these classes of integrals.

%
\section{Outlook}\label{outlook}

The loop integrand for scattering amplitudes is a well-defined object for any gauge theory in the planar limit, and in this paper we have given an explicit recursive prescription for computing it to any loop order in ${\cal N}=4$ SYM, in a way which manifests the full Yangian-invariance of the theory. This provides a complete definition of perturbative scattering amplitudes in planar ${\cal N}=4$ SYM, with no reference to the Lagrangian, gauge redundancies or other off-shell notions. Along the way, we have also seen a new physical picture for how loops can arise purely from on-shell data, associated with removing pairs of particles in a naturally ``entangled" way. From this vantage point, a number of directions for future work immediately suggest themselves.

\subsection{The Origin of Loops}

A few years ago, the tree-level BCFW recursion relations sat at an interesting cross-roads between the usual formulation of field theory, where space-time locality is manifest, and a hoped for dual description, where space-time should be emergent. On the one hand, the recursion relations were directly derived from field theory---without the field-theoretic motivation, it was hard to imagine the motivation for gluing lower-point objects together in the prescribed way. On the other hand, the presentation of the amplitude was very different from anything normally seen in  field theory. The amplitudes could be presented in many different forms, with remarkable identities guaranteeing their equivalence. The simplicity of the answers resulted directly from the presence of non-local poles. These properties, together with the dual super-conformal invariance of all terms in the BCFW expansions, strongly motivated the search for a dual theory which would make these features obvious, and which would furthermore give an intrinsic definition of the tree amplitudes on its own turf.


The Grassmannian duality for leading singularities provides this dual understanding of tree amplitudes in a satisfying way. The Yangian symmetry is manifest (for all leading singularities and not just tree amplitudes). The amplitude can be presented in many forms since it is a contour integral, with many representatives for a given homology class. The remarkable identities guaranteeing cyclic-invariance (together with important physical properties at loop-level) indeed find a new interpretation as higher-dimensional residue theorems. And finally, giving the contour integral over the Grassmannian a ``particle interpretation" poses a natural question, intrinsic to the Grassmannian picture, whose answer yields the tree amplitude, along the way exposing a (still quite mysterious) connection with twistor string theory.
We strongly suspect that a generalization of this picture exists that extends the duality to only to incorporate loop amplitudes but also explain why loops must be computed to begin with.

Our extension of BCFW to all loop orders puts loop amplitudes in the same position at the cross-roads between field theory and a sought-after dual description that tree amplitudes occupied a few years ago. This should set the stage for fully exposing the dual picture, and we have already made some inroads to uncovering its structure. For instance we saw that the remarkable identities guaranteeing cyclic-invariance of the MHV 1-loop amplitude indeed have an origin as a residue theorem in a new Grassmannian integral closely associated to the ``master" integral computing leading singularities/Yangian-invariants. The nature of the ``seed" for loops, arising from removing particles, is also clearly intimately related to the particle interpretation, which has already played a central role in the emergence of locality at tree-level.

Along these lines, here we give another presentation of the 1-loop MHV amplitudes, which differs from the form we obtained using the recursion relation. Consider
the tree-level N$^2$MHV amplitude $M_{n,k=2}({\cal Z}_1, \ldots,
{\cal Z}_n, {\cal Z}_A, {\cal Z}_B)$. The 1-loop MHV amplitude
arises directly from the entangled removal of $A$ and $B$:
\be\vspace{-4cm}
\includegraphics[scale=0.5]{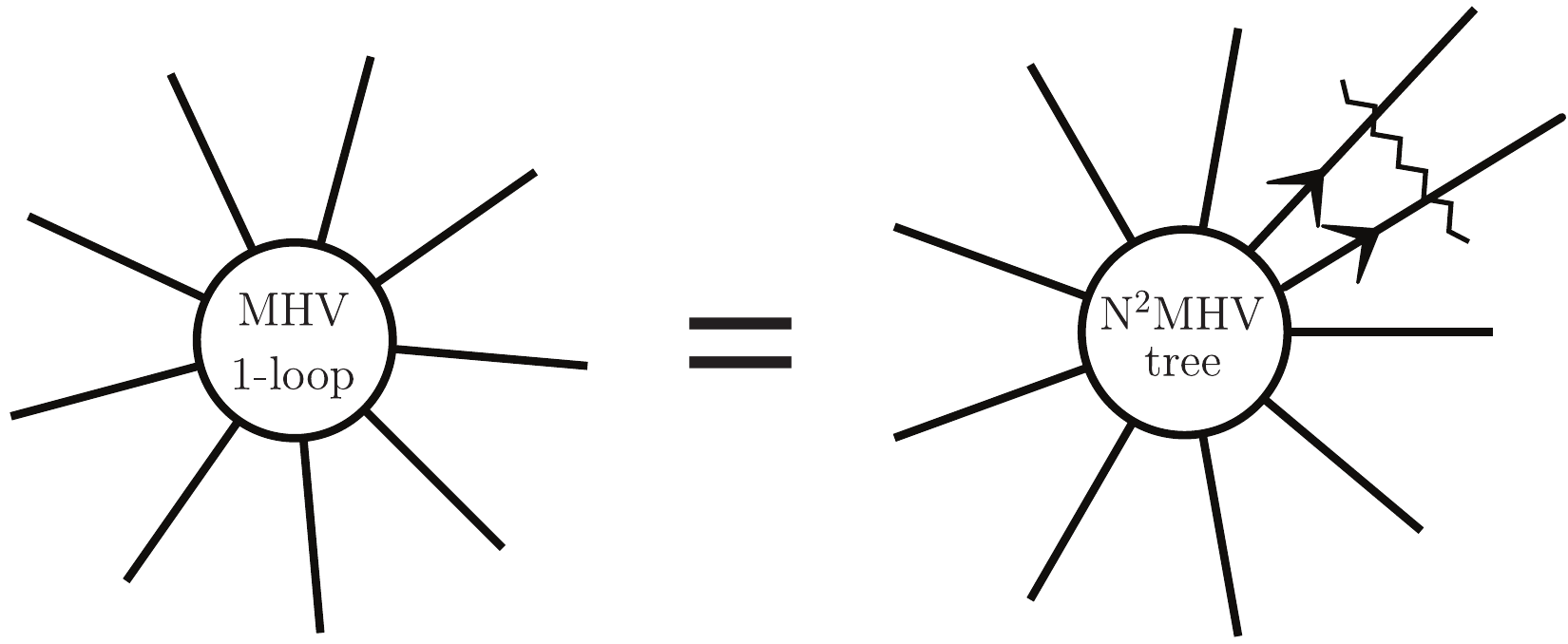} \nonumber\vspace{-3cm}
\ee
Here it is easy to see, using the BCFW form of the tree amplitude, that there is a unique GL(2)-contour of integration associated with each term. This formula differs term-by-term from the BCFW form of this amplitude. We can however recognize all the terms as residues of the same auxiliary Grassmannian integral in equation (\ref{auxGras}), and we have shown that the equivalence to the BCFW form follows from a
residue theorem. While this formula does not directly generalize for other amplitudes, its form is certainly suggestive.

Progress on all these questions would likely be accelerated by finding an explicit solution to the recursion relation for all $(n,k,\ell)$, generalizing the explicit solution already known for tree-amplitudes \cite{Drummond:2008cr}.

As a final comment, our analysis of loops in this paper has been greatly aided by working in momentum-twistor space; these variables allow us to recognize loop integrals in their familiar momentum-space setting. However, given that all the elements in the recursion relation were described  in manifestly Yangian-invariant ways, it must be possible to translate these results into ordinary twistor space. It is likely that the twistor-space formulation will be most fundamental, amongst other things it could offer a natural understanding of non-planar loop amplitudes as well.

The results of this paper also give a renewed hope for extracting loop-information from twistor-string theory. As we have seen, loop amplitudes can easily hide in plain sight in subtle ways, masquerading as a formal way of representing ``1" in terms of IR-divergent integrals in \mbox{$(3,1)$-signature!} It is likely that a deeper understanding of the contours associated with the ``Hodges diagrams" \cite{Hodges:2005aj,ArkaniHamed:2009si}, already for twistor-space tree-amplitudes in \mbox{$(3,1)$-signature}, will be important to make progress here.

\subsection{Simplicity of Integrals and IR-Anomalies}

Putting aside these highbrow issues, we are confronted with a much more urgent question: does our understanding of the integrand help us to carry out the integrations to obtain the physical amplitudes? Are the symmetries of the integrand of any use?

In fact the manifestly Yangian-invariant way of presenting the
integrand {\it does} strongly suggests that the integrals themselves will be
``simple". The ``super" part of super-dual conformal
invariance is already an extremely powerful constraint. Consider MHV
amplitudes  for simplicity. The statement of super-dual conformal
invariance is\vspace{-0.25cm}
\begin{equation}
\sum_a \eta^K_a \frac{\partial}{\partial Z_a^J} M_{{\rm MHV}} = 0
\rightarrow \frac{\partial}{\partial Z^J_a} M_{{\rm MHV}} = 0 \,\; {\rm
for}\; {\rm all} \, a,\vspace{-0.2cm}\end{equation}
where we use the fact that the MHV amplitude has no $\eta_a$
dependence. Thus, the only super-dual conformally invariant amplitude
is forced to be a constant! This reflects the well-known fact
that the only Yangian invariant with $k=0$ is the MHV tree amplitude
(=1 in momentum-twistor space).
Now, we have expressed the integrand for the MHV amplitude in a
manifestly super-dual conformal (indeed Yangian)-invariant way.
Consider for instance the 1-loop amplitude, which has the form\vspace{-0.15cm}
\begin{equation}
M_{{\rm MHV}} = \int\!\!d^{3|4}{\cal Z}_A d^{3|4} {\cal Z}_B \;F({\cal
Z}_A,{\cal Z}_B; {\cal Z}_a),
\end{equation}
with an entangled contour of integration for ${\cal Z}_{A,B}$; we
suppress the explicit expression for $F$.  The statement of super-dual
conformal invariance is perfectly well-defined at the level of the
integrand, which turns into a total derivative:\vspace{-0.1cm}
\begin{equation}
\sum_a \eta^K_a \frac{\partial}{\partial Z_a^J} M_{{\rm
MHV}} = \int  d^{3|4} {\cal Z}_A d^{3|4} {\cal Z}_B \left( \eta_A^K
\frac{\partial}{\partial Z^J_A} + \eta_B^K \frac{\partial}{\partial
Z^J_B} \right) F.
\end{equation}
After doing the $\eta_{A,B}$ and GL(2)-integrals, we have
\be
\frac{\partial}{\partial Z_a^J} M_{{\rm MHV}} = \int \frac{d^4 Z_A d^4
Z_B}{{\rm vol[GL(2)]}} \left( \frac{\partial}{\partial Z^J_A} G^a_A +
\frac{\partial}{\partial Z^J_B} G^a_B \right),\ee
where we suppress the explicit forms of $G^a_{A,B}$. We see that
super-dual conformal-invariance continues to be manifest at the level of
the Bosonic loop integrand in the dual co-ordinate space, also at all loop orders.

This symmetry therefore guarantees that no matter how complicated the integrand looks,
on any contour of integration where the integral is completely
well-defined, it can only integrate to a constant, ``1"!  The integral is not ``1"
{\it only} because we choose a contour of integration over lines $(AB)$ corresponding to real \mbox{$(3,1)$-signature} points in dual spacetime, and this integral is IR-divergent.  We see that IR-divergences are not an annoying side-feature of loop amplitudes, they are the sole reason these amplitudes are non-trivial; in this Yangian-invariant form,  the loop amplitudes are
telling us ``I diverge, therefore I am"\footnote{We thank Peter
Goddard for this remark.}.  This is a powerful statement that should
be turned into an engine to simplify the computation of the loop integrals. Due to the IR-divergences, the
Yangian generators will not quite annihilate the loop amplitude, but they should localize the integral to the IR-divergent regions of loop momentum-space collinear to the external particles. In the
dual co-ordinate space, this is the region localized to the edges of
the null-polygonal Wilson loop.
It seems likely that these IR-anomalies fully control the structure of the amplitude. Amongst other things, they must lie behind the astonishing simplicity recently uncovered in the structure of the remainder function for the 2-loop, 6-particle MHV amplitude \cite{Goncharov:2010jf}. In the same line of thought, it is conceivable that there is a very direct link between the Yangian structure we uncovered and the very beautiful connections made at strong coupling with integrable systems, Y-systems, TBA equations and the Yang-Yang functional \cite{Alday:2009dv,Alday:2010vh}. Already these developments have allowed a bridge to weak coupling by computing sub-leading corrections to collinear limits \cite{Alday:2010ku,Alday:2010zy,Eden:2010zz}.

Having said all of this, there is a very important issue that must be addressed to make progress in directly computing these Yangian-``invariant" but non-local integrals. The question is of course how to handle IR-regularization for these objects. Dimensional regularization has long been the preferred method for regulating IR-divergences in gauge theories, but it does particularly violent damage to the structure of the integrand, and is not useful for our purposes. Fortunately, there is a better regulator, both conceptually and computationally.
Physically, the IR-divergences are removed by moving out on the
Coulomb branch \cite{Alday:2009zm}. This gives a beautifully simple
way to regulate the integrals in momentum-twistor space which is also useful for practical computations \cite{Henn:2010bk, Henn:2010ir}. With the loop integrand written in local form, one simply deforms the local propagators as \mbox{$\ab{AB\,\,j\smallminus1\,\,j}
\mapsto \ab{AB \,\,j\smallminus1\,\,j} + m^2 \langle AB \rangle \langle j\smallminus1 \,\, j \rangle$}. The physics is always four dimensional. The ambiguities in this regulator occur  at an irrelevant level $\mathcal{O}(m^2) ({\rm log}(m^2))^p$. In particular there are no issues with the notorious ``$\mu$-terms" in dimensional regularization, and we don't encounter the ubiquitous $\epsilon/\epsilon$ effects either. This is clearly the physically correct regularization for our set-up.

How should we use this regularization to compute the
non-local integrals of interest? One can glibly regulate all
4-brackets $\ab{AB xy} \mapsto \ab{AB xy} + m^2 \langle AB\rangle \langle
xy \rangle$, but this is not physically correct: the regularization of the local propagators is reflecting the (local!) masses induced by Higgsing; and so it is not clear how the
non-local propagators should be regularized. Indeed, we have checked that for the 1-loop MHV amplitudes, this very
naive regularization of the integrals does not produce the standard result. Of course, since the Yangian invariant form of the full amplitude can be expanded in terms of local integrals, we can in principle work backwards to see how the correct local regulator affects the non-local integrand; the question is whether there is a sensible way of computing these non-local integrals directly. We intend to return to these questions in near future.

We have emphasized that the Yangian-invariant presentation of the loop
integrand strongly suggests that the integrals should be simple.
But as we have seen in a number of examples, even the local
forms of the integrand, when written in terms of the natural chiral
basis of momentum-twistor space integrals with unit leading
singularities, look  surprisingly elegant. In fact, these integrals with unit leading singularities should also be
``simple". The reason is precisely that their leading
singularities are ``1" or ``0"; these are the only possible values of
the integrals on any closed contour of integration, independent
of the kinematic variables. This means that {\it e.g.}~$\partial/\partial
Z_a^I$ acting on these integrals should also be a total derivative with
respect to the loop variables, and that they too should be localized
to regions with collinear singularities. Since these are local integrals their regularization is well defined. Indeed, as we pointed out in our multi-loop examples, the na\"{i}vely ``hardest" integrals are even IR-finite. The integrals for our form of the two-loop 6-pt MHV amplitude have been computed analytically for certain cross-ratios by \cite{JJwip}, passing all non-trivial checks. The simplicity of these partial results strongly 
supports the idea that the full amplitude computed with these integrals are also simple. 
\subsection{Other Planar Theories}

We end by stressing that many of the ideas in this paper are likely to generalize beyond the very special case of ${\cal N}=4$ SYM. Since the integrand is well-defined  in any planar theory, one can try to determine it with recursion relations just as we have done for ${\cal N}=4$ SYM.
In \cite{CaronHuot:2010zt}, it was argued that the single-cuts of the 1-loop amplitude are well-defined for any theory with at least ${\cal N}=1$ SUSY \mbox{(or ${\cal N}=2$ in the presence of massive particles}), so the BCFW recursion determines amplitudes at least up to 1-loop in these theories too, with or without maximal SUSY and Yangian-invariance.
In non-supersymmetric theories, further progress on these questions will require a better understanding of single-cuts.
One difficulty is that the na\"ive forward limit of tree amplitudes is ill-defined.
It is plausible that this is closely related to presence of rational terms in 1-loop amplitudes, which have a beautiful and fascinating structure which is strongly suggestive of a deeper origin.

\section*{Acknowledgments}

We would like to thank Emery Sokatchev, Dave Skinner, Lionel Mason,
James Drummond and Johannes Henn for extensive and inspiring
discussions on loop amplitudes throughout the spring. We are
especially indebted to Andrew Hodges for his deep insights, as well
as continuous discussions and explanations about higher-dimensional
contours of integration, without which this work would not have been
possible. We would also like to thank Fernando Alday, Matt Bullimore, Cliff Cheung,
Louise Dolan, Henriette Elvang, Davide Gaiotto, Peter Goddard, Jared Kaplan, Juan
Maldacena, Amit Sever, Mark Spradlin, Pedro Vieira, Anastasia Volovich
and Edward Witten for stimulating conversations. N.A.-H., J.B. and
F.C. extend a sincere thanks to the librarians and staff of the
Aspen Public Library for their gracious hospitality, and for
providing a truly special environment for stimulating discussions
and research. N.A.-H. is supported by the DOE under grant DE-FG02-91ER40654, F.C.
was supported in part by the NSERC of Canada, MEDT of Ontario and by
The Ambrose Monell Foundation. S.C.-H. is supported by the NSF under grant PHY-0503584.
J.T. is supported by the U.S. Department of State through a Fulbright Science and Technology
Award.

\appendix
\newpage

\section{The BCFW-Form of the 1-Loop 6-Particle NMHV Amplitude}\label{bcfw_6pt}

In this appendix we present the BCFW form of the 1-loop 6-particle NMHV amplitude. The result is
\begin{minipage}[h!]{4cm}\footnotesize\begin{align*}&\phantom{\,+\,}\frac{\delta^{0|4}
\left(\begin{array}{lclclclclcl} 0&+&\eta_{2}\ab{3456}&+&\eta_{3}\ab{4562}&+&\eta_{4}\ab{5623}&+&\eta_{5}\ab{6234}&+&\eta_{6}\ab{2345}\end{array}\right)\ab{AB(561)\bigcap(123)}^2}{\ab{2345} \ab{2356} \ab{3456} \ab{AB12} \ab{AB23} \ab{AB56} \ab{AB61} \ab{AB1(234)\bigcap(56)} \ab{AB1(23)\bigcap(456)}}\\
&+\frac{\delta^{0|4}
\left(\begin{array}{lclclclclcl} \eta_{1}\ab{3456}&+&0&+&\eta_{3}\ab{4561}&+&\eta_{4}\ab{5613}&+&\eta_{5}\ab{6134}&+&\eta_{6}\ab{1345}\end{array}\right)\ab{AB15}^2}{\ab{AB45} \ab{AB56} \ab{AB61} \ab{AB(345)\bigcap(561)} \ab{3451} \ab{AB13} \ab{AB1(34)\bigcap(561)}}\\
&+\frac{\delta^{0|4}
\left(\begin{array}{lclclclclcl} \eta_{1}\ab{3456}&+&0&+&\eta_{3}\ab{4561}&+&\eta_{4}\ab{5613}&+&\eta_{5}\ab{6134}&+&\eta_{6}\ab{1345}\end{array}\right)}{\ab{3456} \ab{4561} \ab{AB34} \ab{AB61} \ab{AB(345)\bigcap(561)} \ab{AB31}}\\
&+\frac{\delta^{0|4}
\left(\begin{array}{lclclclclcl} \eta_{1}\ab{3456}&+&0&+&\eta_{3}\ab{4561}&+&\eta_{4}\ab{5613}&+&\eta_{5}\ab{6134}&+&\eta_{6}\ab{1345}\end{array}\right)\ab{1234}^2}{\ab{3456} \ab{4561} \ab{6134} \ab{AB12} \ab{AB23} \ab{AB34} \ab{1345} \ab{AB1(34)\bigcap(561)}}\\
&+\frac{\delta^{0|4}
\left(\begin{array}{lclclclclcl} \eta_{1}\ab{3456}&+&0&+&\eta_{3}\ab{4561}&+&\eta_{4}\ab{5613}&+&\eta_{5}\ab{6134}&+&\eta_{6}\ab{1345}\end{array}\right)}{\ab{6134} \ab{AB34} \ab{AB45} \ab{AB56} \ab{5613} \ab{AB1(34)\bigcap(561)}}\\
&+\frac{\delta^{0|4}
\left(\begin{array}{lclclclclcl} \eta_{1}\ab{2356}&+&\eta_{2}\ab{3561}&+&\eta_{3}\ab{5612}&+&0&+&\eta_{5}\ab{6123}&+&\eta_{6}\ab{1235}\end{array}\right)\ab{4561}^2}{\ab{5612} \ab{6123} \ab{AB45} \ab{AB56} \ab{AB(561)\bigcap(123)} \ab{3561} \ab{AB4(23)\bigcap(561)}}\\
&+\frac{\delta^{0|4}
\left(\begin{array}{lclclclclcl} \eta_{1}\ab{2356}&+&\eta_{2}\ab{3561}&+&\eta_{3}\ab{5612}&+&0&+&\eta_{5}\ab{6123}&+&\eta_{6}\ab{1235}\end{array}\right)\ab{AB(234)\bigcap(561)}^2}{\ab{5612} \ab{6123} \ab{AB23} \ab{AB34} \ab{AB56} \ab{AB(561)\bigcap(123)} \ab{AB4(23)\bigcap(561)} \ab{AB5(561)\bigcap(23)}}\\
&+\frac{\delta^{0|4}
\left(\begin{array}{lclclclclcl} \eta_{1}\ab{2356}&+&\eta_{2}\ab{3561}&+&\eta_{3}\ab{5612}&+&0&+&\eta_{5}\ab{6123}&+&\eta_{6}\ab{1235}\end{array}\right)\ab{2345}^2}{\ab{2356} \ab{5612} \ab{6123} \ab{AB23} \ab{AB34} \ab{AB45} \ab{1235} \ab{AB5(561)\bigcap(23)}}\\
&+\frac{\delta^{0|4}
\left(\begin{array}{lclclclclcl} \eta_{1}\ab{2345}&+&\eta_{2}\ab{3451}&+&\eta_{3}\ab{4512}&+&\eta_{4}\ab{5123}&+&\eta_{5}\ab{1234}&+&0\end{array}\right)\ab{4561}^2}{\ab{1234} \ab{1245} \ab{2345} \ab{AB45} \ab{AB56} \ab{AB61} \ab{3451} \ab{AB1(123)\bigcap(45)}}\\
&+\frac{\delta^{0|4}
\left(\begin{array}{lclclclclcl} \eta_{1}\ab{2345}&+&\eta_{2}\ab{3451}&+&\eta_{3}\ab{4512}&+&\eta_{4}\ab{5123}&+&\eta_{5}\ab{1234}&+&0\end{array}\right)\ab{AB14}^2}{\ab{1234} \ab{AB12} \ab{AB34} \ab{AB45} \ab{AB15} \ab{AB1(123)\bigcap(45)} \ab{AB4(234)\bigcap(51)}}\\
&+\frac{\delta^{0|4}
\left(\begin{array}{lclclclclcl} \eta_{1}\ab{2345}&+&\eta_{2}\ab{3451}&+&\eta_{3}\ab{4512}&+&\eta_{4}\ab{5123}&+&\eta_{5}\ab{1234}&+&0\end{array}\right)}{\ab{2345} \ab{AB12} \ab{AB23} \ab{3451} \ab{AB15} \ab{AB4(234)\bigcap(51)}}\\
&+\frac{\delta^{0|4}
\left(\begin{array}{lclclclclcl} \eta_{1}\ab{2345}&+&\eta_{2}\ab{3451}&+&\eta_{3}\ab{4512}&+&\eta_{4}\ab{5123}&+&\eta_{5}\ab{1234}&+&0\end{array}\right)}{\ab{1245} \ab{AB23} \ab{AB34} \ab{AB45} \ab{5123} \ab{AB1(123)\bigcap(45)}}\\
&+\frac{\delta^{0|4}
\left(\begin{array}{clclcl}& \eta_{1}\ab{AB(23)\bigcap(456)1}&+&\eta_{2}\ab{4561} \ab{AB13}&+&\eta_{3}\ab{1456} \ab{AB12}\\+&\eta_{4}\ab{AB(123)\bigcap(561)}&+&\eta_{5}\ab{AB(123)\bigcap(46)1}&+&\eta_{6}\ab{AB1(123)\bigcap(45)}\end{array}\right)\ab{AB15}^2}{\ab{AB12} \ab{AB45} \ab{AB56} \ab{AB61} \ab{AB(561)\bigcap(123)} \ab{AB13} \ab{AB14} \ab{AB1(123)\bigcap(45)} \ab{(AB1)\bigcap(45)(AB)\bigcap(561)23}}\\
&+\frac{\delta^{0|4}
\left(\begin{array}{clclcl}& \eta_{1}\ab{AB(23)\bigcap(456)1}&+&\eta_{2}\ab{4561} \ab{AB13}&+&\eta_{3}\ab{1456} \ab{AB12}\\+&\eta_{4}\ab{AB(123)\bigcap(561)}&+&\eta_{5}\ab{AB(123)\bigcap(46)1}&+&\eta_{6}\ab{AB1(123)\bigcap(45)}\end{array}\right)}{\ab{4561} \ab{AB12} \ab{AB23} \ab{AB61} \ab{AB13} \ab{AB14} \ab{AB1(23)\bigcap(456)} \ab{(AB1)\bigcap(45)(AB)\bigcap(561)23}}\\
&+\frac{\delta^{0|4}
\left(\begin{array}{clclcl}& \eta_{1}\ab{AB1(234)\bigcap(56)}&+&\eta_{2}\ab{AB(34)\bigcap(561)1}&+&\eta_{3}\ab{AB1(24)\bigcap(561)}\\+&\eta_{4}\ab{AB(561)\bigcap(123)}&+&\eta_{5}\ab{1234} \ab{AB61}&+&\eta_{6}\ab{1234} \ab{AB15}\end{array}\right)}{\ab{1234} \ab{AB12} \ab{AB34} \ab{AB56} \ab{AB61} \ab{AB(234)\bigcap(561)} \ab{AB(561)\bigcap(123)} \ab{AB14} \ab{AB15}}\\
&+\frac{\delta^{0|4}
\left(\begin{array}{clclcl}& \eta_{1}\ab{AB1(234)\bigcap(56)}&+&\eta_{2}\ab{AB(34)\bigcap(561)1}&+&\eta_{3}\ab{AB1(24)\bigcap(561)}\\+&\eta_{4}\ab{AB(561)\bigcap(123)}&+&\eta_{5}\ab{1234} \ab{AB61}&+&\eta_{6}\ab{1234} \ab{AB15}\end{array}\right)}{\ab{AB12} \ab{AB23} \ab{AB61} \ab{AB(234)\bigcap(561)} \ab{AB14} \ab{AB15} \ab{AB1(234)\bigcap(56)} \ab{AB1(34)\bigcap(561)}}
\end{align*}
\end{minipage}
\vskip .1in
\noindent A note on notation: the expression $\ab{AB1|(56)\!\cap\!(234)}$ refers to $\ab{AB1X}$ where $X = (56)\cap(234)$ is the point where the line (56) intersects the plane (234), namely, $Z_5\ab{6\,2\,3\,4}+Z_6\ab{2\,3\,4\,5}=-\left(Z_2\ab{3\,4\,5\,6}+Z_3\ab{4\,5\,6\,2}+Z_4\ab{5\,6\,2\,3}\right)$; similarly, `$(123)\bigcap(456)$' is $Z_{12}\ab{3\,4\,5\,6}+Z_{23}\ab{1\,4\,5\,6}+Z_{31}\ab{2\,4\,5\,6}$.

\section{The Local 2-Loop 7-Particle NMHV Amplitude}\label{local_7pt}
Here we give the explicit formula for the 2-loop 7-particle NMHV amplitude. We find it most convenient to give a formula for $M^{{\rm 2 loop}}_{{\rm NMHV}} - M^{{\rm tree}} M^{{\rm 2 loop}}_{{\rm MHV}}$. We can expand this in three cyclic classes as $[(7)(1) C_{7,1}] + [(7)(2) C_{7,2}] + [(7)(3) C_{7,3}]$ + cyclic. We give the expression for the coefficients $C_{7,1},C_{7,2},C_{7,3}$ in the tables below. Here $``g"$ refers to the operation $i \mapsto i+1$, and $P$ is a parity flip, that exchanges wavy- and dashed-lines (together with their corresponding normalization), and $r$ is the reflection operation $i \mapsto(8 - i)$.
\begin{table}[h!]\centering\caption{Coefficients of residue $(7)(1)=[2\,3\,4\,5\,6]$.}\mbox{\scriptsize\begin{tabular}{|l@{\hspace{-1.cm}}c|l@{\hspace{-1.25cm}}c|}
\hline\multirow{2}{*}{\mbox{$  1$}}  &  \raisebox{-1.45cm}{\includegraphics[scale=0.5]{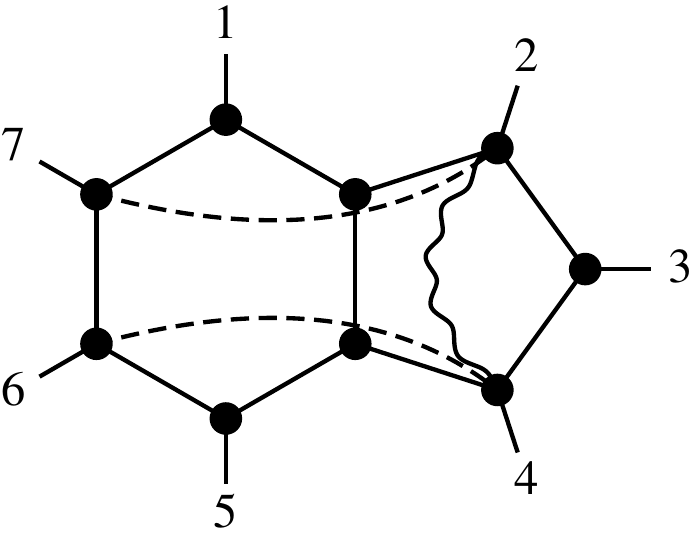}} & \multirow{2}{*}{\mbox{$  -(1 - g)$}} &  \raisebox{-1.45cm}{\includegraphics[scale=0.45]{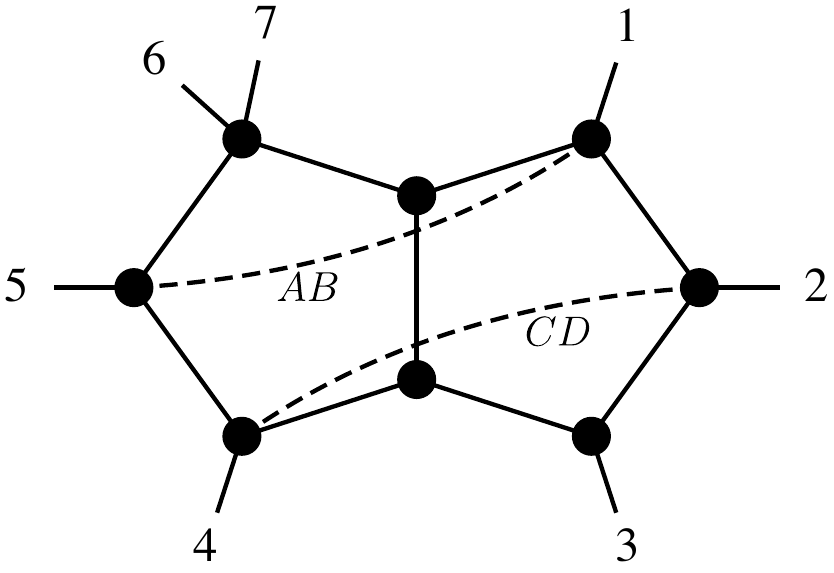}} \\&  \mbox{$  \ab{4512} \ab{5671} \ab{AB|(123)\!\cap\!(345)} \ab{CD64} \ab{CD72}  $} &&  \mbox{$\ab{4563} \ab{4713} \ab{7123} \ab{AB51} \ab{CD24}  $} \\
\hline\multirow{2}{*}{\mbox{$  1$}}  &  \raisebox{-1.25cm}{\includegraphics[scale= 0.385]{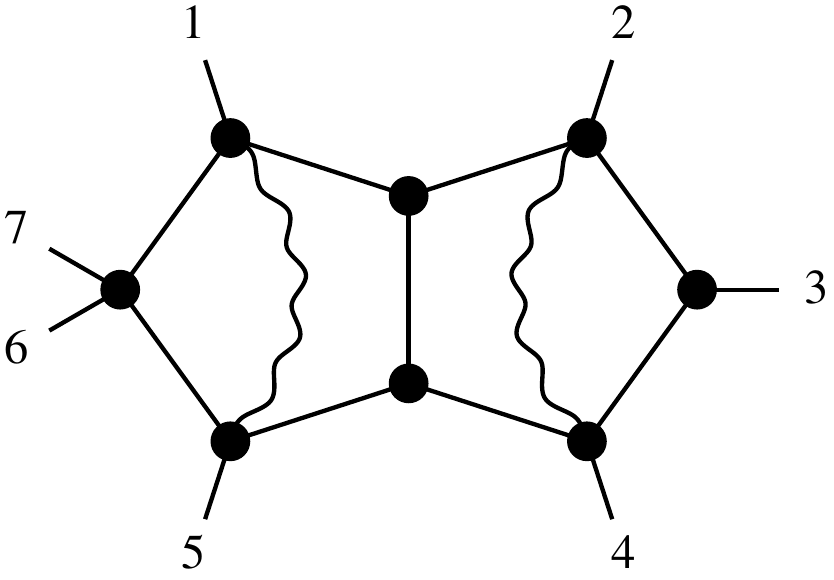}} & \multirow{2}{*}{\mbox{$  -(1 + g^{2} + g^{4})$}} &  \raisebox{-1.25cm}{\includegraphics[scale= 0.385]{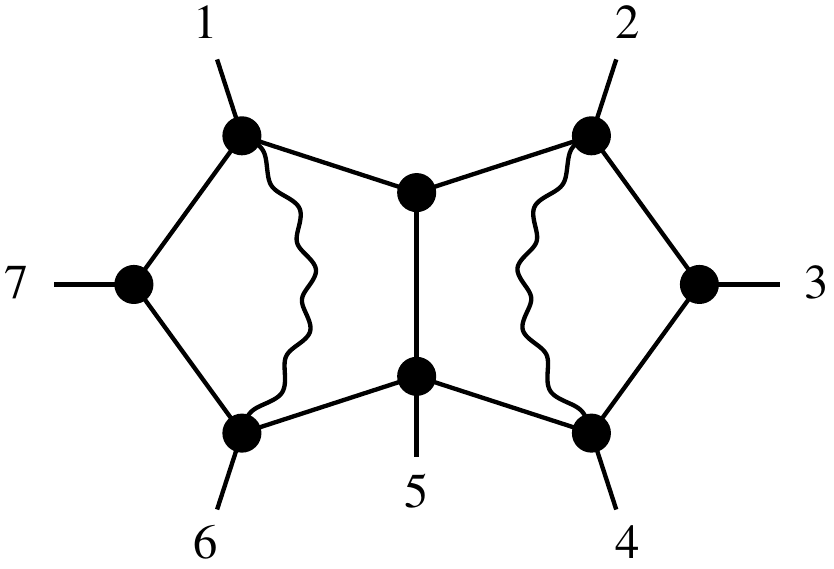}} \\&  \mbox{$  \ab{5124} \ab{AB|(456)\!\cap\!(712)} \ab{CD|(123)\!\cap\!(345)}  $} &&  \mbox{$\ab{2461} \ab{AB|(567)\!\cap\!(712)} \ab{CD|(123)\!\cap\!(345)}  $} \\
\hline\multirow{2}{*}{\mbox{$  -(1 + g^{4} r)$}}  &  \raisebox{-1.25cm}{\includegraphics[scale= 0.385]{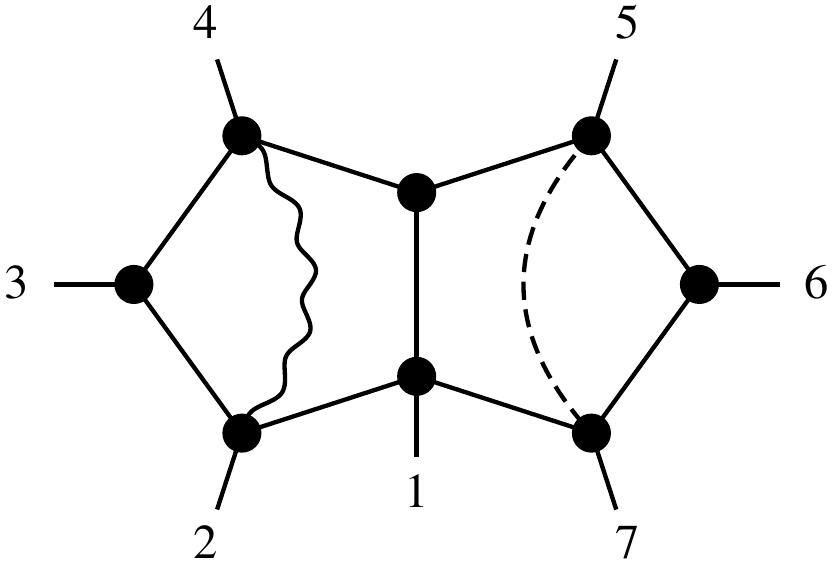}} & \multirow{2}{*}{\mbox{$  1$}} &  \raisebox{-1.25cm}{\includegraphics[scale= 0.385]{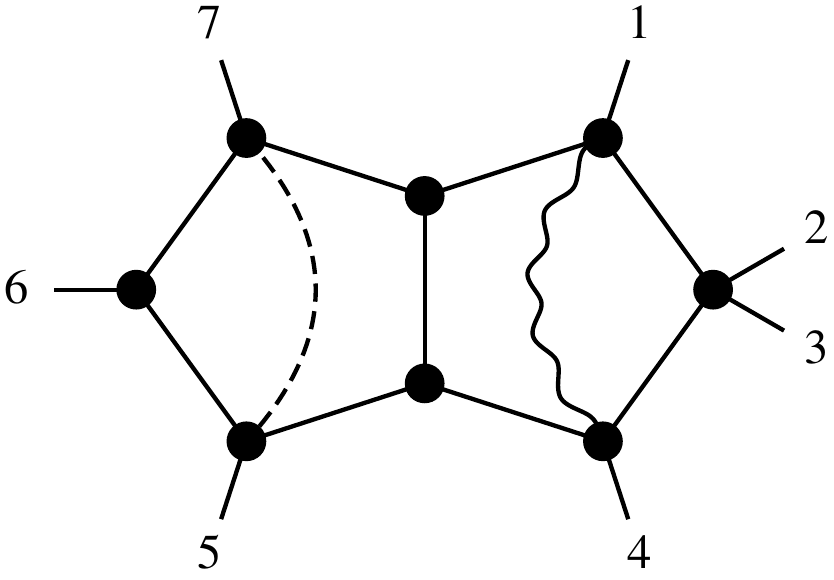}} \\&  \mbox{$  \ab{5624} \ab{6714} \ab{AB|(123)\!\cap\!(345)} \ab{CD57}  $} &&  \mbox{$\ab{5614} \ab{6714} \ab{AB|(712)\!\cap\!(345)} \ab{CD57}  $} \\
\hline\multirow{2}{*}{\mbox{$  -(1 - g)$}}  &  \raisebox{-1.25cm}{\includegraphics[scale= 0.385]{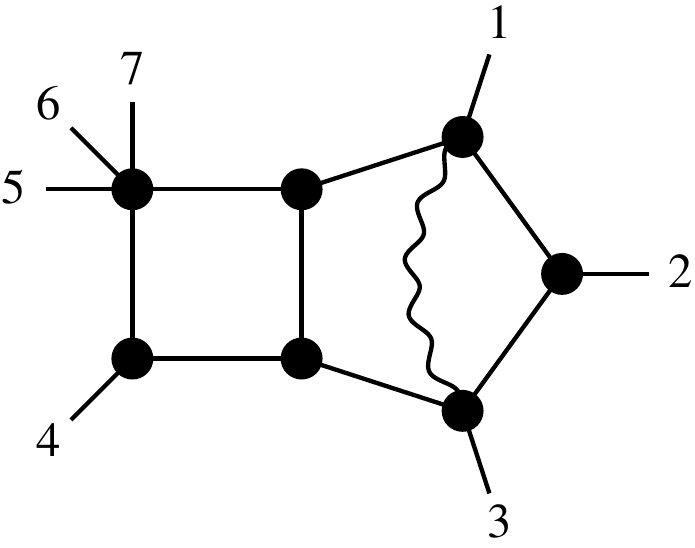}} & \multirow{2}{*}{\mbox{$\begin{array}{l}(1 + g^{2}) (1 - g^{2} r)\\+ g^{2} P (1 - g^{4} r)\end{array}$}} &  \raisebox{-1.25cm}{\includegraphics[scale=0.4]{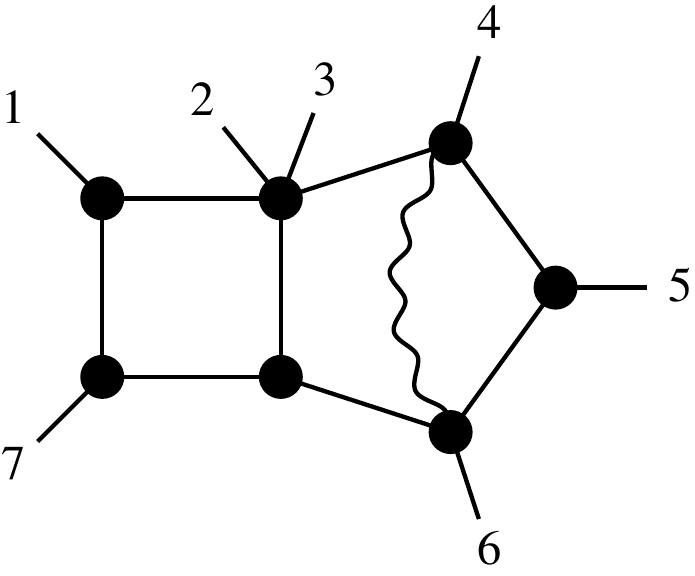}} \\&  \mbox{$  \ab{1345} \ab{1347} \ab{AB|(712)\!\cap\!(234)}  $} &&  \mbox{$\ab{4671} \ab{6712} \ab{AB|(345)\!\cap\!(567)}  $} \\
\hline\multirow{2}{*}{\mbox{$  -(1 + g^{2} - g P)$}}  &  \raisebox{-1.25cm}{\includegraphics[scale= 0.385]{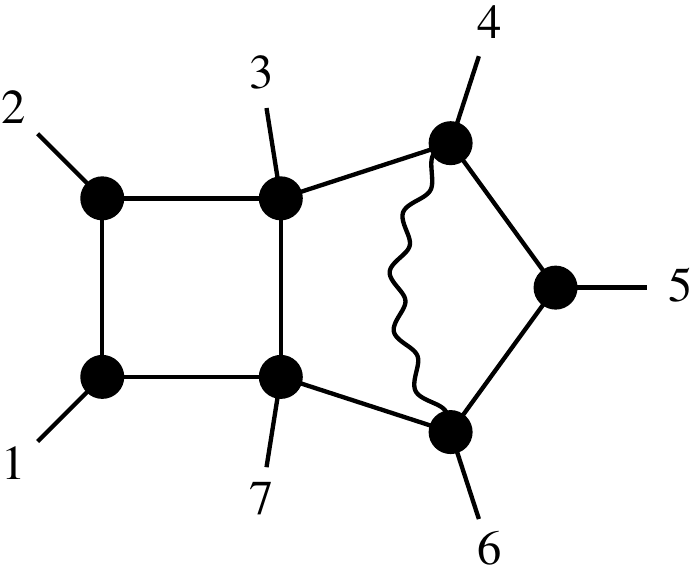}} & \multirow{2}{*}{\mbox{$  1 - g^{4} r$}} &  \raisebox{-1.25cm}{\includegraphics[scale=0.385]{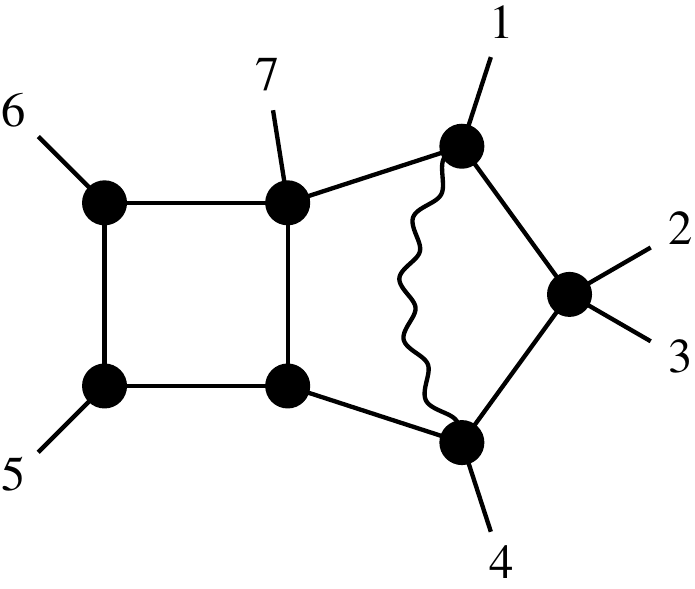}} \\&  \mbox{$  \ab{4612} \ab{7123} \ab{AB|(345)\!\cap\!(567)}  $} &&  \mbox{$\ab{1456} \ab{4567} \ab{AB|(712)\!\cap\!(345)}  $} \\\hline
\end{tabular}}\end{table}
\begin{table}[h!]\centering\caption{Coefficients of residue $(7)(1)=[2\,3\,4\,5\,6],$ continued}\mbox{\hspace{-0.0cm}\scriptsize\begin{tabular}{|l@{\hspace{-0.25cm}}c|l@{\hspace{-.25cm}}c|}
\hline\multirow{2}{*}{\mbox{$  1 + g^{2}$}}  &  \raisebox{-1.25cm}{\includegraphics[scale= 0.385]{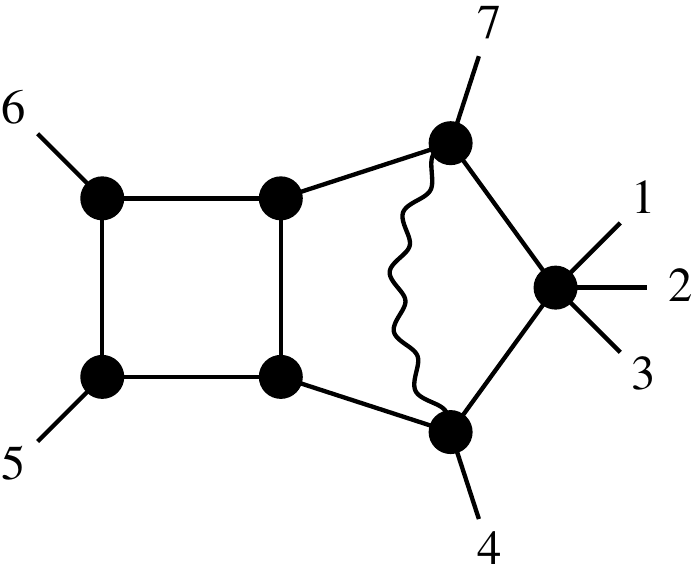}} & \multirow{2}{*}{\mbox{$  (1 + g P) (1 + g^{4} r)$}} &  \raisebox{-1.25cm}{\includegraphics[scale= 0.385]{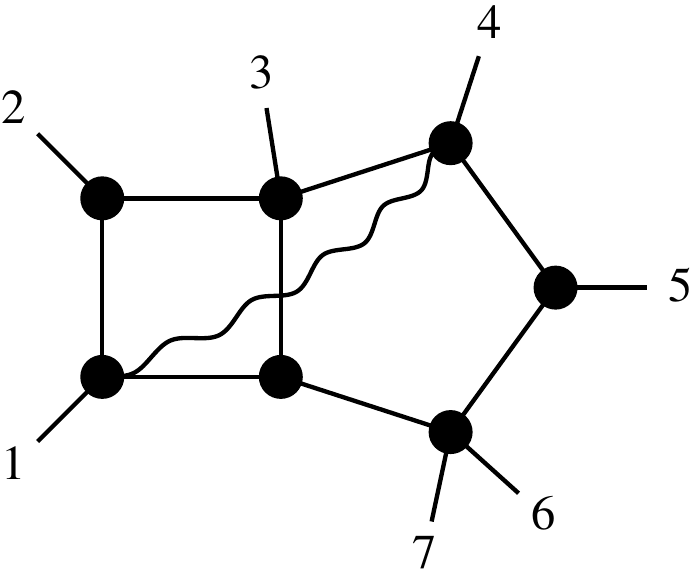}} \\&  \mbox{$  \ab{7456}^{2} \ab{AB|(671)\!\cap\!(345)}  $} &&  \mbox{$\ab{4561} \ab{7123} \ab{AB|(345)\!\cap\!(712)}  $} \\
\hline\multirow{2}{*}{\mbox{$  -(1 - g) (1 + P)$}}  &  \raisebox{-1.25cm}{\includegraphics[scale= 0.385]{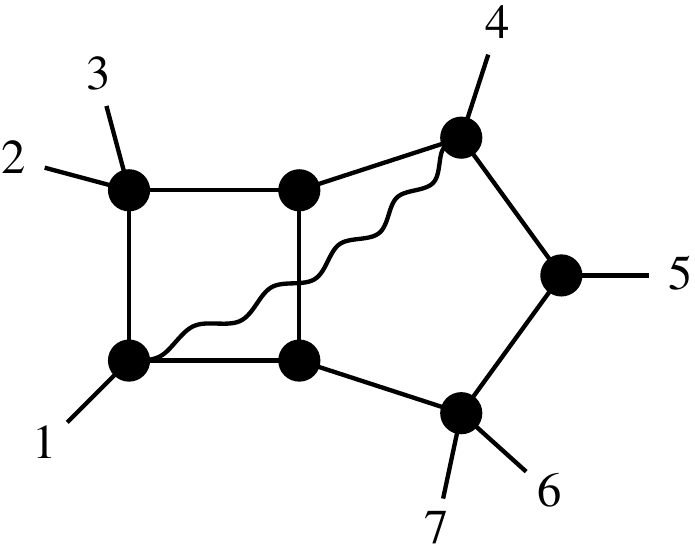}} & \multirow{2}{*}{\mbox{$  1 - g^{3} - g r$}} &  \raisebox{-1.25cm}{\includegraphics[scale= 0.385]{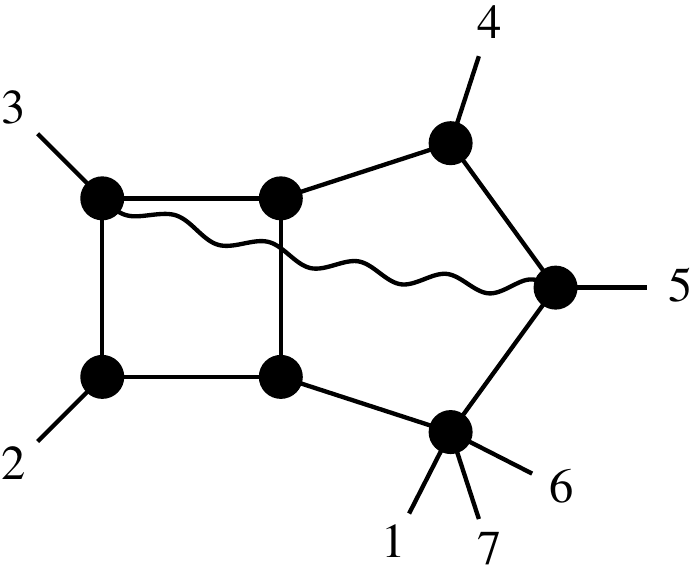}} \\&  \mbox{$  \ab{4561} \ab{4713} \ab{AB|(345)\!\cap\!(712)}  $} &&  \mbox{$\ab{4123} \ab{5123} \ab{AB|(234)\!\cap\!(456)}  $} \\
\hline\multirow{2}{*}{\mbox{$  -1$}}  &  \raisebox{-1.25cm}{\includegraphics[scale= 0.385]{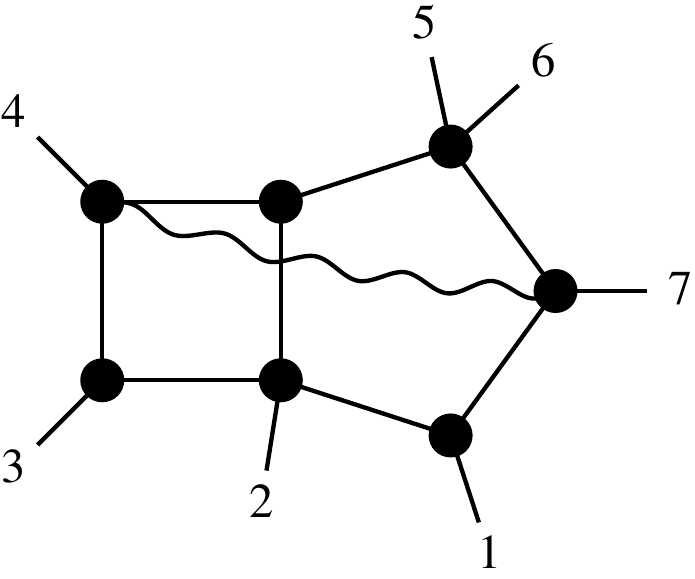}} & \multirow{2}{*}{\mbox{$  -(1 - g^{2} r)$}} &  \raisebox{-1.25cm}{\includegraphics[scale= 0.385]{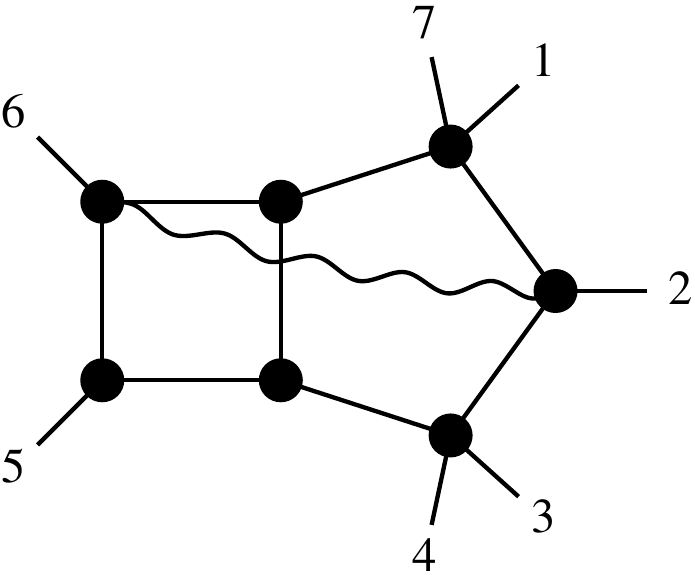}} \\&  \mbox{$  \ab{5234} \ab{7124} \ab{AB|(345)\!\cap\!(671)}  $} &&  \mbox{$\ab{2456} \ab{7456} \ab{AB|(567)\!\cap\!(123)}  $} \\
\hline\multirow{2}{*}{\mbox{$ -(1 + g^{2})$}}  &  \raisebox{-1.25cm}{\includegraphics[scale= 0.385]{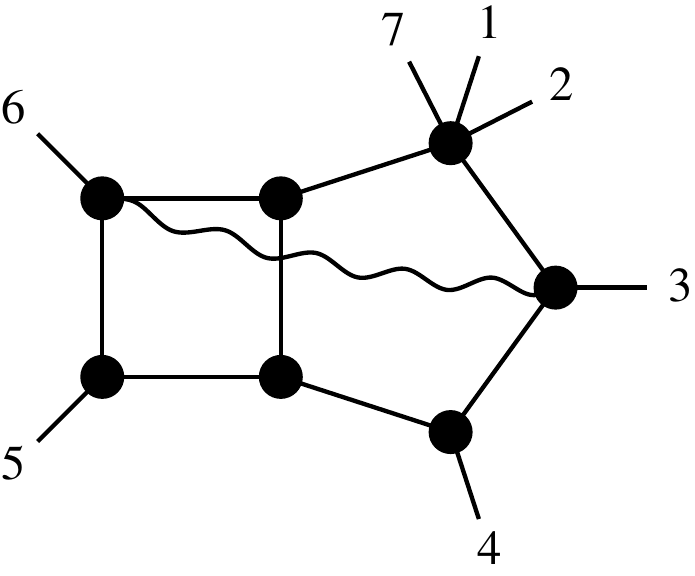}} & \multirow{2}{*}{\mbox{$  1 - g$}} &  \raisebox{-1.25cm}{\includegraphics[scale= 0.385]{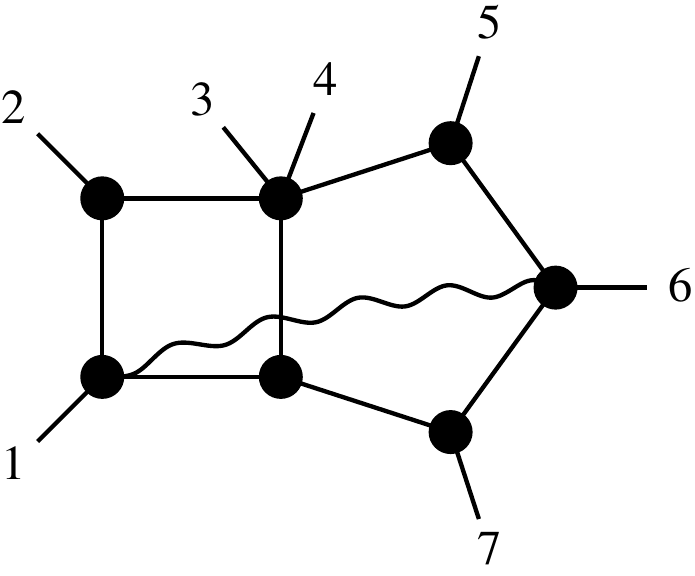}} \\&  \mbox{$  \ab{3456} \ab{7456} \ab{AB|(567)\!\cap\!(234)}  $} &&  \mbox{$\ab{5614} \ab{7123} \ab{AB|(567)\!\cap\!(712)}  $} \\
\hline\multirow{2}{*}{\mbox{$  1$}}  &  \raisebox{-1.25cm}{\includegraphics[scale= 0.385]{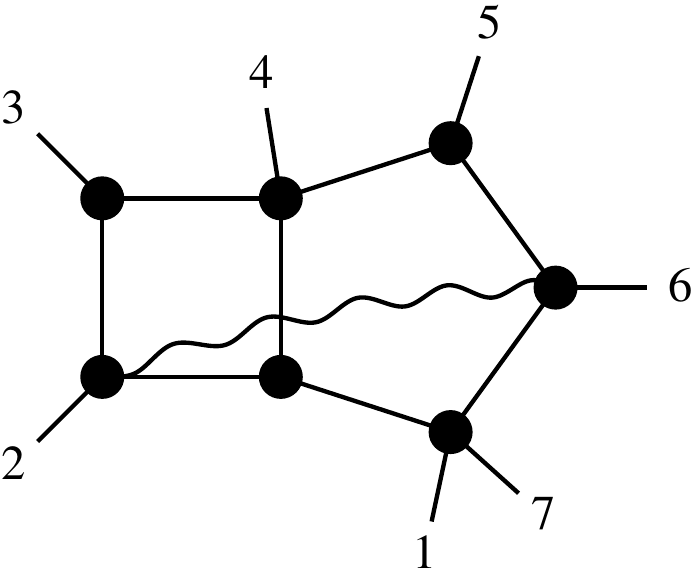}} & \multirow{2}{*}{\mbox{$  -(1 - g)$}} &  \raisebox{-1.25cm}{\includegraphics[scale= 0.385]{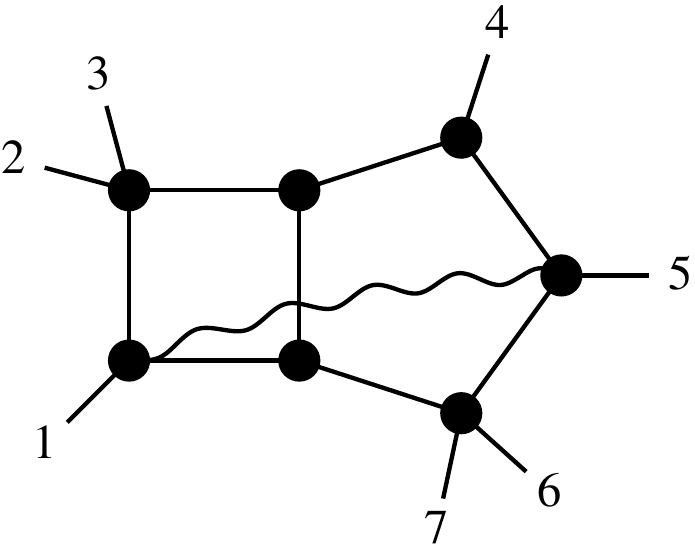}} \\&  \mbox{$  \ab{1234} \ab{5624} \ab{AB|(567)\!\cap\!(123)}  $} &&  \mbox{$\ab{4513} \ab{4713} \ab{AB|(456)\!\cap\!(712)}  $} \\
\hline\multirow{2}{*}{\mbox{$  1 + g^{4} r$}}  &  \raisebox{-1.25cm}{\includegraphics[scale= 0.385]{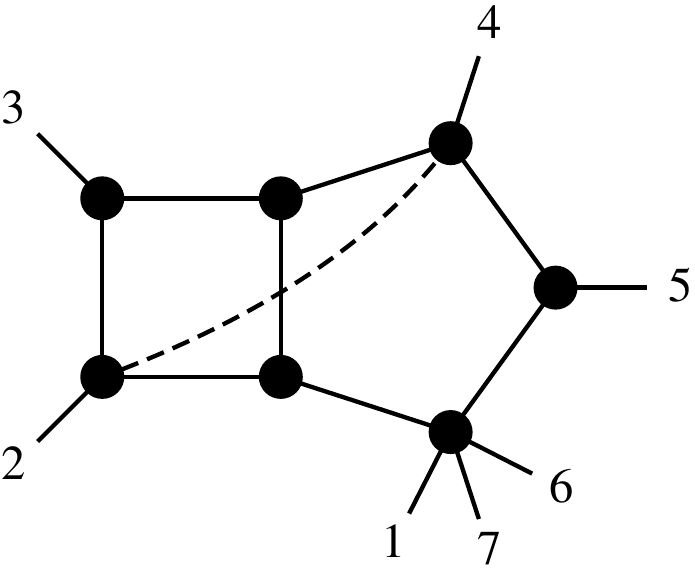}} & \multirow{2}{*}{\mbox{$  1 + g^{4} r$}} &  \raisebox{-1.25cm}{\includegraphics[scale= 0.385]{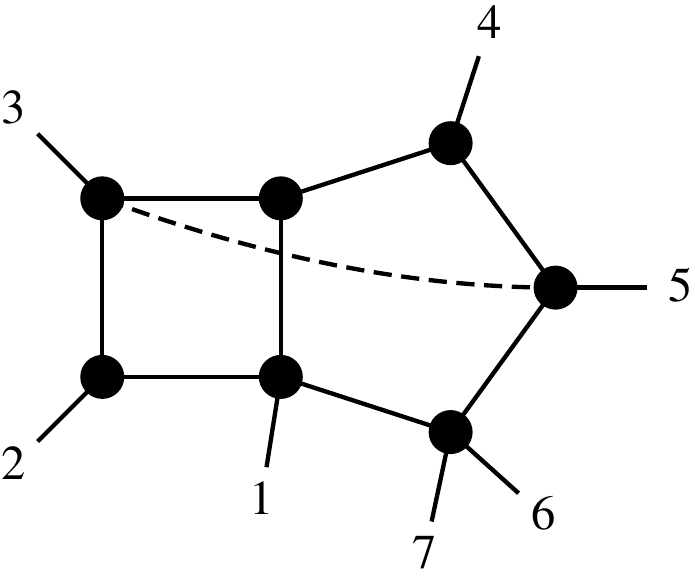}} \\&  \mbox{$  \ab{4123} \ab{4563} \ab{5123} \ab{AB42}  $} &&  \mbox{$\ab{4123}\ab{71|(234)\!\cap\!(456)}\ab{AB53}  $} \\
\hline\multirow{2}{*}{\mbox{$ -(1 + g^{2})$}}  &  \raisebox{-1.05cm}{\includegraphics[scale= 0.345]{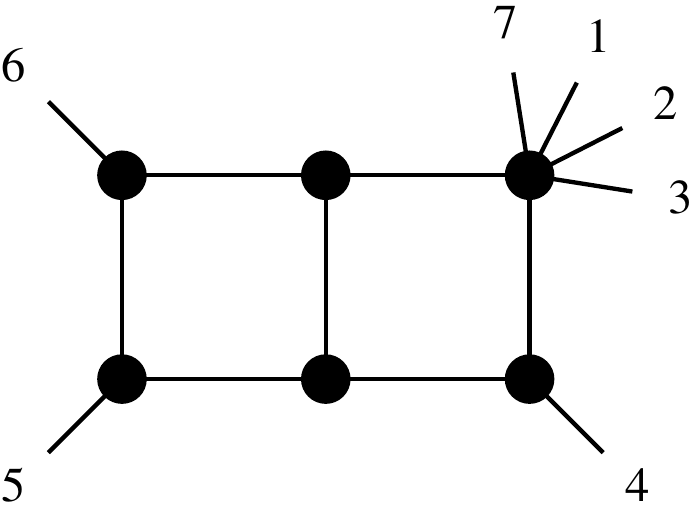}} & &\\&  \mbox{$  \ab{4563} \ab{4567}^{2}  $} && \\
\hline
\end{tabular}}\end{table}
\begin{table}[h!]\centering\caption{Coefficients of residue $(7)(2)=[1\,3\,4\,5\,6]$}\mbox{\hspace{-0.0cm}\scriptsize\begin{tabular}{|l@{\hspace{-0.25cm}}c|l@{\hspace{-.25cm}}c|}
\hline\multirow{2}{*}{\mbox{$  1$}}  &  \raisebox{-1.25cm}{\includegraphics[scale=0.4]{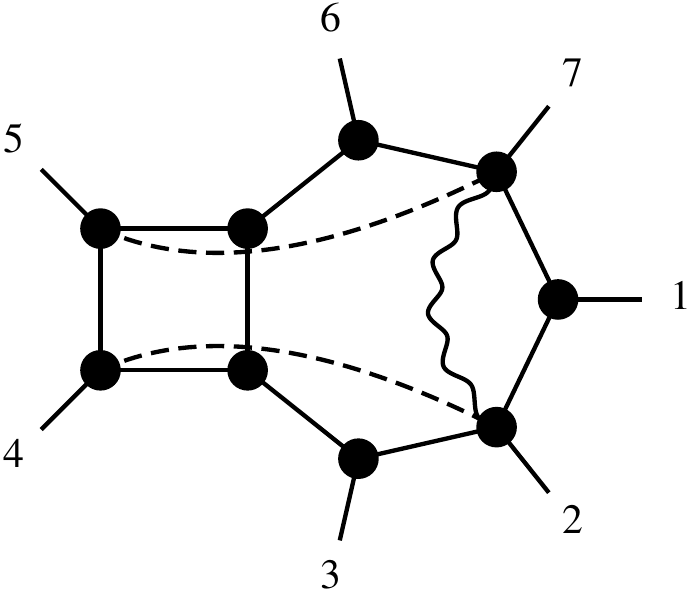}} & \multirow{2}{*}{\mbox{$  -1$}} &  \raisebox{-1.25cm}{\includegraphics[scale=0.38]{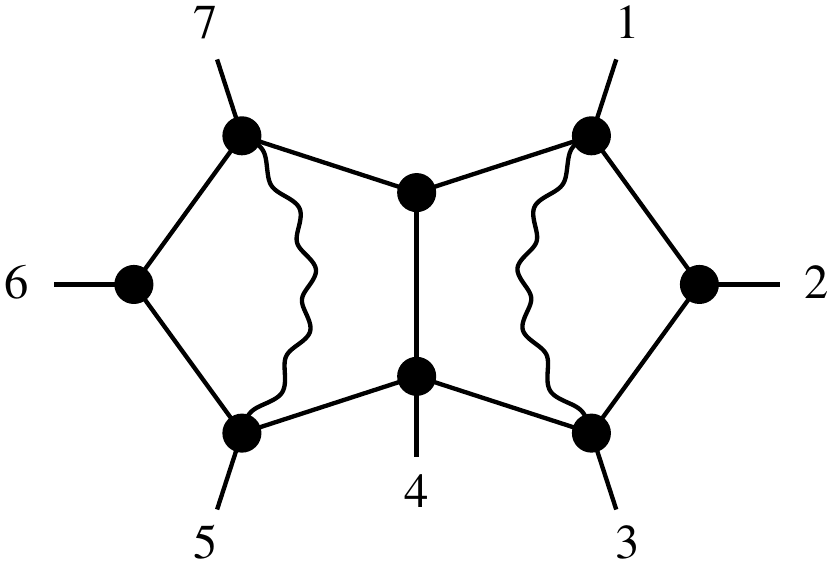}} \\&  \mbox{$  \ab{3456}^{2} \ab{AB24} \ab{AB57} \ab{AB|(671)\!\cap\!(123)}  $} &&  \mbox{$\ab{1357} \ab{AB|(712)\!\cap\!(234)} \ab{CD|(456)\!\cap\!(671)}  $} \\
\hline\multirow{2}{*}{\mbox{$  -1$}}  &  \raisebox{-1.25cm}{\includegraphics[scale=0.4]{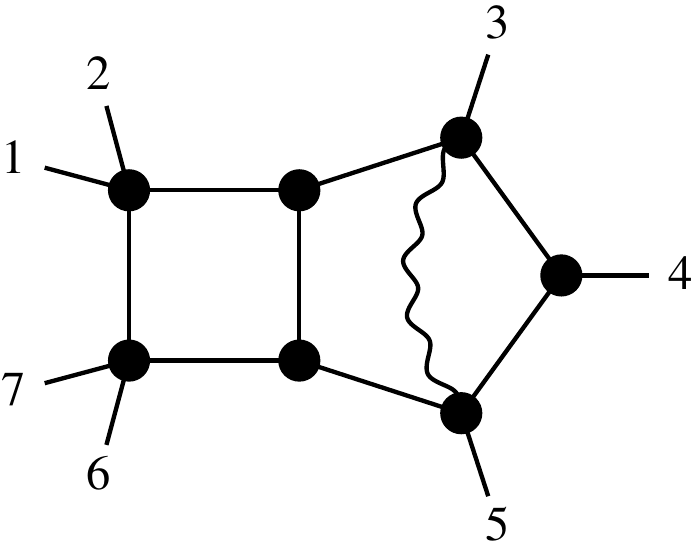}} & \multirow{2}{*}{\mbox{$  -(1 - g^{3})$}} &  \raisebox{-1.25cm}{\includegraphics[scale=0.375]{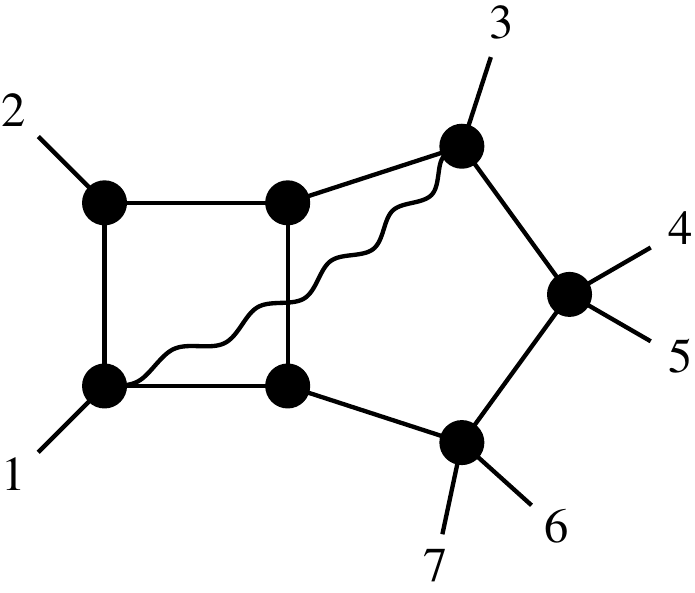}} \\&  \mbox{$  \ab{3562} \ab{3571} \ab{AB|(234)\!\cap\!(456)}  $} &&  \mbox{$\ab{3561} \ab{3712} \ab{AB|(234)\!\cap\!(712)}  $} \\
\hline\multirow{2}{*}{\mbox{$  1 - g^{6} r$}}  &  \raisebox{-1.25cm}{\includegraphics[scale=0.4]{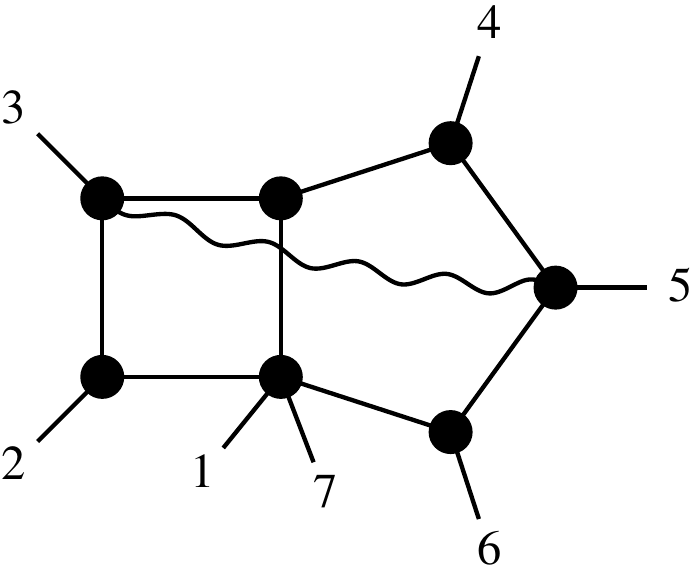}} & \multirow{2}{*}{\mbox{$  1$}} &  \raisebox{-1.25cm}{\includegraphics[scale= 0.375]{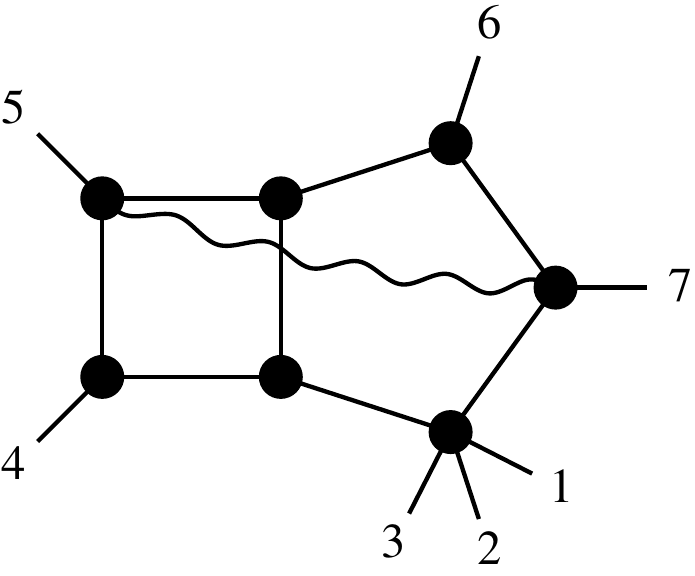}} \\&  \mbox{$  \ab{4123} \ab{5673} \ab{AB|(234)\!\cap\!(456)}  $} &&  \mbox{$\ab{6345} \ab{7345} \ab{AB|(456)\!\cap\!(671)}  $} \\
\hline\multirow{2}{*}{\mbox{$  -(1 - g^{6} r)$}}  &  \raisebox{-1.25cm}{\includegraphics[scale=0.4]{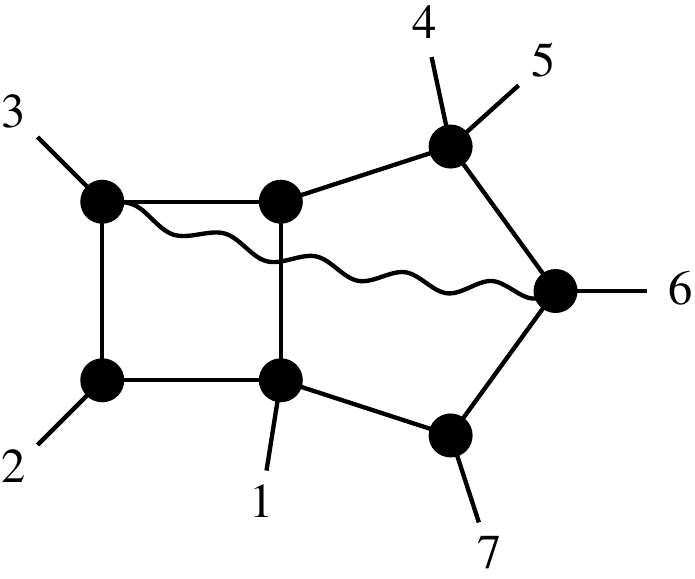}} & \multirow{2}{*}{\mbox{$  -(1 - g r P)$}} &  \raisebox{-1.25cm}{\includegraphics[scale=0.4]{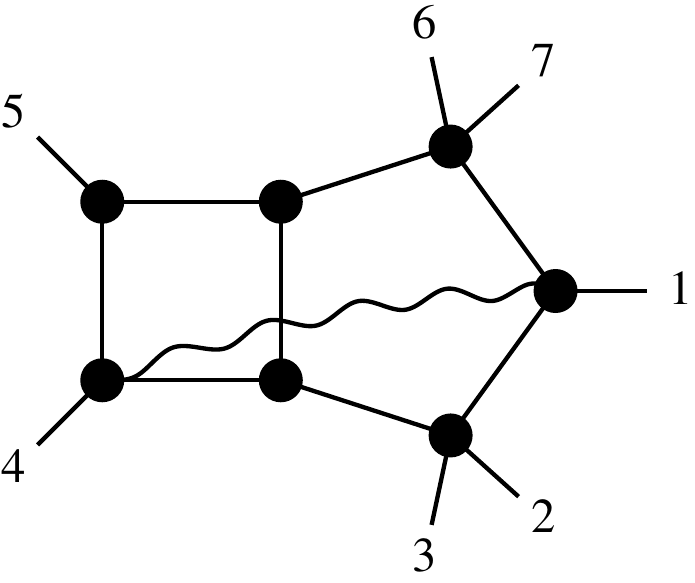}} \\&  \mbox{$  \ab{4123} \ab{6713} \ab{AB|(234)\!\cap\!(567)}  $} &&  \mbox{$\ab{6145} \ab{6345} \ab{AB|(712)\!\cap\!(345)}  $} \\
\hline\multirow{2}{*}{\mbox{$  -(1 - g r)$}}  &  \raisebox{-1.25cm}{\includegraphics[scale=0.4]{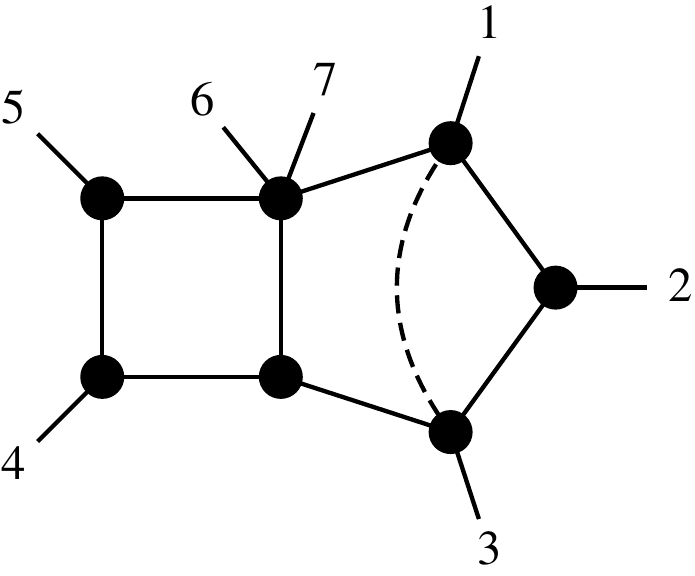}} & \multirow{2}{*}{\mbox{$  1 - g r$}} &  \raisebox{-1.25cm}{\includegraphics[scale= 0.375]{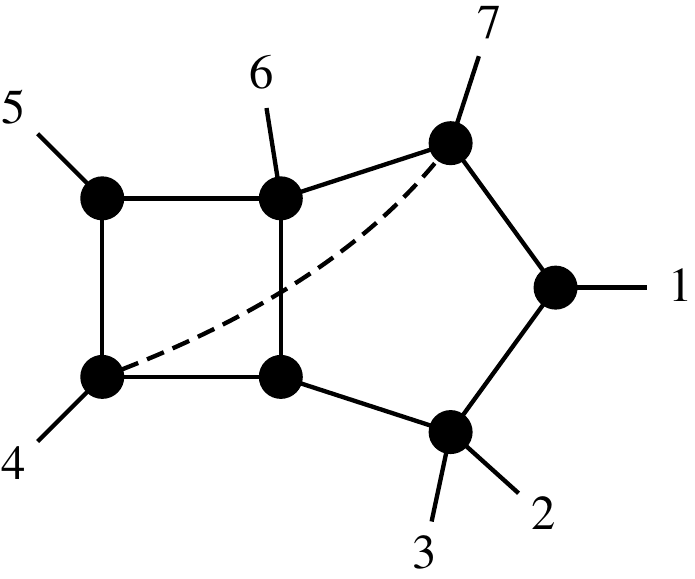}} \\&  \mbox{$  \ab{1247} \ab{2345} \ab{3456} \ab{AB13}  $} &&  \mbox{$\ab{1345} \ab{3456} \ab{7126} \ab{AB74}  $} \\
\hline\multirow{2}{*}{\mbox{$  1 + g r$}}  &  \raisebox{-1.25cm}{\includegraphics[scale=0.4]{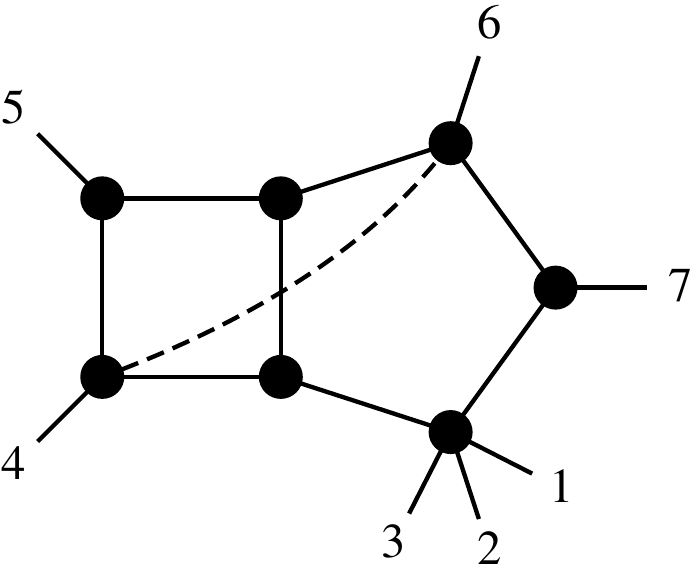}} &&  \\&  \mbox{$  \ab{6345} \ab{6715} \ab{7345} \ab{AB64}  $} &&  \\
\hline
\end{tabular}}\end{table}
\begin{table}[h!]\centering\vspace{-1.5cm}\caption{Coefficients of residue $(7)(3)=[1\,2\,4\,5\,6]$}\mbox{\hspace{-0.0cm}\scriptsize\begin{tabular}{|l@{\hspace{-0.25cm}}c|l@{\hspace{-.25cm}}c|}
\hline\multirow{2}{*}{\mbox{$  1 + g^{6} r$}}  &  \raisebox{-1.25cm}{\includegraphics[scale= 0.38]{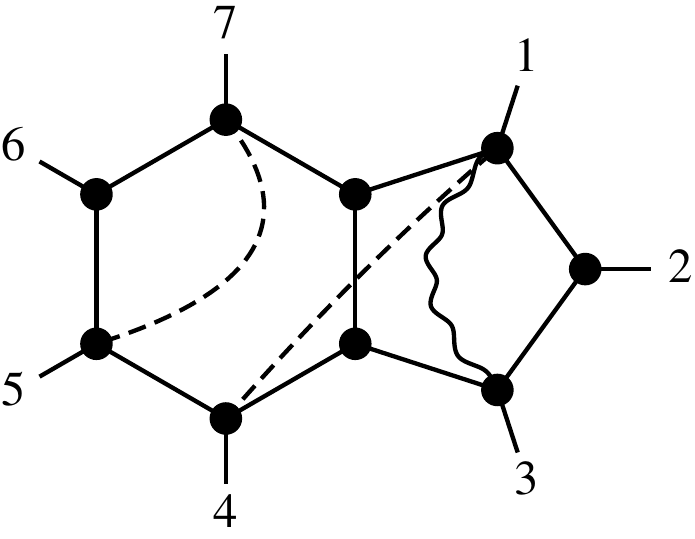}} & \multirow{2}{*}{\mbox{$  -1$}} &  \raisebox{-1.25cm}{\includegraphics[scale= 0.38]{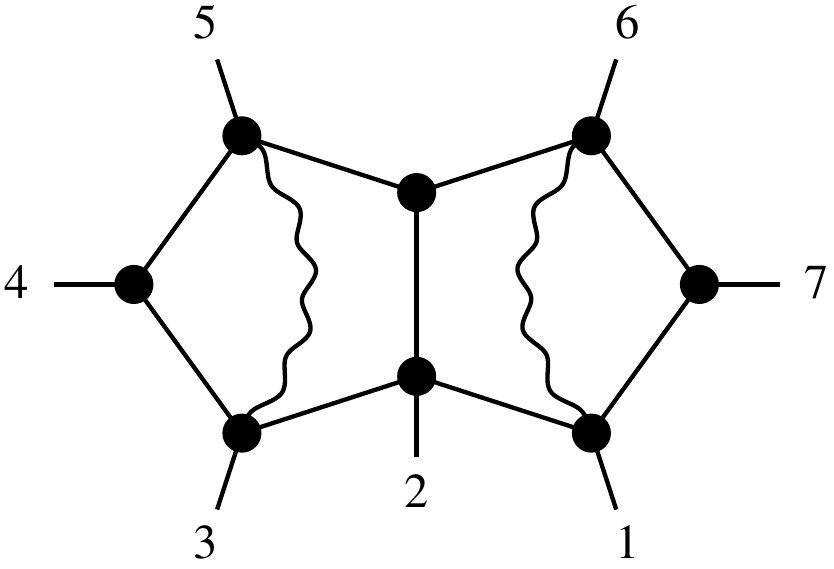}} \\&  \mbox{$  \ab{3456} \ab{6371} \ab{AB|(712)\!\cap\!(234)} \ab{CD41} \ab{CD57}  $} &&  \mbox{$\ab{6135} \ab{AB|(567)\!\cap\!(712)} \ab{CD|(234)\!\cap\!(456)}  $} \\
\hline\multirow{2}{*}{\mbox{$  -(1 + g^{3})$}}  &  \raisebox{-1.25cm}{\includegraphics[scale= 0.38]{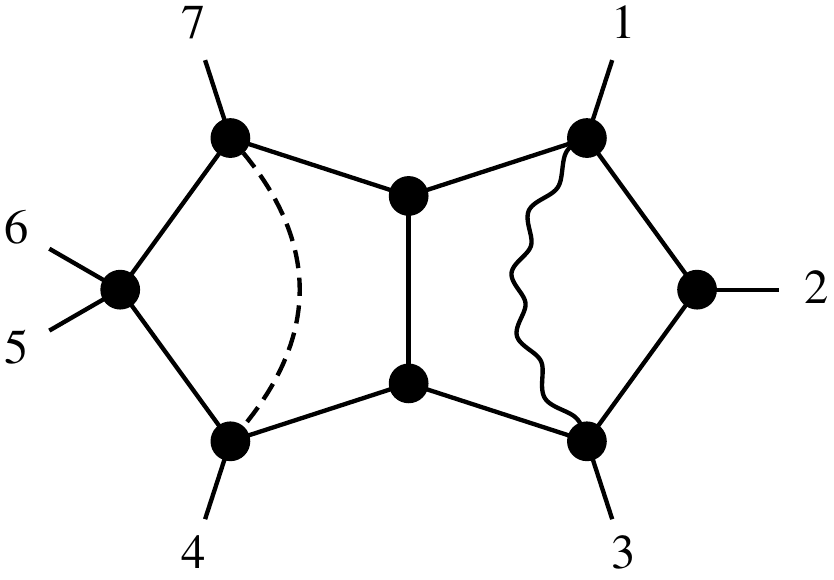}} & \multirow{2}{*}{\mbox{$  -(1 + g^{2} P r)$}} &  \raisebox{-1.25cm}{\includegraphics[scale= 0.38]{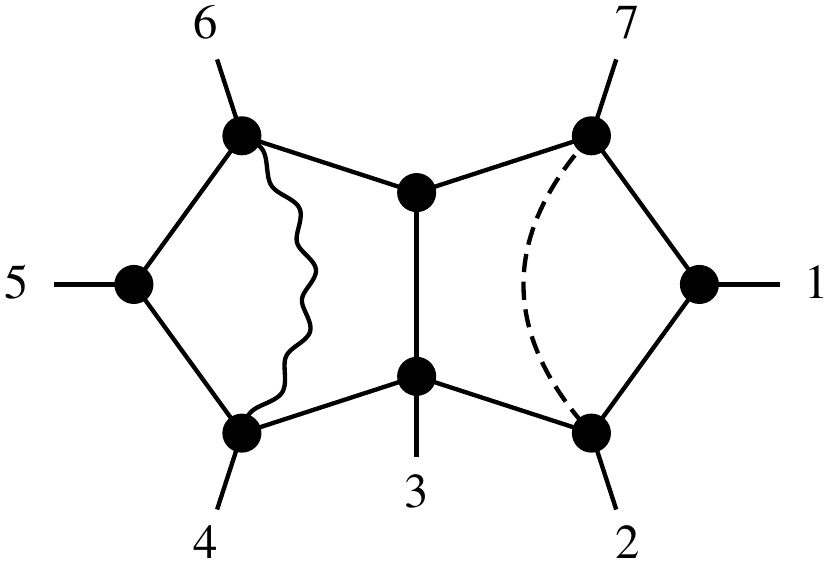}} \\&  \mbox{$  \ab{4513} \ab{6713} \ab{AB|(712)\!\cap\!(234)} \ab{CD47}  $} &&  \mbox{$\ab{1236} \ab{7146} \ab{AB|(345)\!\cap\!(567)} \ab{CD72}  $} \\
\hline\multirow{2}{*}{\mbox{$  1$}}  &  \raisebox{-1.25cm}{\includegraphics[scale= 0.38]{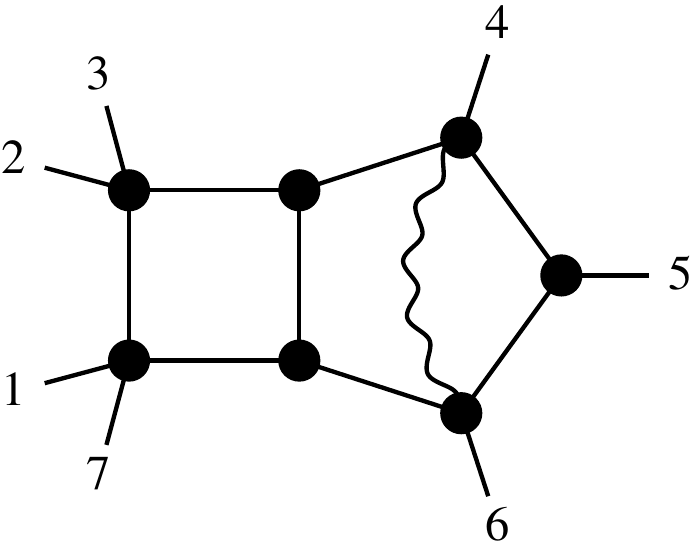}} & \multirow{2}{*}{\mbox{$  -(1 - g^{6} r)$}} &  \raisebox{-1.25cm}{\includegraphics[scale=0.38]{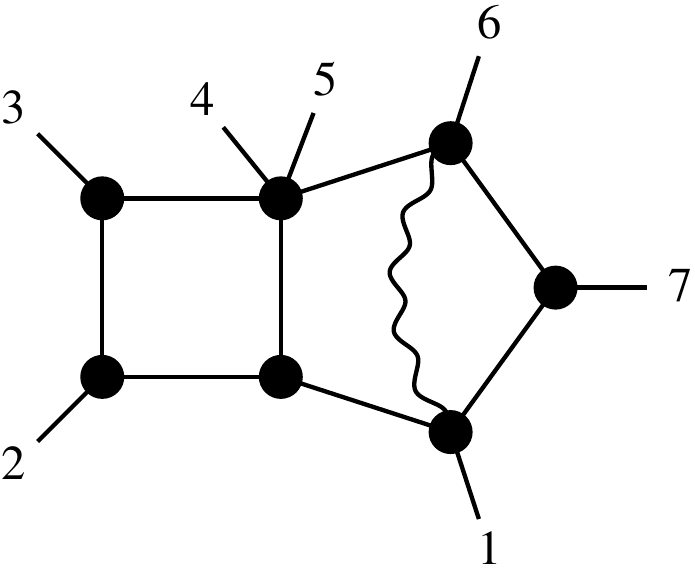}} \\&  \mbox{$  \ab{4612} \ab{4673} \ab{AB|(345)\!\cap\!(567)}  $} &&  \mbox{$\ab{1234} \ab{6123} \ab{AB|(567)\!\cap\!(712)}  $} \\
\hline\multirow{2}{*}{\mbox{$  -1$}}  &  \raisebox{-1.25cm}{\includegraphics[scale= 0.38]{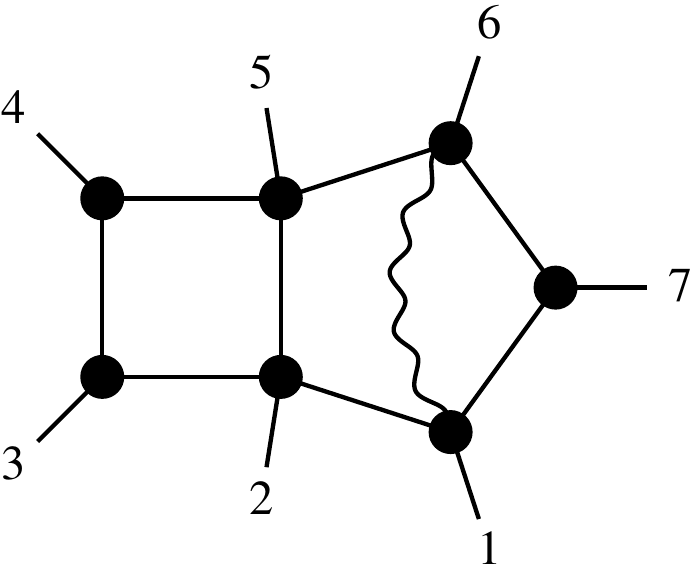}} & \multirow{2}{*}{\mbox{$  1 + g^{2}$}} &  \raisebox{-1.25cm}{\includegraphics[scale=0.38]{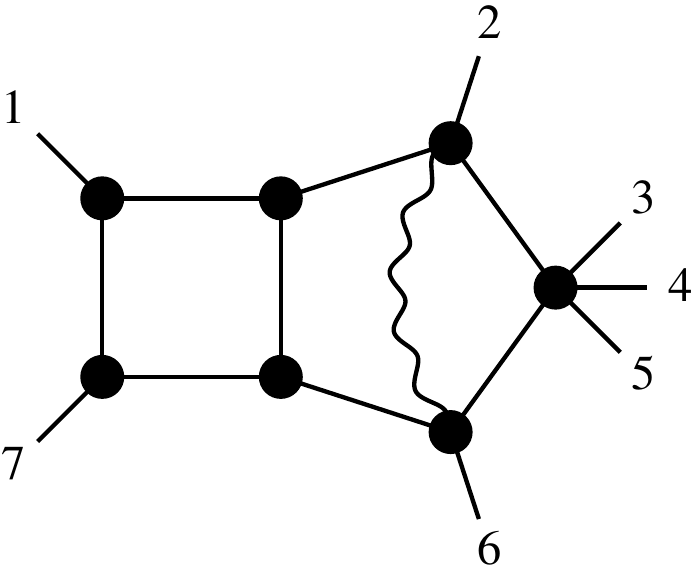}} \\&  \mbox{$  \ab{2345} \ab{6134} \ab{AB|(567)\!\cap\!(712)}  $} &&  \mbox{$\ab{2671}^{2} \ab{AB|(123)\!\cap\!(567)}  $} \\
\hline\multirow{2}{*}{\mbox{$  1 + g^{6} r$}}  &  \raisebox{-1.25cm}{\includegraphics[scale= 0.38]{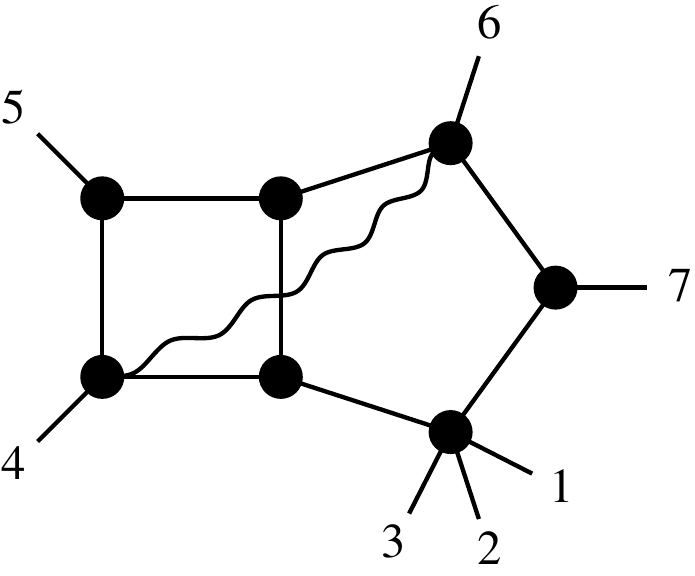}} & \multirow{2}{*}{\mbox{$  -(1 - g^{3})$}} &  \raisebox{-1.25cm}{\includegraphics[scale=0.38]{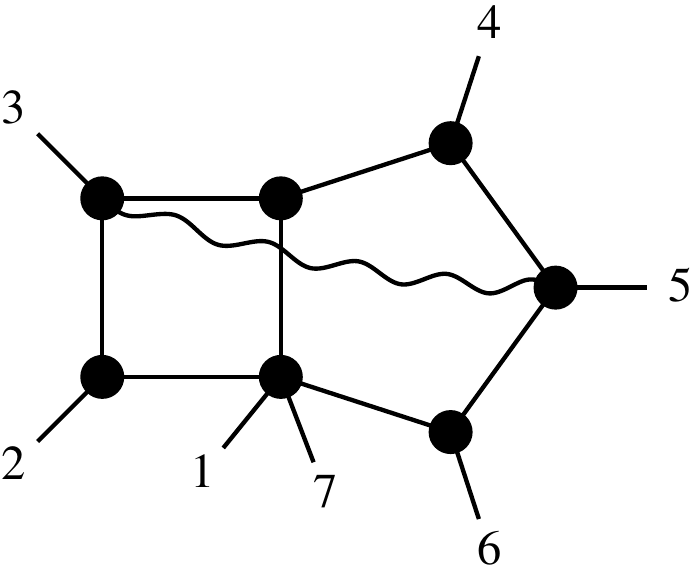}} \\&  \mbox{$  \ab{6345} \ab{6714} \ab{AB|(567)\!\cap\!(345)}  $} &&  \mbox{$\ab{4123} \ab{5673} \ab{AB|(234)\!\cap\!(456)}  $} \\
\hline\multirow{2}{*}{\mbox{$  1 + g^{6} r$}}  &  \raisebox{-1.25cm}{\includegraphics[scale= 0.38]{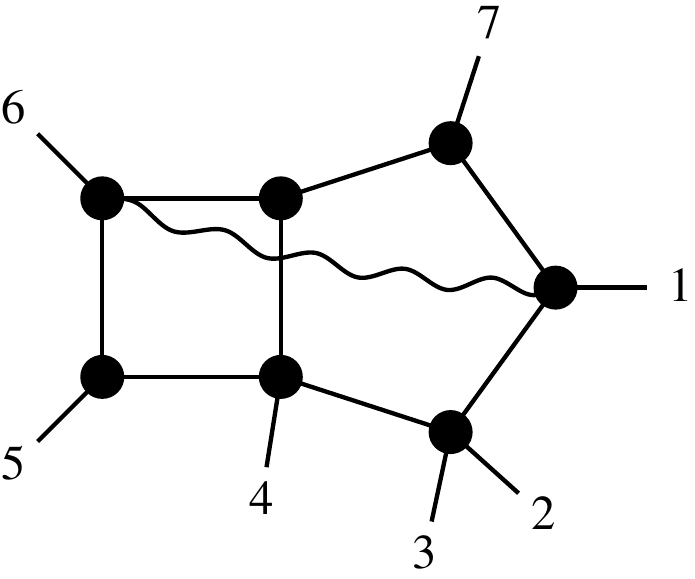}} & \multirow{2}{*}{\mbox{$  -(1 - g^{2} r)$}} &  \raisebox{-1.25cm}{\includegraphics[scale= 0.38]{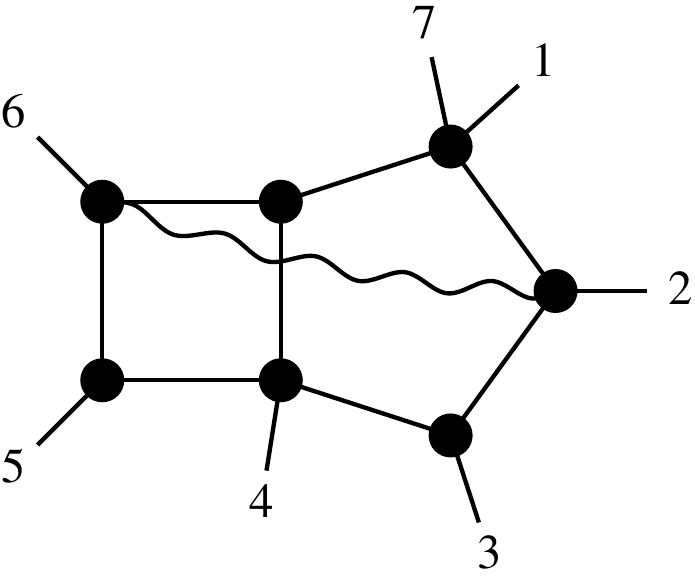}} \\&  \mbox{$  \ab{1346} \ab{7456} \ab{AB|(567)\!\cap\!(712)}  $} &&  \mbox{$\ab{2346} \ab{7456} \ab{AB|(567)\!\cap\!(123)}  $} \\
\hline\multirow{2}{*}{\mbox{$  -(1 - g^{4} r)$}}  &  \raisebox{-1.25cm}{\includegraphics[scale= 0.38]{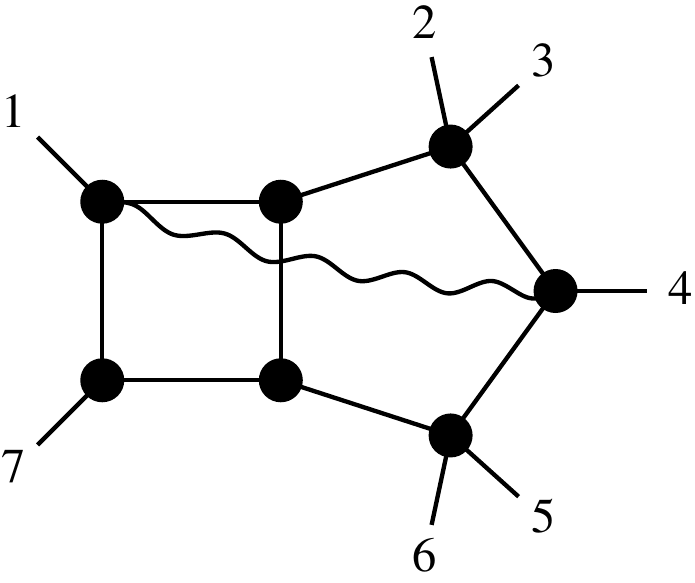}} & \multirow{2}{*}{\mbox{$  -(1 + g^{2})$}} &  \raisebox{-1.25cm}{\includegraphics[scale= 0.38]{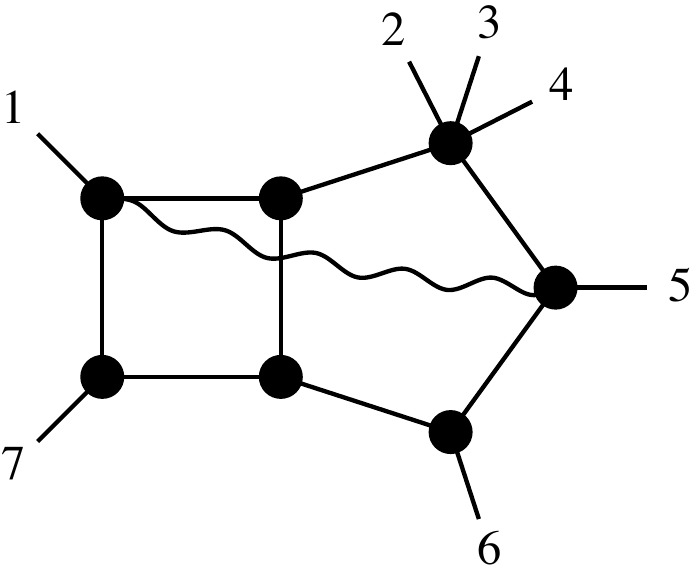}} \\&  \mbox{$  \ab{2671} \ab{4671} \ab{AB|(712)\!\cap\!(345)}  $} &&  \mbox{$\ab{2671} \ab{5671} \ab{AB|(712)\!\cap\!(456)}  $} \\
\hline\multirow{2}{*}{\mbox{$  -(1 + g^{2})$}}  &  \raisebox{-1.05cm}{\includegraphics[scale= 0.38]{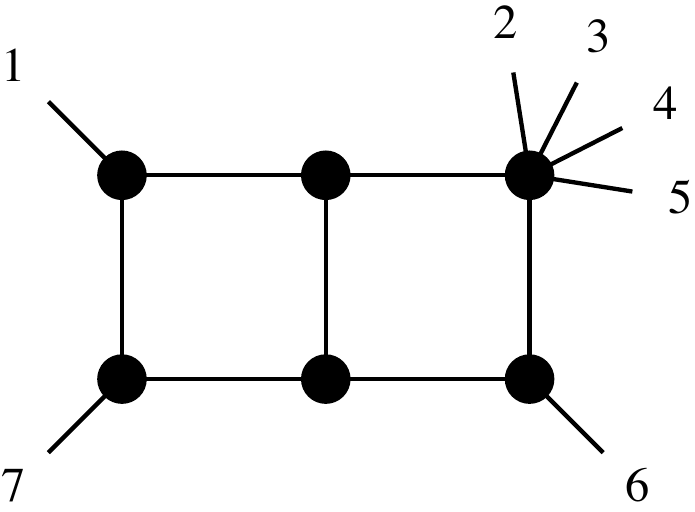}} & &  \\&  \mbox{$  \ab{6712}^{2} \ab{6715}  $} &&  \\
\hline
\end{tabular}}\end{table}

\begin{table}[h!]\centering\caption{Coefficients of $A_{\mathrm{tree}}$ (in addition to the 2-loop MHV amplitude)}\mbox{\hspace{-0.0cm}\scriptsize\begin{tabular}{|l@{\hspace{-0.25cm}}c|l@{\hspace{-.25cm}}c|}
\hline\multirow{2}{*}{\mbox{$  -1 $}}  &  \raisebox{-1.25cm}{\includegraphics[scale= 0.38]{732loopFig12_3.pdf}} & \multirow{2}{*}{\mbox{$ -(1-g^4 r)$}} &  \raisebox{-1.25cm}{\includegraphics[scale= 0.38]{732loopFig12_10.pdf}} \\&  \mbox{$  \ab{5124} \ab{AB|(456)\!\cap\!(712)} \ab{CD|(123)\!\cap\!(345)}  $} && \mbox{$\ab{1456} \ab{4567} \ab{AB|(712)\!\cap\!(345)}  $} \\
\hline\multirow{2}{*}{\mbox{$  1 $}}  &  \raisebox{-1.25cm}{\includegraphics[scale= 0.38]{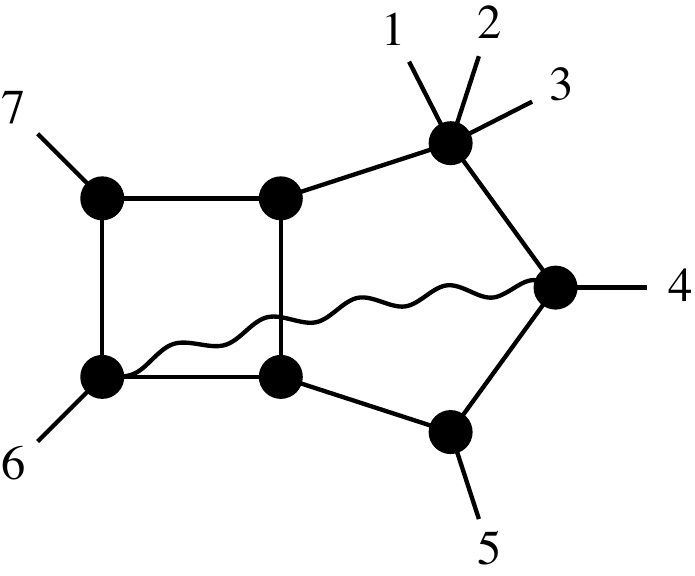}} &&\\
&\mbox{$\ab{1467} \ab{1567} \ab{AB|(345)\!\cap\!(567)}$}&&\\\hline
\end{tabular}}\end{table}

~\newpage~\newpage
~\vspace{0.2cm}~

%

\providecommand{\href}[2]{#2}\begingroup\raggedright\endgroup

\end{document}